\documentclass[stsy,nonblindrev]{informs-stsy_2}
\usepackage{graphicx} 

\usepackage{natbib}
 \bibpunct[, ]{(}{)}{,}{a}{}{,}%
 \def\bibfont{\small}%

\usepackage{algorithm,algorithmic}
\usepackage{bbm}
\usepackage{tikz}

\usepackage{amsfonts,mathrsfs}
\usepackage{hyperref}
\usepackage{mathtools}
\usepackage{graphicx}
\usepackage{subfig}
\usepackage{enumitem}

\usepackage{cleveref}

\TheoremsNumberedThrough 
\newtheorem{Lemma}{Lemma}
\newtheorem{Theorem}{Theorem}
\newtheorem{Definition}{Definition}
\newtheorem{Proposition}{Proposition}
\newtheorem{Remark}{Remark}
\newtheorem{Corollary}{Corollary}

\usepackage{xcolor}
\renewcommand{\textcolor}[2]{#2}
\renewcommand{\color}[1]{}

\DeclareMathOperator*{\E}{\mathbb{E}}
\let\Pr\relax
\DeclareMathOperator*{\Pr}{\mathbb{P}}
\DeclareMathOperator*{\Bc}{\bigm|}
\DeclarePairedDelimiter{\ceil}{\lceil}{\rceil}

\renewcommand{\algorithmicreturn}{\textbf{Input:}}

\begin{document}


\TITLE{Learning to Admit Optimally in an $M/M/k/k+N$ Queueing System with Unknown Service Rate}

\ARTICLEAUTHORS{%
\AUTHOR{Saghar Adler}
\AFF{TikTok, \EMAIL{saghar.adler@gmail.com}} 
\AUTHOR{Mehrdad Moharrami}
\AFF{University of Iowa, \EMAIL{mehrdad-moharrami@uiowa.edu}}
\AUTHOR{Vijay G. Subramanian}
\AFF{University of Michigan, \EMAIL{vgsubram@umich.edu}}
} 

\ABSTRACT{Motivated by applications of the Erlang-B blocking model {\color{blue} and the extended $M/M/k/k+N$ model that allows for some queueing,} beyond communication networks 
to sizing and pricing in production, 
messaging, 
and app-based parking systems,
we study admission control for {\color{blue} such systems}
with unknown {\color{blue} service rate}. 
In our model, a dispatcher {\color{blue} either admits every arrival into the system (when there is room) or blocks it.} 
Every served job yields a fixed reward but incurs a per unit time holding cost {\color{blue}which includes the waiting time in the queue to get service if there is any.} 
We aim to design a dispatching policy that maximizes the long-term average reward 
by observing arrival times and system state at arrivals, a realistic {\color{blue}decision-event driven} sampling of such systems. The dispatcher observes neither service times nor departure epochs, which excludes the use of reward-based reinforcement learning approaches. 
We develop our learning-based dispatch scheme as a parametric learning problem \textit{a'la} self-tuning adaptive control. In our problem, certainty equivalent control switches between \textit{always admit if room} (explore infinitely often), and \textit{never admit} (terminate learning), so at judiciously chosen times we avoid the \text{never admit} recommendation. 
We prove that
our proposed policy asymptotically converges to the optimal policy and present finite-time regret guarantees. The extreme contrast in the control policies shows up in our regret bounds for different parameter regimes: constant 
in one versus logarithmic 
in another.
}


\maketitle

\section{Introduction}
\label{sec:introduction}


Queueing systems are widely applicable models used to study resource allocation problems in communication networks, distributed computing systems,  semiconductor manufacturing, supply chains, and many other {\color{blue}dynamical} systems. 
Queueing models are analyzed under various system information settings, but a common assumption is that the core system parameters like arrival rates, service rates and distributions are available to the system designer---see [\cite{srikant2013communication,harchol2013performance}].  However, there are many applications where these parameters are unknown, and the designer needs to learn them to be able to 
optimally assign jobs to the servers or block them. For example, the service rate of every server in large-scale server farms may be unknown, or the treatment times in hospitals may be unpredictable and time-varying. 


{\color{blue} The focus of this paper is learning based admission control in an $M/M/k/k+N$ system [\cite{kelly2011reversibility,srikant2013communication,asmussen2003applied}] with $N\geq 0$ but finite; we refer to this system as an Erlang-B with finite waiting room system. This system is widely applied for sizing related questions in telecommunications, network systems, call centers, manufacturing systems, healthcare, and transportation systems. The traditional use of these systems has been in telecommunication applications---for sizing and analyzing voice and circuit-switched systems, i.e., loss systems [\cite{kelly2011reversibility,srikant2013communication}]. In addition, it has also been used to study packetized data systems [\cite{suter1999buffer,roberts2004internet}], and multiple-access schemes in wireless networks (\cite{marbach2011asynchronous}). These models are also used in sizing, managing, and analyzing call-center systems [\cite{gans2003telephone}], where the finite queue represents the limited capacity to handle incoming calls. Such models are also used in production systems, where the finite buffer models the limited inventory for ongoing jobs [\cite{ammar1980modelling,berman1982efficiency,gershwin1983modeling,dallery1988efficient,gershwin2018future}]. Finally, in transportation systems [\cite{restrepo2009erlang}] and healthcare [\cite{green2002many,fomundam2007survey,de2007modeling}], these models have been used to answer sizing questions related to ambulances, transit vehicles or hospital beds.}

Motivated by these 
applications and to highlight challenges in learning-based optimal control,
we study optimal admission control in an Erlang-B {\color{blue} with finite waiting room} queueing system with exponentially distributed service times, and {\color{blue}known arrival rate, but unknown service rate}, denoted by $\lambda$ and $\mu$, {\color{blue}respectively,}
with the goal of designing an optimal learning-based dispatching policy. 
At every arrival, the dispatcher can accept or block the arrival.  
Accepted jobs incur a {\color{blue}holding} cost $c$ per unit time, 
and yield a fixed reward $R$ {\color{blue}(upon completion of service)}. {\color{blue}To highlight learning-related issues in a simple manner, we will consider two specific alternatives---either block all the arrivals or accept all the arrivals subject to available room. When the waiting room $N=0$, then the optimal admission control algorithm which can use the system occupancy information only needs to choose between these two options; this is not true when $N>0$ where the dispatcher can use the system occupancy too. However, such a decision choice helps with determining whether such a service is worth it or not. Then,} assuming 
that the service rate $\mu$ is known, the dispatcher  
can maximize its expected reward using a threshold policy: if the service rate exceeds a value $\mu^*$, all arrivals are admitted subject to availability; otherwise, 
all arrivals are rejected, and when the service rate equals $\mu^*$, the dispatcher is indifferent between admitting or rejecting arrivals. 

A key aspect of our problem setting is that the information available to the dispatcher consists only of the inter-arrival times and the number of busy servers at each arrival, as the system is sampled at arrivals. Contrarily, the service rate, departure {\color{blue}epochs}, and service times are not known to the dispatcher. 
Hence, the dispatcher cannot form a direct estimate of the service rate (e.g., by taking an empirical average of the observed service times)
to then choose its policy, and instead has to use the queueing dynamics to estimate the service time for policy determination. This facet of the problem brings it closer to practice {\color{blue}(since continuous monitoring is memory and processing intensive)}, but also complicates the analysis. Based on this information structure, our focus 
is to design an optimal policy that maximizes the long-term average reward.

We study the problem of learning the service rate in the framework of parametric learning of a stochastic dynamical system. Specifically, consider a stochastic system governed by parameter $\theta$: 
\begin{equation} \label{eq:MDP}
    X_{t+1}=\mathscr{F}_t(X_t,U_t,W_t;\theta), \quad t=0,1, \ldots
\end{equation}
where $X_t \in \mathcal{X}$, $U_t \in \mathcal{U}$,  $W_t \in \mathcal{W}$ are the state of the system, control input, and noise at time $t$
and $\mathscr{F}_t$ is any measurable function. Further, $\theta \in \Theta$ is a fixed but unknown parameter, and the initial state and noise process are 
mutually independent. 
In line with the literature, we study a system where
our controller \textit{perfectly observes} the state $X_t$ and uses its history of observations to choose the control $U_t$.    
For a specified reward function $r_t (x,u)$ for $(x,u) \in \mathcal X \times \mathcal U$,
the objective is to maximize the long-term reward. We also assume that the optimal policy $\mathscr{G}^*(.;\theta)$ is \textit{known} for each $\theta \in \Theta $.

To achieve the optimization objective whilst learning the unknown parameter $\theta$, an adaptive control law is applied: using past 
observations $X_{1:t}$, an estimate $\hat{\theta}_{t+1}$ is formed, and then by {\color{blue}the certainty equivalent control principle}, the optimal policy according to $\hat{\theta}_{t+1}$, or $\mathscr{G}^*(.;\hat{\theta}_{t+1})$,  is applied.
One approach to form the estimate $\hat{\theta}_{t+1}$ is to use the maximum likelihood estimate (MLE). \cite{mandl1974estimation} prove that under identifiability, the MLE converges to the true parameter. 
When these conditions do not hold, 
\cite{kumar1982new,kumar1982optimal} use reward bias-based exploration schemes to ensure asymptotic optimality.
Our problem fits the above paradigm: the system 
state $X_t$ is the number of busy servers at time $t$ with the dispatcher observing the (continuous-time) system state at arrivals, and the unknown parameter is the service rate $\mu$, so $\Theta= \mathbb{R}_+$~\footnote{More generally, we can take both the arrival and service rates, $\lambda$ and $\mu$, to be the unknown parameters.}. Using an adaptive control law with {\color{blue}(necessitated)} forced exploration, we propose a dispatching policy to maximize the long-term average reward. Our main analysis-related contributions are:\\
\textbf{1. Asymptotic optimality.} 
We prove the convergence of our learning-based policy to the optimal policy. 
{\color{blue}We first focus on the Erlang-B queueing system, i.e., an $M/M/k/k+N$ system with $N=0$,---see  \Cref{subsec:asymp_optimal_multi_server}---, where using an intricate argument based on martingale sequences, we establish asymptotic optimality for our learning rule. In \Cref{subsec:asymp_optimal_multi_server_MB}, we show that the argument and the result generalize to the $M/M/k/k+N$ system with $N>0$.}\\
\textbf{2. Finite-time performance analysis.} {\color{blue} Next we characterize the finite-time regret of our learning in the two distinct service rate regimes relevant to our system. Once again, we start by focusing on the Erlang-B queueing system---see \Cref{subsec:regret_analysis_multi_server}. In the high service-rate regime, we show finite regret, and in the low service-rate regime, the exploration done by our policy leads to a regret upper bound that scales as $\log(n)$, where $n$ is the number of arrivals. The analysis  for the multi-server setting is based on Doob's decomposition 
and concentration inequalities for martingale sequences. Further, on \Cref{subsec:regret_analysis_multi_server_MB}, using the same proof methodology, we show that the results generalize to the $M/M/k/k+N$ system with $N>0$.}

	\begin{figure*}[t]
	\centering
	\subfloat[$\mu=2.05, \mu \in (c/R,+\infty)$  \label{fig:compare_diff_f_a}]{\includegraphics[width=0.4\linewidth]{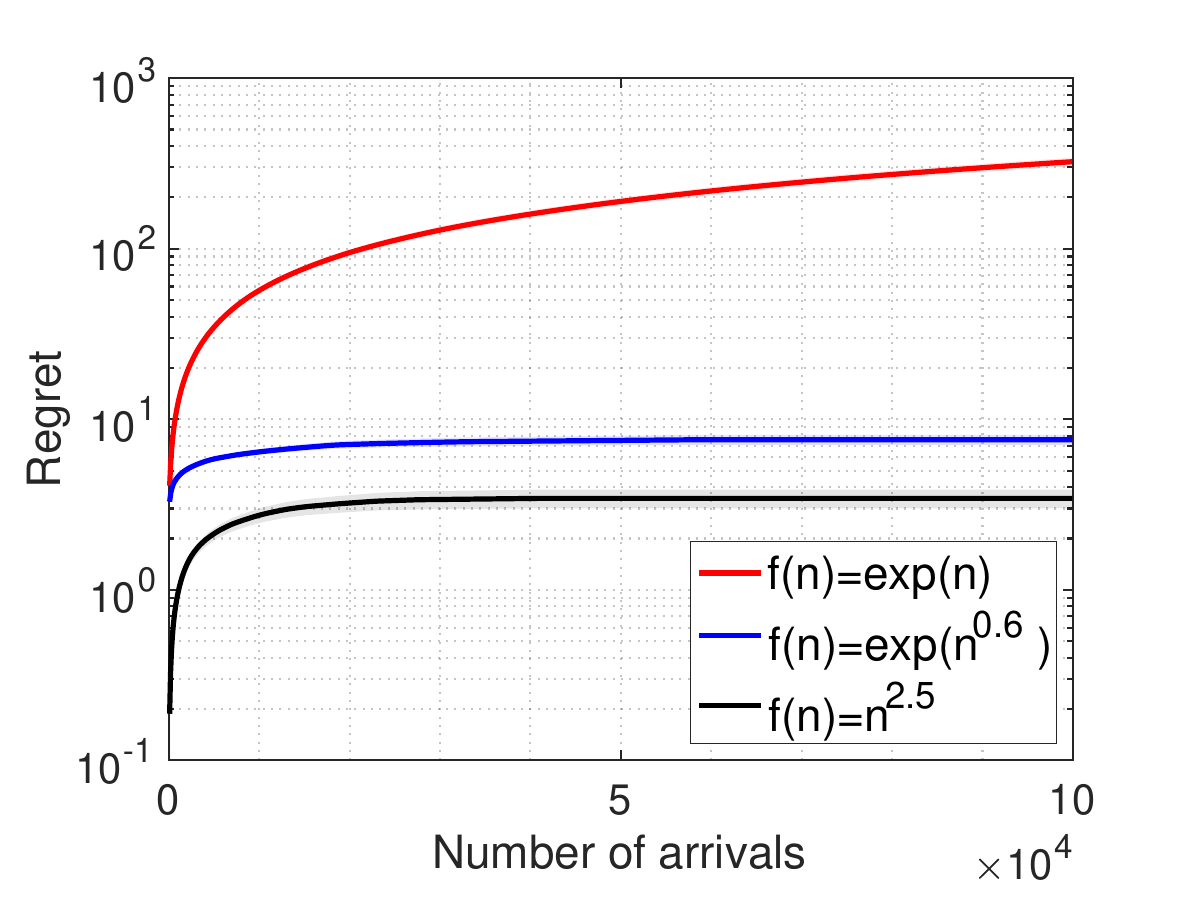}} \hspace{1cm}
	\subfloat[$\mu=1.05, \mu \in (0, c/R)$  \label{fig:compare_diff_f_b}]{\includegraphics[width=0.4\linewidth]{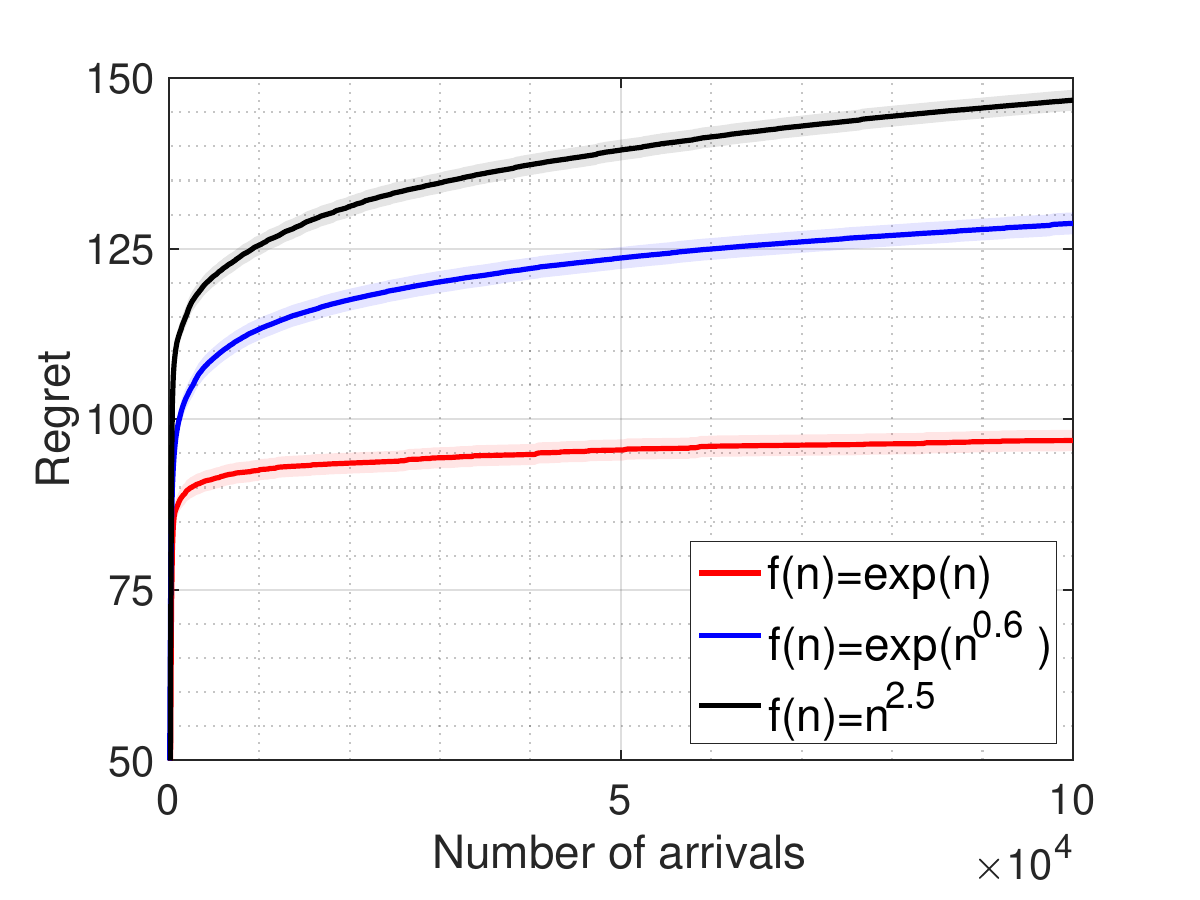}}
	\caption{Comparison of regret performance of \Cref{alg1} for different functions $f(n)$ in a 5 server system with $\lambda=5$ and $c/R=1.3$. 
	The shaded region indicate the  $\pm \sigma$ area of mean regret. 
	}
	 \label{fig:compare_diff_f}
	\end{figure*}

{\color{blue}
\begin{Remark}\label{rem:singleserver}
When the number of servers $k=1$ and $N=0$---that is, an Erlang-B system with one server---, then our analysis is particularly simple as there is an underlying independence structure which leads to a simpler analysis than using an elaborate martingale argument---asymptotic optimality using the strong law of large numbers, and finite-time regret using simpler concentration inequalities. See Appendix~\Cref{app:Appendix_single_server} for details.
\end{Remark}
}

We end by contrasting our work
with the literature on learning in stochastic dynamical systems.
We study an example of a parametric learning problem for which we do not expect a single policy to achieve minimum regret in all regions of the parameter space. Whereas we don't have an explicit proof of such a claim, the contrasting behavior an optimal adaptive control scheme must exhibit in different parameter regimes---quickly converging to always admitting arrivals if room versus quickly rejecting all arrivals---gives credence to the claim. We discuss the above point
in \Cref{fig:compare_diff_f}, which depicts the 
performance of our algorithm for functions 
$f(n) \in \{n^{2.5}, \exp\left(n^{0.6}\right), \exp(n)\}$ where $1/f(n)$ is proportional to the (forced) exploration probability.
For $f(n)=n^{2.5}$, exploration is employed  aggressively, causing better performance for  
$\mu \in (c/R,+\infty)$, and higher regret in the other regime.
Conversely, when $f(n)=\exp(n)$, aggressive exploitation is enforced, 
leading to {the opposite behavior.}
For $\mu \in (c/R, +\infty)$, we show finite regret for $f(n)\in\{n^{2.5},\exp\left(n^{0.6}\right)\}$ in \Cref{subsec:regret_analysis_multi_server}, but finite regret is not guaranteed for $f(n)=\exp(n)$ in our analysis.  In  \Cref{subsec:regret_analysis_multi_server}, when $\mu \in (0,c/R)$, we establish an $O(\log^{5/3}(n))$ regret bound for $f(n)=\exp\left(n^{0.6}\right)$. Similar arguments lead to a $O(\log(n))$ upper bound for $f(n)=\exp(n)$ in the same regime. From this discussion, we expect big differences in performance of any algorithm based on the parameter regime.
Based on our numerical results, we also conjecture that for $\mu \in (0,c/R)$, there is an $\Omega(\log(n))$ regret lower bound. This is consistent with the lower bound on the asymptotic growth of regret from the literature on learning in unknown stochastic systems under the assumption that the transition kernels of the underlying controlled Markov chains are strictly bounded away from $0$; see  \cite{agrawal1989asymptotically,graves1997asymptotically}.

Furthermore, our simulation results in \Cref{sec:experimental results} 
provide evidence that depending on the relationship between the arrival rate and the service rate, sampling our continuous-time system at a faster rate than the arrivals could reduce the regret.  
We also show that subtle differences in  
variable updates in the learning scheme have a substantial impact on the regret achieved. Thus, the choice of the trade-off of regret between the different parameter regimes determines the learning scheme.

\subsection{Related Work}

\textbf{Adaptive control.} 
The self-tuning adaptive control literature studies asymptotic learning in the parametric or non-parametric 
version of the problem described in \eqref{eq:MDP}, and the study was initiated by Mandl. 
\cite{mandl1974estimation} showed that the MLE converges 
to the true parameter 
under an identifiability condition. {Since then, the adaptive control problem has been vastly studied in great generality; see  \cite{borkar1979adaptive,kumar1982new,kumar1982optimal,agrawal1989asymptotically,graves1997asymptotically,gopalan2015thompson}.
} 
Learning in queueing systems is one of the applications in this literature; see \cite{lai1995machine,kumar2015stochastic}. 

A core assumption in the above literature  is that 
the transition kernels of the underlying controlled Markov chains are strictly bounded away from $0$ and $1$, with the bound 
uniform in the parameter and the class of (optimal) policies. 
This core assumption does not hold in our problem: the controlled Markov chain found by sampling the queueing system at arrivals has drastically different behavior under the available class of policies---admit if room or never admit---, and thus the conclusions of this literature do not apply. 
Furthermore, in the above  literature,
most of the results are on asymptotic learning, and only recently, finite-time regret guarantees have been obtained. The existing finite-time regret guarantees 
are largely for certain {discrete-time} queueing systems with {geometrically distributed} service times 
and unknown parameters, which we will discuss below.

\textbf{Queueing systems.} There is a growing body of work on learning-based control in discrete-time queueing systems; see  \cite{walton2021learning}. 
\cite{krishnasamy2018learning,krishnasamy2021learning} studied a discrete-time multi-class, multi-server queueing system with unknown service rates. After imposing stability conditions, 
\cite{krishnasamy2018learning} used a forced exploration-based learning 
scheme to prove finite regret compared to the $c\mu$ rule in a system with service rates known.  
\cite{krishnasamy2021learning} used UCB and Thompson sampling-based algorithms to prove a polylogarithmic regret bound. \cite{choudhury2021job} proved an $\Tilde{O}(\sqrt T)$ regret over time horizon $T$ 
 using a queue-length agnostic randomized-routing-based algorithm for a multi-server discrete-time queueing system. 
All of these works form empirical service rate estimates
by observing 
service successes and failures.
    
\cite{stahlbuhk2021learning} studied the problem of finding the optimum server for service in a discrete-time multi-server system with unknown service rates and a single queue and proves constant regret  
by sampling service rates during idle periods. \cite{ojeda2021learning} employed generative adversarial networks to numerically learn the unknown service time distributions in a $G/G/\infty$ queuing system.   { \cite{zhong2022learning} studied  scheduling in a multi-class queue with abandonment with unknown arrival, service, and abandonment rates. 
 By using service and patience times and forming estimates of the service and abandonment rates, 
 logarithmic regret is shown 
 using an exploration-exploitation based scheme. } 
\cite{zhang2022learning,cohen2024learning} studied social-welfare maximizing admission control in an $M/M/1$ queuing system with unknown service and arrival rates; with system parameters known a threshold-based admission control scheme is optimal. 
By observing the queueing system at all times, they proposed 
a dispatching algorithm that achieves constant regret for one set of parameters, and $O(\log^{1+\epsilon}(n))$ regret for any $\epsilon > 0$ for another set of parameters  ($n$ is the number of arrivals).
 
In all of these works, all completed service times or the entire queueing processes are observed and used for learning. Such observations may not be feasible in real-world queueing systems due to increased computation and memory requirements: see \cite{stidham1985optimal,harchol2013performance}. Multi-server settings introduce other complications: to correctly identify completed service times, server assignments need to be tracked from the entire process history (even for homogeneous servers). 
In our work, observations are the (minimal) Markov state of the system at each arrival, which despite being a nonlinear function of service times, aligns better with real-world systems. In \Cref{sec:experimental results}, using simulations, we also show that the sampling of such continuous-time systems requires careful design. 

Learning-based decision-making has also been studied in inventory control and dynamic pricing with the goal of maximizing the  expected total profit (\cite{agrawal2022learning,chen2023online,jia2022online}).  Another line of work focuses on the use of pricing strategies to regulate queue sizes and studies differences between individually optimal and  socially optimal rules with known model; see \cite{Naor,chr1972individual,lippman1977individual}. These works show that an individually optimal policy has more arrivals than a socially optimal rule leading to 
congestion.

\textbf{Reinforcement learning (RL).} Recently, RL methods have been applied to queueing problems with the goal of finding the average cost optimal policy,  in both known model and cost parameter cases (\cite{dai2022queueing}), and unknown parameter cases with available rewards (\cite{massaro2019optimal}). 
These methods do not apply to our setting as we neither observe the reward sequence nor know the expected rewards: the random reward is a linear function of the service times of accepted jobs which are not observed, and the expected reward is a function of the unknown arrival and service rates. 
We only observe the system state: a nonlinear and complex 
function of the reward. {\color{blue} Similarly, other reward-based schemes used in the bandit literature~(\cite{lattimore2020bandit})---such as UCB or ``estimate and then commit with expanding horizons for commitment"---do not directly apply since the reward signal is not available. One could potentially use model knowledge (with parameters unknown) and then a combination of Poisson Arrivals See Time Averages (PASTA) and Little's law---see \cite{srikant2013communication,harchol2013performance}---to estimate the reward for each policy being used, but getting statistical performance guarantees (based on duration of use of each policy) from such estimates, especially coming from the use of PASTA, is an open question.} In contrast to the model-agnostic viewpoint in RL, we use the knowledge of the queueing dynamics to design an algorithm matched to our setting. Although RL methods do not apply to our setting, in \Cref{sec:experimental results}, we consider a fictitious setup wherein the service times are observed 
and implement
an average reward RL algorithm, R-learning (\cite{schwartz1993reinforcement,sutton2018reinforcement}), {\color{blue}as a representative of reward-based RL algorithms}.
{Despite not observing the service times,  
our policy outperforms  R-learning, 
providing evidence that model-class knowledge can be as effective as 
observing the reward signal; 
see \Cref{fig:compare_RL}. } In \Cref{fig:compare_RL}, we also compare our algorithm to a Thompson sampling-based algorithm~(\cite{gopalan2015thompson}) {\color{blue} used in the frequentist sense~(see \cite{agrawaljia2017})}, showing that our algorithm using model-class knowledge is again as effective as Thompson sampling {\color{blue} (which essentially encodes model knowledge in the prior)}. 

{\color{blue} In a recent work, \cite{weber2024reinforcement} studied the problem of admission control in an \(M/M/k/k+N\) queueing system with \(m\) job classes, assuming a known service rate but unknown arrival rates for each class. They implement an algorithm inspired by UCRL2 and, using the same cost and reward structure as ours, demonstrate a regret of \(O(\sqrt{n})\). In their setting, knowing the service rate implies that rewards are fully specified, leaving the transition kernel as the only unknown component, since arrival rates are unknown. In contrast, in our setting where the service rate is unknown, both the transition probabilities and the rewards are unknown. In addition, the work considers a continuous-time MDP formulation, which implies that the decision-maker has knowledge of the entire past of the system when making decisions. Again, in contrast, we have a much coarser information structure based on sampling the system state only on arrivals.}

\subsection{Organization}

{\color{blue} The paper is organized as follows. In \Cref{sec:PROBLEM FORMULATION}, we introduce the problem and the learning objective. \Cref{sec:Proposedpolicy} presents our learning-based dispatching policy. 
In \Cref{sec:multi_server_queueing_model_sec}, we demonstrate the asymptotic optimality of the proposed policy in a multi-server Erlang-B system and characterize its regret compared to a system with a known service rate. 
 \Cref{sec:queue_sys_fin_buffer} extends the results of \Cref{sec:multi_server_queueing_model_sec} to a queueing system with a finite buffer. In \Cref{sec:experimental results}, we study the performance of our proposed policy through experiments and verify our theoretical analysis.} 


 \section{Problem Formulation}
\label{sec:PROBLEM FORMULATION}
We consider an $M/M/k/k+N$ queueing system with $k$ identical servers and a finite waiting room of size $N\geq 0$. Arrivals to the system are according to a Poisson process with rate $\lambda$, and at each arrival, a dispatcher decides between admitting the arrival or blocking it. If admitted, the arrival is dispatched to the first available server and serviced with exponentially distributed service times
with parameter $\mu$. Otherwise, if blocked, it leaves the system. Each time an arrival is accepted,
the dispatcher receives a fixed reward $R$ (after service completion), but incurs a cost of $c$ per unit time service; 
we assume that rejecting an arrival has no penalty. In our setting, we assume that the dispatcher knows the arrival rate $\lambda$ and  parameters $R$ and $c$; but does not know the service rate $\mu$. We also assume that the dispatcher observes the arrival times to the system and the system state upon arrivals. 
In contrast to the inter-arrival times, the service times of completed services are unknown. 

Consider the queueing system sampled at arrival $i$ for $i \in \{0,1,\ldots\}$, and let $A_i$ denote the action of the dispatcher to admit or block arrival $i$. 
If arrival $i$ is blocked, 
$A_i=0$; otherwise, if arrival $i$ is admitted (when there's room), 
$A_i=1$. 
We define $N_i$ as {the number of total jobs waiting for service in the server-queue pair before arrival $i$, and let $N_0=0$.} 
Let $T_{i}$ be the inter-arrival time between arrival $i-1$ and $i$, and  $M_i$ be the number of departures during 
$T_i$. Notice that $N_{i-1}+A_{i-1}=M_{i}+N_i$
and the value of $M_i$ can be found with the knowledge of $\{N_{i-1}, N_i, A_{i-1} \}$. The dispatcher chooses $A_i$ based on past observations up to arrival $i$, 
 i.e., $\mathcal{H}_i=\{T_1,\dots,T_i,A_0,A_1,\dots,A_{i-1},N_0,N_1,\dots,N_{i}\}$.
 {\color{blue}Consider the policy class $\Pi = \{\pi_a, \pi_b\}$, where $\pi_b$ is the policy that blocks all arrivals and $\pi_a$ is the policy that accepts all arrivals, subject to available room. Using the history of observations, the dispatcher’s goal is to identify the optimal policy within the class $\Pi$ that maximizes the expected average reward per unit time.
}
We note that by PASTA (\cite{srikant2013communication}), the expected average reward per unit time is $\limsup\limits_{n\rightarrow\infty}\frac{1}{n}\sum_{i=0}^{n-1} \mathbb{E}[K(A_i,\sigma_i)]$,
where $\sigma_i$ is the {\color{blue} sojourn} time of arrival $i$, and the reward function $K\left(\cdot,\cdot\right)$ is given by $K(a,s)=a(R-cs)$. 

{\color{blue}
In an Erlang-B {system---that is, the system with buffer size $N=0$---}with known service rate $\mu$, the optimal policy of the dispatcher is to accept all arrivals if $\mu>c/R$ (subject to availability) and block all arrivals if $\mu<c/R$; see \Cref{sec:Proposedpolicy}.
The dispatcher is indifferent between accepting or rejecting when $\mu=c/R$. {\color{blue} Based on this observation, we see that when the buffer size $N$ is zero, identifying the best-in-class policy in $\Pi$ is equivalent to finding the optimal policy.  }
However, when $N>0$, the optimal policy that maximizes the long-term average reward is a threshold-based admission policy, which does not belong to the set $\Pi$. {\color{blue}The optimal threshold---see \cite{chr1972individual}---is a complex function of arrival and service rates, and also the cost and reward parameters. Using the optimal threshold to obtain low-regret whilst learning unknown system parameters is complicated, and left for future work.} 
Consequently, in this scenario, we also focus on learning whether it is better to accept all arrivals or reject all.} In \Cref{sec:problem_form_MB}, we argue that when the service rate is known,  the best-in-class policy of the dispatcher is to accept all arrivals if $\mu>\mu^*$ (subject to availability) and block all arrivals if $\mu<\mu^*$, for some positive $\mu^*$ derived in \Cref{sec:problem_form_MB}.
{ We evaluate the performance of a candidate policy with respect to the best-in-class policy, denoted by $\pi^*$.} 
  In Sections \ref{sec:Proposedpolicy} and \ref{sec:problem_form_MB}, we propose a dispatching policy that uses past observations to learn the best-in-class policy, 
  and in Sections \ref{subsec:asymp_optimal_multi_server} and \ref{subsec:asymp_optimal_multi_server_MB}, we show the asymptotic optimality of our policy by proving its convergence to  $\pi^*$. Further, in Sections \ref{subsec:regret_analysis_multi_server} and \ref{subsec:regret_analysis_multi_server_MB}, 
   the finite-time performance of our policy is evaluated using the following definition. {\color{blue}Our metric focuses on the inaccuracies in the decisions instead of the payoffs, as the latter will likely be continuous around the indifference parameter $\mu^*$.} 
    
\begin{Definition} \label{def:regret}
Set $A^{\pi}_i$ as the action taken at arrival $i$ in a system that follows policy $\pi$. The expected regret of policy $\pi$ with respect to the {best-in-class} policy $\pi^*$ after $n$ arrivals is given by 
\begin{equation*}
\E \left[\mathcal{R} \left(n\right); \pi \right]=
\Big| \E \Big[ \sum_{i=0}^{n-1} (A^{\pi}_i-A^{\pi^*}_i) \Big] \Big|.
\end{equation*}
\end{Definition} 
\section{Multi-server Queueing Model with No Waiting Room}
\label{sec:multi_server_queueing_model_sec}

\subsection{Proposed Maximum Likelihood Estimate-based Dispatching Policy}
\label{sec:Proposedpolicy}
When $N=0$, for the optimal dispatching policy it is sufficient to estimate the service rate. In other words, for the $M/M/k/k$ queueing system,  knowledge of the arrival rate $\lambda$ is not necessary.
We would like a dispatching policy that (asymptotically) performs optimally, and further, (if possible) 
we want to minimize the regret of this system with respect to the system with known $\mu$. As mentioned in \Cref{sec:introduction}, we take a self-tuning adaptive control viewpoint: 
we  consider the system as being driven by parameter $\mu$, and the learning problem as a parameter estimation problem using system measurements given by the sequence of policies chosen. Specifically, 
we use maximum likelihood (ML) estimation to estimate parameter $\mu$, and then select the certainty equivalent control but with forced exploration. 

{\color{blue} As outlined in \Cref{sec:PROBLEM FORMULATION}, our objective is to identify  the best policy within the policy class $\Pi = \{\pi_a, \pi_b\}$. In the following arguments, we show that when $N$ equals zero, the optimal policy is either $\pi_a$ or $\pi_b$, and hence, the best-in-class policy coincides with the optimal policy. 
In \cite{arapostathis1993discrete}, it is shown that there exists a stationary deterministic policy that achieves the optimal average reward. In our model, for every stationary deterministic policy $\pi:\{0,1,\ldots,k\}\rightarrow  \{0,1\}$ such that $\pi(k)=0$, the discrete-time Markov chain attained by sampling the queueing system at job arrivals forms a unichain process  \cite{puterman1990markov}; in other words, it consists of a single recurrent class and a possibly non-empty set of transient states. Let $i$ denote the smallest state such that $\pi(i)=0$, i.e., the action taken according to policy $\pi$ at state $i$ is to reject the arrival. The resulting Markov Chain forms a single recurrent class $\{0,1,\ldots,i\}$ and states $\{i+1,\ldots,k\}$ are transient. Each $0\leq i \leq k$ corresponds to a different class of stationary deterministic policies. Denote the class of stationary deterministic policies corresponding to threshold $i$ by $\Pi^i$; notice that $\Pi^k = \{ \pi_a\}$ and $\Pi^0 = \{ \pi_b\}$. In each of the $k+1 $ different classes of the stationary deterministic policies, the underlying Markov process has a unique stationary distribution. Let $\eta^i$ be the corresponding unique stationary distribution of a Markov chain found by following a stationary deterministic policy in class $\Pi^i$. $\eta^i$ is given by
\begin{equation} \label{eq:stat_dist}
 \eta^i(j)=
 \begin{cases}
 \frac{(\frac{\lambda}{\mu})^j\frac{1}{j!}}{\sum_{l=0}^{i} (\frac{\lambda}{\mu})^l\frac{1}{l!}}, & 0 \leq j \leq i \\
 0. & i+1 \leq j \leq k
 \end{cases} 
\end{equation}
As the state and action space are finite and the Markov process is unichain, from \cite{arapostathis1993discrete}, for every deterministic stationary policy $\pi \in \Pi^{i}$,  the limit in \eqref{eq:limit_avr_rew} exists, is independent of the initial state, and equals 
\begin{equation} \label{eq:limit_avr_rew}
\lim\limits_{n\rightarrow\infty}\frac{1}{n}\sum_{l=0}^{n-1} \mathbb{E}^\pi [K(A_l,\sigma_l)]=\sum_{j=0}^{i-1}\left(R-\frac{c(j+1)}{\lambda+\mu}\right)\eta^{i}(j)-\frac{c\,i}{\lambda+\mu}\eta^{i}(i) ,
\end{equation}
and the problem of finding the optimal stationary deterministic policy is equivalent to finding the optimal threshold $i^*$ such that the right-hand side of \eqref{eq:limit_avr_rew} is maximized. Using \eqref{eq:stat_dist}, we can simplify \eqref{eq:limit_avr_rew} to get
\begin{equation*}
\lim\limits_{n\rightarrow\infty}\frac{1}{n}\sum_{l=0}^{n-1} \mathbb{E}^d [K(A_l,\sigma_l)]=\left(R-\frac{c}{\mu}\right) \left( 1-\eta^{i}(i) \right).
\end{equation*}
The expression above is intuitive and follows a different interpretation of the long-term expected reward---the total expected reward of each accepted arrival is $R-\tfrac{c}{\mu}$, and as we accept until $i-1$ (for $i\geq 1$) but reject at $i$, the above expression holds. To find the optimal threshold $i^*$, notice that the Erlang-B blocking probability $\eta^{i}(i)$ is a decreasing function of $i$. As a result, the optimal policy depends only on the sign of $R-\tfrac{c}{\mu}$ and belongs to the policy class $\Pi = \{\pi_a, \pi_b\}$.
}

\subsubsection{Maximum Likelihood Estimate Derivation}\label{subsec:MLE_der}

In this section, we derive the log-likelihood function and the corresponding MLE.  
The probability of $m_i$ departures and $n_i$ incomplete services in inter-arrival duration $t_i$ and given $m_i+n_i=N_{i-1}+A_{i-1}$ is 
\begin{equation} \label{prob_base}
p \left(m_i,n_i,t_i;\mu\right)=\binom{n_i+m_i}{n_i}  \left(1-\exp\left(-\mu t_i\right)\right)^{m_i}\left(\exp\left(-\mu t_i\right)\right)^{n_i}.
\end{equation} 
From \eqref{prob_base}, the conditional probability of observing sequences $\{m_i\}_{i=1}^n$ and $\{n_i\}_{i=1}^n$ for a fixed $\mu$ given the inter-arrival sequence $\{t_i\}_{i=1}^n$ is given by 
\begin{equation} \label{MLE_just}
\Pr \left( M_1=m_1,\ldots,M_n=m_n,N_1=n_1,\ldots,N_n=n_n \Bc \mu, \{t_i\}_{i=1}^n \right)= \prod_{i=1}^n p \left(m_i,n_i,t_i;\mu\right).
\end{equation}
In our problem formulation, no prior distribution is assumed for $\mu$, 
and thus, the posterior probability of a fixed $\mu$ given observations of $\{m_i\}_{i=1}^n$,$\{n_i\}_{i=1}^n$ and $\{t_i\}_{i=1}^n$ is proportional to \eqref{MLE_just}. From \eqref{prob_base} and \eqref{MLE_just}, we form the likelihood function of the past observations $\mathcal{H}_n$ under parameter $\mu$ as 
\begin{equation}\label{eq:likelihood}
L\left(\mathcal{H}_n;\mu\right):=c_b \prod_{i=1}^n \left(1-\exp\left(-\mu T_i\right)\right)^{M_i}\left(\exp\left(-\mu T_i\right)\right)^{N_i},
\end{equation}
where $c_b$ is the product of the binomial coefficients found in \eqref{prob_base} and independent of $\mu$. Maximization of 
function $L\left(\mathcal{H}_n;\mu\right)$ is equivalent to maximization of log-likelihood function $l\left(\mathcal{H}_n;\mu\right)$ defined as
\begin{equation}\label{eq:loglikelihood}
l\left(\mathcal{H}_n;\mu\right):=\log L\left(\mathcal{H}_n;\mu\right) =\log c_b+\sum_{i=1}^{n} M_i \log \left(1-\exp\left(-\mu T_i\right)\right)-\mu \sum_{i=1}^{n} N_i T_i.
\end{equation}
If $M_i=0$ for all $i$, the maximum of $l\left(\mathcal{H}_n;\mu\right)$ in $[0,+\infty)$ is obtained at $\mu=0$, and if $N_i=0$ for all $i$, the maximum is reached as $\mu\to +\infty$. Otherwise, from differentiability and strict concavity of the log-likelihood function, 
it has at most one maximizer, and as $\lim_{\mu \to 0 } \;  l\left(\mathcal{H}_n;\mu\right)=\lim_{\mu \to + \infty }  l\left(\mathcal{H}_n;\mu\right)= -\infty$,
there exists a unique $\hat{\mu}_n>0$ that maximizes
$l\left(\mathcal{H}_n;\mu\right)$, which can be determined by solving the first-order condition. The derivative of $l\left(\mathcal{H}_n;\mu\right)$ is given by
\begin{equation} \label{solution_MLE}
l'\left(\mathcal{H}_n;\mu\right)=\sum_{i=1}^{n}  \frac{M_iT_i\exp\left(-\mu T_i\right)}{1-\exp\left(-\mu T_i\right)}
-\sum_{i=1}^{n} N_i T_i.
\end{equation}
From 
\eqref{solution_MLE}, the maximum likelihood estimate $\hat{\mu}_n$ is the solution to the following equation:
\begin{equation} \label{sol_2}
\sum_{i=1}^{n} g\left(T_i,M_i,\hat{\mu}_n\right)=\sum_{i=1}^{n} h\left(T_i,N_i,\hat{\mu}_n\right),
\end{equation}
where 
$g\left(t,m,\mu\right):=\tfrac{mt\exp\left(-\mu t\right)}{1-\exp\left(-\mu t\right)}$ and $h\left(t,n,\mu\right):=nt$. It is easy to verify that $\sum_{i=1}^{n} g\left(T_i,M_i,{\mu}\right)$ is a positive and decreasing function of $\mu$. Moreover, $\lim_{\mu \to 0 } \; \sum_{i=1}^{n}  g\left(T_i,M_i,{\mu}\right)= +\infty$ and $\lim_{\mu \to + \infty } \sum_{i=1}^{n}  g\left(T_i,M_i,{\mu}\right)= 0$.
Since $\sum_{i=1}^{n}  h\left(T_i,N_i,\mu\right)$ is a positive constant independent of $\mu$, \Cref{sol_2} has a unique positive solution $\hat{\mu}_n$. However, given the simple set of optimal policies for our problem, we do not need to solve this equation to determine our policy. For a given estimate $\hat{\mu}_n$, the optimal policy only requires a comparison of $\hat{\mu}_n$ and $c/R$, and, based on the properties of $g$ and $h$,
 to compare $\hat{\mu}_n$ with $c/R$, it suffices to compare $\sum_{i=1}^{n} g\left(T_i,M_i,c/R\right)$  with $\sum_{i=1}^{n} h\left(T_i,N_i,c/R\right)$. 

 \subsubsection{The Learning Algorithm}

\begin{algorithm}[t]
	\caption{Proposed ML estimate-based Policy for Learning the Optimal Dispatching Policy }
	\label{alg1}
	\begin{algorithmic}[1]
	    \RETURN $\mu^*$ and $f: \mathbb{N} \cup \{0\} \rightarrow  \left[1,\infty \right) $, increasing, and $	\lim_{n \to +\infty} f \left(n\right) =+\infty$. \label{line_1_alg}
	    \renewcommand{\algorithmicreturn}{\textbf{Initialize}}
	    \RETURN $N_0=0, \alpha_0=0$.
		\STATE At arrival $n\geq 0$, \algorithmicdo 
		\STATE Update $\alpha_n$ using \eqref{update_alpha_n_multi}, and find $S(n)=\max\{0 \leq i\leq n: N_i=0\}$. 
		\IF{$N_n=k+N$}
		\STATE Block the arrival. 
		\ELSIF{$N_n<k+N$ and $\sum_{i=1}^{S\left(n\right)}g\left(T_i,M_i,\mu^*\right)> \sum_{i=1}^{S\left(n\right)}  h\left(T_i,N_i,\mu^*\right)$}
		\STATE Admit the arrival.
		\ELSIF{$N_n<k+N$ and $\sum_{i=1}^{S\left(n\right)}g\left(T_i,M_i,\mu^*\right)\leq \sum_{i=1}^{S\left(n\right)}  h\left(T_i,N_i,\mu^*\right)$}
		\STATE Admit the arrival with probability $p_{\alpha_n}=1/f\left(\alpha_n\right)$. \label{prob_line}
		\ENDIF
	\end{algorithmic}
\end{algorithm}
 The discussion at the end of the previous subsection leads to the following two cases:
\begin{enumerate}[wide, labelindent=0pt]
\item  
$\sum_{i=1}^{n} g\left(T_i,M_i,c/R\right)>\sum_{i=1}^{n} h\left(T_i,N_i,c/R\right)$ implies that $\hat{\mu}_n> c/R$. \label{case_1_MLE}
\item 
$\sum_{i=1}^{n} g\left(T_i,M_i,c/R\right)\leq \sum_{i=1}^{n} h\left(T_i,N_i,c/R\right)$ implies that $\hat{\mu}_n\leq c/R$. \label{case_2_MLE}
\end{enumerate}
In Case \ref{case_1_MLE}, the MLE indicates the \textit{always admit if room} policy is optimal. In our proposed policy, we 
follow the  MLE whenever Case \ref{case_1_MLE} applies and 
admit the arrival (if there is a free server). In contrast to Case \ref{case_1_MLE}, the MLE in Case \ref{case_2_MLE} suggests blocking all arrivals. However, if we follow the MLE in both cases,  we may falsely identify the service rate and incur linear regret. Notably, using the optimal policy in Case \ref{case_2_MLE} results in no arrivals 
 and new system samples.  
Thus, 
in Case \ref{case_2_MLE}, 
our policy will not use the certainty equivalent control with a small probability that converges to 0. 
Finally, we introduce
\Cref{alg1} for optimal dispatch in an 
Erlang-B system with unknown service rate. Notice that when $N=0$, the boundary value $\mu^*$ is equal to $c/R$.

We label the policy in \Cref{alg1} as $\pi_{\mathrm{Alg1}}$. Then $S(n)$ is defined as the last arrival instance before or at arrival $n$ when the system is empty. 
The probability of using the sub-optimal policy in Case \ref{case_2_MLE} is equal to $p_{\alpha_n}=1/f\left(\alpha_n\right)$, where a valid function $f: \mathbb{N} \cup \{0\} \rightarrow  \left[1,\infty \right) $ is increasing and converges to infinity as $\alpha_n$ goes to infinity. Further, $\alpha_0=0$ and $\alpha_n$ is defined as below for $n\geq 1$
\begin{equation} \label{update_alpha_n_multi}
\alpha_n= \begin{cases}
\alpha_{n-1}+1, &  \text{if } \sum_{i=1}^{n-1} g\left(T_i,M_i,c/R\right) \leq \sum_{i=1}^{n-1}  h\left(T_i,N_i,c/R\right) \text{, } A_{n-1}=1, N_{n-1}=0,\\
\alpha_{n-1}, & \text{otherwise}.
\end{cases}
\end{equation}
In other words, $\alpha_n$ is the number of accepted arrivals $0 \leq l<n$ such that $\sum_{i=1}^{l} g_i\left(c/R\right)\leq \sum_{i=1}^{l}  h_i$ and the system is empty right before arrival $l$. 
We also note that 
any choice of  $f \geq 1 $
that increases to infinity leads to asymptotic optimality of $ \pi_{\mathrm{Alg1}}$, 
as proved in \Cref{subsec:asymp_optimal_multi_server}. 
However, the class of admissible functions is restricted in \Cref{subsec:regret_analysis_multi_server} to provide finite-time guarantees.

The parameters of  policy $\pi_{\mathrm{Alg1}}$ are only updated when the system becomes empty, 
rather than at all arrivals.  
The reason for this modification is that the busy period boundary is a regenerative epoch that provides sufficient independence 
needed in the analysis, whereas the regret of the policy that updates its parameters at all arrivals is hard to analyze. 
However, this alternate policy, called $\pi_{\mathrm{Alg2}}$, is also asymptotically optimal, and we empirically compare it to $\pi_{\mathrm{Alg1}}$ in \Cref{sec:experimental results}. We also note that in the single-server setting, the two policies $\pi_{\mathrm{Alg1}}$ and $\pi_{\mathrm{Alg2}}$ coincide.


\subsection{Analysis}

In this section, we focus on the Erlang-B queueing system to provide a simpler pathway to the queueing system with a non-zero waiting room.
In \Cref{subsec:asymp_optimal_multi_server}, the convergence of  $\pi_{\mathrm{Alg1}}$ to the optimal policy is shown by a martingale-based analysis coupled with the SLLN for martingale sequences. Then, in \Cref{subsec:regret_analysis_multi_server}, we evaluate the finite-time performance of our proposed policy in terms of the expected regret defined in \Cref{def:regret} using martingale concentration inequalities.

\subsubsection{Asymptotic Optimality}\label{subsec:asymp_optimal_multi_server}

First, we describe a stochastic process whose limiting behavior will determine the performance of our learning scheme. 
Define 
$\{\tilde{X}_n\}_{n=0}^\infty$ as
\begin{equation} \label{state_MC_original_klarger}
\tilde{X}_n=\left(X_n,N_n,\alpha_n\right)=\Big(\sum_{i=1}^{n} \Big(g\left(T_i,M_i,{c}/{R}\right)-   h\left(T_i,N_i,{c}/{R}\right)\Big),N_n,\alpha_n\Big).
\end{equation} 
We note that the action at arrival $n$ defined by $\pi_{\mathrm{Alg1}}$ 
is uniquely determined by $\tilde{X}_{S(n)}$: if a server is available and $X_{S(n)}>0$, the arrival will be accepted. Otherwise, if  $X_{S(n)}\leq0$, the arrival will be admitted with probability $p_{\alpha_n}$. 
To prove asymptotic optimality, we show that eventually, $X_n$ will always be positive for  $\mu>\tfrac{c}{R}$, and negative for $\mu<\tfrac{c}{R}$. In  the process $\{\tilde{X}_n\}_{n=0}^\infty$, $X_n$ is updated as 
\begin{equation} \label{delta_x}
X_{n}-X_{n-1}=g\left(T_n,M_n,{c}/{R}\right)-   h\left(T_n,N_n,{c}/{R}\right).
\end{equation}
In \eqref{delta_x}, random variables $N_n$ and $M_n$ only depend on the history through the previous state $\tilde{X}_{n-1}$ and the sign of $\tilde{X}_{S(n)}$, and $\alpha_n$ is  updated from 
$\tilde{X}_{n-1}$ by \eqref{update_alpha_n_multi}. 
Thus,the stochastic process $\{\tilde{X}_n\}_{n=0}^\infty$ is not a Markov process. 
Random variables $\{{X}_{n }-X_{n-1}\}_{n=1}^\infty$ are not independent since values of $N_n$ and $M_n$ depend on 
$\tilde{X}_{n-1}$. 
Hence, it is not straightforward to analyze the asymptotic behavior of $\{\tilde{X}_n\}_{n=0}^\infty$. We will define a new stochastic process that will simplify the analysis
and establish convergence results for this process. 
Define $\{\beta_n\}_{n=0}^\infty$ as the sequence of the indices of accepted arrivals when the system is empty and $Y_n:=X_{\beta_n}$. We down-sample 
$\{\tilde{X}_n\}_{n=0}^\infty$ using  sequence $\{\beta_n\}_{n=0}^\infty$ to get the 
process $\{\tilde{Y}_n\}_{n=0}^\infty$ given by
\begin{equation} \label{state_MC_sampled}
\tilde{Y}_n=\tilde{X}_{\beta_n}=\left(X_{\beta_n},N_{\beta_n},\alpha_{\beta_n}\right)=:\left(Y_{n},0,\alpha_{\beta_n}\right).
\end{equation}
Note that $N_{\beta_n}=0$ as the system is empty just before a arrival is accepted. {To ensure  
process $\{\tilde{Y}_n\}_{n=0}^\infty$ is well-defined, in \Cref{lem:infinite_acceptance}, we prove that the number of accepted arrivals following $\pi_{\mathrm{Alg1}}$ 
is almost surely infinite; see \Cref{sec:proof_lem_inf_acc}.}
Random variables $\{{Y}_n-Y_{n-1}\}_{n=1}^\infty$ are not independent 
as ${Y}_n-Y_{n-1}$ depends on the acceptance probabilities.
We will argue that process $\{{Y}_n\}_{n=0}^\infty$ is a submartingale (or supermartingale), and using this result, we will analyze its convergence. 
We define random variable $D_{i}$ as the change in $X_i$ at inter-arrival $T_{i}$, i.e., $ D_{i}:=X_i-X_{i-1}$. 
Next,  
for any $n \geq 0$, we define process $\{W_{n,m}\}_{m=0}^{\infty}$ as 
\begin{equation} \label{eq:def_W}
W_{n,m}=Y_{n}+\sum_{i=1}^{m} D_{\beta_n+i}=X_{\beta_n+m}.
\end{equation}
We define 
$\tau_n$ as the index of the first arrival after $\beta_n$ that finds the system empty, 
i.e., $\tau_n=\min \left\{i \geq 1 : N_{\beta_n+i}=0\right\}$.
Note that by \eqref{eq:def_W}, $W_{n,\tau_n}=X_{\beta_n+\tau_n}$. We claim that  process $\{{X}_n\}_{n=0}^\infty$ at the first arrival acceptance after $\tau_n$, i.e., $X_{\beta_{n+1}}$, is equal to $ W_{n,\tau_n}$. 
Indeed, process $\{{X}_n\}_{n=0}^\infty$ does not change when there are no departures or ongoing services. Hence, $W_{n,0}=Y_{n}$ and $W_{n,\tau_n}=X_{\beta_{n+1}}=Y_{n+1}$.
Thus, to analyze the convergence of 
$\{{Y}_n\}_{n=0}^\infty$, we study the properties of process $\{W_{n,m}\}_{m=0}^{\infty}$ and random variable $\tau_n$ for 
$n\geq 1$. We determine the behavior of $\tau_n$ by coupling the system that runs \Cref{alg1} with a system that accepts all arrivals (subject to availability)
as follows.

\paragraph{\color{blue} Coupling of two systems:} Let $Q^{(n)}$ denote the system that accepts all arrivals as long as it has at least one available server. We also define random variable $\zeta_n$ as the first arrival after arrival $\beta_n$ that finds $Q^{(n)}$ empty, starting from an empty state. Starting from arrival $\beta_n$, we couple this system with the  system that follows \Cref{alg1} such that at each arrival, the number of busy servers in $Q^{(n)}$ is greater than or equal to our system.
We couple the arrival sequences in both systems such that the inter-arrival times  are equal. Moreover, when an arrival is accepted in both systems, we assume that its service time is identical in both. System $Q^{(n)}$ will accept all arrivals unless none of its servers are available. Suppose all of the servers of $Q^{(n)}$ are busy, and our system accepts an arrival. In this case, we assume that the service time of the accepted arrival in our system equals the remaining service time of the $k^\mathrm{th}$ server in $Q^{(n)}$, which has an exponential distribution with parameter $\mu$ due to the memoryless property. 
Using this coupling, we verify that all moments of $\tau_n$ are finite in \Cref{lem:tau_finite_mean}. 

\begin{Lemma} \label{lem:tau_finite_mean}
All moments of random variable $\tau_n$ are bounded by a constant independent of $n$.
\end{Lemma}

\proof{Proof of \Cref{lem:tau_finite_mean}.}
By the above coupling of $Q^{(n)}$ with the system that follows our proposed policy, we ensure that at each arrival, the number of busy servers in $Q^{(n)}$ is greater than or equal to our system. Hence, the moments of $\tau_n$ are bounded by the moments of  $\zeta_n$. In system $Q^{(n)}$, the number of busy servers just before each arrival forms a finite-state irreducible Markov chain, and random variable $\zeta_n$ is the first passage time of the state zero starting from  zero, and has moments bounded by a constant which only depends on $\lambda$, $\mu$ and the number of servers.
\Halmos
\endproof

After characterizing the behavior of  $\tau_n$, in \Cref{lem:W_sub},  we show that the process $\{W_{n,m}\}_{m=0}^{\infty}$ is a submartingale or supermartingale depending on the sign of $\mu-c/R$.  

\begin{Lemma}\label{lem:W_sub}
Fix $n \geq 0$. For $\mu\in (c/R,+\infty)$, the stochastic process $\{W_{n,m}\}_{m=0}^{\infty}$ forms a submartingale  sequence with respect to the filtration  $\{\mathcal{G}_{n,m}\}_{m=0}^{\infty}$, wherein the $\sigma$-algebra $\mathcal{G}_{n,m}$ is defined as 
$	\mathcal{G}_{n,m}:=\sigma\left(T_{\beta_n+1},\ldots, T_{\beta_n+m},N_{\beta_n+1},\ldots, N_{\beta_n+m},\alpha_{\beta_n},\ldots, \alpha_{\beta_n+m},A_{\beta_n+1},\ldots, A_{\beta_n+m},Y_{n}\right).$
	For $\mu\in (0,c/R)$, the process $\{W_{n,m}\}_{m=0}^{\infty}$ is a supermartingale with respect to filtration
	$\{\mathcal{G}_{n,m}\}_{m=0}^{\infty}$.
\end{Lemma}

\proof{Proof of \Cref{lem:W_sub}.}  
	We show the proof for the case of $\mu>c/R$. The other region follows similarly. 
 To prove $\{W_{n,m}\}_{m=0}^{\infty}$ is a submartingale sequence, we first show 
	$\mathbb{E} \left[\left| W_{n,m}\right|\right]<\infty$. From 
	\eqref{eq:def_W}, 
	\begin{align} \label{upp_mart_finite}
	\mathbb{E} \left[\left| W_{n,m}\right|\right]  &\leq \mathbb{E} \Big[\left|Y_{n}\right|+\sum_{i=1}^{m} |D_{\beta_n+i}|\Big] \leq \mathbb{E} \Big[|Y_{n}|+\sum_{i=1}^{m} \Big| g\big(T_{\beta_n+i},M_{\beta_n+i},\frac{c}{R}\big)-   h\big(T_{\beta_n+i},N_{\beta_n+i},\frac{c}{R}\big) \Big|
	\Big] \nonumber\\
 &\leq \mathbb{E} \left[|Y_{n}|\right]+k\sum_{i=1}^{m} \Big( \mathbb{E} \left[ g\big(T_{\beta_n+i},1,\frac{c}{R}\big)\right]
	+\mathbb{E}\left[T_{\beta_n+i}\right]\Big),
	\end{align}  
	where \eqref{upp_mart_finite} holds as $0 \leq M_{\beta_n+i},N_{\beta_n+i}\leq k$. 
	For $t>0$, we have $g\left(t,1,x\right) \leq \frac{1}{x},$ 
	and thus, the summation in \eqref{upp_mart_finite} is finite. 
 To show that $\mathbb{E} \left[\left|Y_{n}\right|\right]<\infty$, it suffices to show $\mathbb{E}\left[\left|Y_{n+1}-Y_{n}\right|\right]$ is finite for all $n$: 
	\begin{align} \label{eq:first_line}
	\mathbb{E}\left[\left|Y_{n+1}-Y_{n}\right|\right] & = \mathbb{E}\left[\left|W_{n,\tau_n}-Y_n\right|\right] = \mathbb{E} \Big[\Big|\sum_{i=1}^{\tau_n} D_{\beta_n+i}\Big|\Big]  
	\leq k \mathbb{E} \Big[\sum_{i=1}^{\tau_n}\left( T_{\beta_n+i}+{g}\left(T_{\beta_n+i},1,\frac{c}{R}\right) \right)\Big]\\
		&\leq k \mathbb{E} \Big[\sum_{i=1}^{\zeta_n} \left(T_{\beta_n+i}+{g}\left(T_{\beta_n+i},1,\frac{c}{R}\right) \right) \Big]  \label{coupling} 
	= k \mathbb{E} \left[\zeta_n\right]\mathbb{E} \left[T_{\beta_n+1}+{g}\left(T_{\beta_n+1},1,\frac{c}{R}\right) \right],
	\end{align}  
 where \eqref{eq:first_line} is derived similar to \eqref{upp_mart_finite} and \eqref{coupling} follows from coupling $Q^{(n)}$ with the system that runs \Cref{alg1}. Hitting time $\zeta_n$ is a stopping time for the finite-state irreducible Markov chain found by sampling $Q^{(n)}$ at arrivals and $\E[\zeta_n]<\infty$. 
 Hence,  \eqref{coupling} follows from Wald's equation (\cite{durrett2019probability}), and $	\mathbb{E}\left[\left|Y_{n+1}-Y_{n}\right|\right]<\infty,$
	which implies that $\mathbb{E}\left[|Y_{n}|\right]<\infty$, and by \eqref{upp_mart_finite}, $\mathbb{E} \left[ |W_{n,m}|\right]<\infty$. 
	We next verify the submartingale property of  $\{W_{n,m}\}_{m=0}^{\infty}$. From the Markov property of $\{\tilde{X}_n\}_{n=0}^\infty$,
	\begin{align} \label{eq:exp_W_submart}
	\mathbb{E}\left[W_{n,m+1}-W_{n,m}\Bc \mathcal{G}_{n,m}\right]
	&=\mathbb{E}\left[X_{\beta_n+m+1}-X_{\beta_n+m}\Bc X_{\beta_n+m},N_{\beta_n+m},\alpha_{\beta_n+m},A_{\beta_n+m}\right],
	\end{align}
which is equal to the expected change in $X_i$ during inter-arrival $T_{\beta_n+m+1}$. To show $\mathbb{E}[W_{n,m+1}-W_{n,m}\Bc \mathcal{G}_{n,m}]\geq 0$, we argue that $\mathbb{E}[X_{i+1}-X_{i}\Bc X_{i},N_{i},\alpha_i ,A_i]$ is non-negative for all $i$ as follows,
\begin{align} \label{eq:X_exp_g_h}
&\mathbb{E}\big[X_{i+1}-X_{i}\big| X_{i},N_{i},\alpha_i ,A_i \big]\nonumber\\ &= \mathbb{E}\big[g\big(T_{i+1},N_i+A_i-N_{i+1},\frac{c}{R}\big)- h\big(T_{i+1},N_{i+1},\frac{c}{R}\big)\big| N_{i},A_i \big]\nonumber  \\
&=\mathbb{E}\big[\big( N_i+A_i-N_{i+1}\big)g\big(T_{i+1},1,\frac{c}{R}\big) \big| N_{i},A_i \big]-\mathbb{E}\left[ T_{i+1}N_{i+1}\Bc N_{i},A_i \right]
\nonumber	\\ &=\big( N_i+A_i\big)\mathbb{E}\big[g\big(T_{i+1},1,\frac{c}{R}\big) \big]-\mathbb{E}\big[ N_{i+1}g\big(T_{i+1},1,\frac{c}{R}\big) \big| N_{i},A_i \big] -\left(N_{i}+A_i\right)\mathbb{E} \left [T_{i+1}\mathbbm{1}_A\right],
\end{align}
where $A$ is the event that a fixed server from the $N_i+A_i$ busy servers remains busy during inter-arrival  $T_{i+1}$.
  The second term of \eqref{eq:X_exp_g_h} can be simplified as follows
	\begin{align}
	\mathbb{E}\big[ N_{i+1}g\big(T_{i+1},1,\frac{c}{R}\big)  \Bc N_{i},A_i \big]  &= \left(N_{i}+A_i\right)\mathbb{E} \big [g\big(T_{i+1},1,\frac{c}{R}\big)\mathbbm{1}_A
	\big] \nonumber \\
	&= \left(N_{i}+A_i\right)\int_{t=0}^{+\infty} \frac{t\exp\left(-t\frac{c}{R}\right)}{1-\exp\left(-t\frac{c}{R}\right)}   \lambda \exp\left(-\lambda t \right)\exp\left(-\mu t \right) dt \nonumber   \\
& \textcolor{blue}{= 
 \left(N_{i}+A_i\right)\int_{t=0}^{+\infty} {t\exp\left(-t\frac{c}{R}\right)}  \lambda \exp\left(-\lambda t \right)\exp\left(-\mu t \right)\Big(\sum_{s=0}^{+\infty}\exp\left(-st\frac{c}{R}\right)\Big) 	 dt \nonumber  } \\
 &= \left(N_{i}+A_i\right) \sum_{j=0}^{\infty}  \frac{\lambda }{\big(\lambda+\mu+\left(j+1\right)\frac{c}{R}\big)^2}. \label{eq:second_g_part}
	\end{align}
 Furthermore, we derive $\mathbb{E}\left[g\left(T_{i+1},1,{c}/{R}\right)  \right]$ using similar calculations as above,
	\begin{align} 
	\mathbb{E}\big[g\big(T_{i+1},1,\frac{c}{R}\big)  \big] &=\int_{t=0}^{+\infty} \frac{t\exp\left(-t\frac{c}{R} \right)}{1-\exp\left(-t\frac{c}{R}\right)}   \lambda \exp\left(-\lambda t \right) dt 
	= \sum_{j=0}^{\infty}  \frac{\lambda }{\big(\lambda+\left(j+1\right)\frac{c}{R}\big)^2}. \label{eq:first_g_part}
	\end{align}	
	 Next, we simplify the third term of \eqref{eq:X_exp_g_h}:
	\begin{align} 
 \left(N_{i}+A_i\right)\mathbb{E} \left [T_{i+1}\mathbbm{1}_A
	\right] \nonumber
	=\left(N_{i}+A_i\right)\int_{t=0}^{+\infty}\int_{x=t}^{+\infty}  t \mu \exp\left(-\mu x \right) \lambda \exp\left(-\lambda t \right) dx dt \nonumber 
	= \left(N_{i}+A_i\right)\frac{\lambda}{\left(\lambda+\mu\right)^2} . \label{eq:third_h_part}
	\end{align}
	Substituting the terms found in the above equation,  \eqref{eq:first_g_part}, and \eqref{eq:second_g_part}, in \Cref{eq:X_exp_g_h}, 	we have $\mathbb{E}\big[X_{i+1}-X_{i}\Bc X_{i},N_{i},\alpha_i ,A_i \big]=\tilde{\delta}\left(N_{i}+A_i\right)$
	where  $\tilde{\delta}:=-\frac{\lambda}{\left(\lambda+\mu\right)^2}+\sum_{j=0}^{\infty} \frac{\lambda }{(\lambda+\left(j+1\right)\tfrac{c}{R})^2} -\frac{\lambda }{(\lambda+\mu+\left(j+1\right)\tfrac{c}{R})^2}$ 
	and is positive for $\mu \in (c/R,+\infty)$. Hence, from \eqref{eq:exp_W_submart}, 
		\begin{equation} \label{eq:exp_W_pos}
	 \mathbb{E}\left[W_{n,m+1}-W_{n,m}\Bc \mathcal{G}_{n,m}\right]=\tilde{\delta}\left(N_{\beta_n+m}+A_{\beta_n+m}\right) \geq 0,
	\end{equation}
	and we conclude that $\{W_{n,m}\}_{m=0}^{\infty}$ is a submartingale sequence with respect to  $\{\mathcal{G}_{n,m}\}_{m=0}^{\infty}$.
\Halmos
\endproof

Next, in \Cref{thm:martingale}  we argue that the stopped sequence $\{W_{n,\tau_n}\}_{n=0}^{\infty}$ or $\{Y_n\}_{n=0}^{\infty}$ also forms a submartingale or supermartingale sequence depending on the problem parameters. 

\begin{Proposition} \label{thm:martingale}
	Sequence 
	$\{Y_n\}_{n=0}^{\infty}$  forms a submartingale or supermartingale 
	(depending on the sign of $\mu-c/R$) with respect to filtration $\{\mathcal{F}_n\}_{n=0}^{\infty}$ defined as $\mathcal{F}_n=\sigma \left(Y_0,\ldots,Y_n, \alpha_{\beta_0},\ldots, \alpha_{ \beta_n}\right)$.
 Specifically, $\{Y_n\}_{n=0}^{\infty}$ is a submartingale sequence 
	if $\mu>c/R$ and a supermartingale 
	otherwise.
\end{Proposition}

\proof{Proof of \Cref{thm:martingale}.}
	We show the proof for the case of $\mu>{c}/{R}$, and the other regime follows similarly.	
	Note that $ Y_{n+1}$ is equal to  submartingale $\{W_{n,m}\}_{m=0}^{\infty}$ stopped at $\tau_n$; in other words, $Y_{n+1}= W_{n,\tau_n}=Y_{n}+\sum_{i=1}^{\tau_n} D_{\beta_n+i}.$
	In \Cref{lem:tau_finite_mean},  we argued 
 that $	\mathbb{E} \left[\tau_n\right]    < \infty.$
    Moreover, 
	\begin{equation} \label{bdd}
	\mathbb{E}\left[\left|W_{n,m+1}-W_{n,m}\right| \Bc \mathcal{G}_{n,m}\right]=  \mathbb{E}\left[\left|D_{\beta_n+m+1}\right| \Bc \mathcal{G}_{n,m}\right] \leq k \mathbb{E} \Big[ g\Big(T_{\beta_n+1},1,\frac{c}{R}\Big)\Big]+k\mathbb{E}\left[T_{\beta_n+1}\right].
	\end{equation}
 As $g$ is bounded, the RHS of \eqref{bdd} is also finite. Hence, we can use Doob's optional stopping theorem \cite[Theorem 4.8.5]{durrett2019probability} for submartingale $\{W_{n,m}\}_{m=0}^{\infty}$ and stopping time $\tau_n$ 
 to get
	\begin{equation*}
	\mathbb{E} \big[Y_{n+1} \Bc \mathcal{G}_{n,0}\big] = \mathbb{E} \big[W_{n,\tau_n}\Bc \mathcal{G}_{n,0}\big]  \geq  \mathbb{E} \big[W_{n,0}\Bc \mathcal{G}_{n,0}\big] = Y_{n}.
	\end{equation*}
	Thus, we have 
	\begin{equation*}
	\mathbb{E} \big[Y_{n+1}-Y_{n} \Bc \mathcal{G}_{n,0}\big]=\mathbb{E} \big[Y_{n+1}-Y_{n} \Bc \mathcal{F}_{n}\big] \geq 0.
	\end{equation*}
	As $\mathbb{E} \left[|Y_n|\right]$ is finite, $\{Y_n\}_{n=0}^{\infty}$ is a submartingale sequence with respect to $\{\mathcal{F}_n\}_{n=0}^{\infty}$.
\Halmos \endproof

Now that we proved the submartingale (or supermartingale) property of $\{Y_n\}_{n=0}^{\infty}$, we can examine the convergence of this process. 
From \Cref{thm:martingale}  and Doob's decomposition of $\{Y_n\}_{n=0}^{\infty}$, we have $Y_n=Y_n^A+Y_n^M$,
where $Y_n^M$ is a martingale sequence, and $Y_n^A$ is a predictable and almost surely increasing (or decreasing) sequence with $Y_0^A=0$. In Lemmas \ref{lem:incr_seq} and \ref{lem:mart_seq}, we examine the limiting behavior of sequences $\{Y_n^A\}_{n=0}^{\infty}$ and $\{Y_n^M\}_{n=0}^{\infty}$. The basic idea is to show that $\{Y_n^A\}_{n=0}^{\infty}$ 
converges to infinity, and $\{Y_n^M\}_{n=0}^{\infty}$ 
is well-behaved in a way that their sum,   $\{Y_n\}_{n=0}^{\infty}$, converges to infinity. 

\begin{Lemma} \label{lem:incr_seq}
For $\mu \in (c/R,+\infty)$, there exists a positive problem-dependent constant $\tilde{\delta_1}$ 
such that the 
process $\{Y_n^A\}_{n=0}^{\infty}$ from Doob's decomposition of $\{Y_n\}_{n=0}^{\infty}$ satisfies
    $Y_n^A \geq \tilde{\delta_1}  n$ 
    \emph{a.s.}, 
 and for $\mu \in (0,c/R)$, there exists a negative constant $\tilde{\delta_2}$ such that the 
 process $\{Y_n^A\}_{n=0}^{\infty}$ satisfies
    $Y_n^A \leq \tilde{\delta_2} n$ 
    \emph{a.s.}
\end{Lemma}

\proof{Proof of \Cref{lem:incr_seq}.}
WLOG,  we assume $\mu \in (c/R,+\infty)$. By \Cref{thm:martingale},  sequence $\{Y_n\}_{n=0}^{\infty}$ is a submartingale with respect to filtration $\{\mathcal{F}_n\}_{n=0}^{\infty}$. Hence, the increasing sequence is given as below 
	\begin{equation} \label{incr_sequence_X}
	Y_n^A=\sum_{m=0}^{n-1} \mathbb{E} \left[Y_{m+1}-Y_{m} \Bc \mathcal{F}_{m}\right]=\sum_{m=0}^{n-1} \left( \mathbb{E} \left[W_{m,\tau_m} \Bc \mathcal{F}_{m}\right] -Y_{m} \right).
	\end{equation}
     In \Cref{lem:W_sub},  we argued $\{W_{n,m}\}_{m=0}^{\infty}$ is a submartingale with respect to $\{\mathcal{G}_{n,m}\}_{m=0}^{\infty}$. From Doob's decomposition, 
     we get $	W_{n,m}=W_{n,m}^A+W_{n,m}^M.$ 
	For the predictable process $\{W_{n,m}^A\}_{m=0}^{\infty}$, from \eqref{eq:exp_W_pos}, 
	\begin{align} \label{eq:incr_seq_W_tild}
	W_{n,m}^A=\sum_{i=0}^{m-1} \mathbb{E} \left[W_{n,i+1}-W_{n,i} \Bc \mathcal{G}_{n,i}\right] = \sum_{i=0}^{m-1} \tilde{\delta}\left(N_{\beta_n+i}+A_{\beta_n+i}\right). 
	\end{align}
	Next, we use Doob's optional stopping theorem 
 for the martingale sequence $\{W_{n,m}^M\}_{m=0}^{\infty}$ to find $\mathbb{E} \left[W_{n,\tau_n}^M \Bc \mathcal{F}_{n}\right] $. The stopping time $\tau_n$ has finite expectation as argued in \Cref{lem:tau_finite_mean},  and
	\begin{align}
	\mathbb{E}\left[\left|W_{n,i+1}^M-W_{n,i}^M\right| \Bc \mathcal{G}_{n,i}\right]&=  \mathbb{E}\left[\left|W_{n,i+1}-W_{n,i}-\left(W_{n,i+1}^A-W_{n,i}^A\right)\right| \Bc \mathcal{G}_{n,i}\right]  \nonumber \\
	&= \mathbb{E}\left[\left|D_{\beta_n+i+1}-\mathbb{E} \left[D_{\beta_n+i+1} \Bc \mathcal{G}_{n,i}\right] \right]\Bc \mathcal{G}_{n,i}\right]
\leq  \mathbb{E} \left[ \left|2 D_{\beta_n+i+1} \right| \Bc \mathcal{G}_{n,i}\right],
	\label{doob}
	\end{align}
	 where \eqref{doob} is bounded by a constant, as argued in \eqref{bdd}. After verifying the conditions of the optional stopping theorem, we  are able to use this theorem to get $\mathbb{E} \big[W_{n,\tau_n}^M \Bc \mathcal{F}_{n}\big] =
	\mathbb{E} \big[W_{n,0}^M \Bc \mathcal{F}_{n}\big]
	=Y_{n}.$
 From \eqref{incr_sequence_X} and \eqref{eq:incr_seq_W_tild},  we can find $Y_n^A$ as follows
	\begin{equation} \label{eq:incr_seq_Y}
	Y_n^A=\tilde{\delta} \sum_{m=0}^{n-1} \mathbb{E} \Big[\sum_{i=0}^{\tau_m-1} \left(N_{\beta_m+i}+A_{\beta_m+i}\right) \Bc \mathcal{F}_{m}\Big].
	\end{equation}
	Note that $A_{\beta_m}=1$, as arrival $\beta_n$ is accepted by the definition of the sampling times $\{\beta_n\}_{n=0}^\infty$. Hence,
$	\mathbb{E} \Big[\sum_{i=0}^{\tau_m-1} \left(N_{\beta_m+i}+A_{\beta_m+i}\right) \Bc \mathcal{F}_{m}\Big] \geq 1$, which gives 	
	$Y_n^A \geq \tilde{\delta} n$.
\Halmos \endproof
We next state the strong law of large numbers for martingale sequences in \Cref{thm:SLLN} and then, using this result, prove \Cref{lem:mart_seq}.  
\begin{Theorem}\cite[Corollary 7.3.2]{prob_shiryaev} \label{thm:SLLN}
	let $\{M_n\}_{n=0}^\infty$ be a martingale sequence with $M_0=0$ and $\mathbb{E} \big[\left|M_n\right|^{2r}\big]<\infty$
	for some $r\geq 1 $, and it satisfies $\sum_{n=1}^{\infty} {n^{-(1+r)}}{\mathbb{E} \left[| M_n-M_{n-1}|^{2r}\right]} < \infty.$
	Then, 
	\begin{equation*}
	\lim_{n \to \infty}\frac{M_n}{n}= 0. \quad \emph{a.s.}
	\end{equation*}
\end{Theorem}
\begin{Lemma} \label{lem:mart_seq}
The martingale process $\{Y_n^M\}_{n=0}^{\infty}$ found by Doob's decomposition of $\{Y_n\}_{n=0}^{\infty}$ satisfies
	\begin{align*}
	\lim_{n \to \infty} \frac{Y_n^M}{n}=0. \quad \emph{a.s.}
	\end{align*}
\end{Lemma}
\proof{Proof of \Cref{lem:mart_seq}.}
We prove \Cref{lem:mart_seq} for $\mu>c/R$.  
We first derive upper and lower bounds for the martingale difference sequence $Y_{n+1}^M-Y_{n}^M$. We have 
	\begin{align}
	Y_{n+1}^M-Y_{n}^M  &= Y_{n+1}-Y_{n}-\left(Y_{n+1}^A-Y_{n}^A\right)  = \sum_{i=1}^{\tau_n} D_{\beta_n+i}-\mathbb{E} \Big[ \tilde{\delta}\sum_{i=0}^{\tau_n-1} \left(N_{\beta_n+i}+A_{\beta_n+i}\right) \Bc \mathcal{F}_{n}\Big]  \label{eq:incr_MDS}\\
	&= \sum_{i=1}^{\tau_n} \Big( g\Big(T_{\beta_n+i},M_{\beta_n+i},\frac{c}{R}\Big)-   h\Big(T_{\beta_n+i},N_{\beta_n+i},\frac{c}{R}\Big) \Big)-\mathbb{E} \Big[ \tilde{\delta}\sum_{i=0}^{\tau_n-1} \left(N_{\beta_n+i}+A_{\beta_n+i}\right) \Bc \mathcal{F}_{n}\Big]  \label{eq:def_D_MDS} ,
	\end{align}
	where \eqref{eq:incr_MDS} is true by \eqref{eq:incr_seq_Y}, and \eqref{eq:def_D_MDS} follows from the definition of $D_i$. 
	To derive an upper bound for the martingale difference sequence, we only consider the non-negative terms in \eqref{eq:def_D_MDS} as below
	\begin{equation} \label{upper_bound}
	Y_{n+1}^M-Y_{n}^M \leq \sum_{i=1}^{\tau_n} g\Big(T_{\beta_n+i},M_{\beta_n+i},\frac{c}{R}\Big)  
	\leq k \frac{R}{c} \tau_n ,
	\end{equation}
	 which holds as for $t>0$, we have $g\left(t,1 ,x\right) \leq \frac{1}{x}$. 
	To find a lower bound, using the non-positive terms, 
	\begin{align} \label{lower_bound}
	Y_{n+1}^M-Y_{n}^M &\geq -\sum_{i=1}^{\tau_n} h\Big(T_{\beta_n+i},N_{\beta_n+i},\frac{c}{R}\Big) -\mathbb{E} \Big[ \tilde{\delta}\sum_{i=0}^{\tau_n-1} \left(N_{\beta_n+i}+A_{\beta_n+i}\right) \Bc \mathcal{F}_{n}\Big] \nonumber\\
 & \geq -k\sum_{i=1}^{\tau_n} T_{\beta_n+i}-\tilde{\delta} k \mathbb{E} \left[\tau_n \Bc \mathcal{F}_{n}\right],
	\end{align}
 where we have used  the definition of function $h$. From \Cref{lem:tau_finite_mean}, $\tilde{\delta} k \mathbb{E} \left[\tau_n \Bc \mathcal{F}_{n}\right]$ is bounded by a constant, which we call $c_{\tilde{\delta}}$.
	By \eqref{upper_bound} and \eqref{lower_bound}, we have 
	\begin{align} \label{bounds}
	-k\sum_{i=1}^{\tau_n} T_{\beta_n+i} -c_{\tilde{\delta}}\leq Y_{n+1}^M-Y_{n}^M \leq k\frac{R}{c} \tau_n.
	\end{align}
 We next verify the conditions of  \Cref{thm:SLLN} for the martingale sequence $Y_n^M$ with $r=1$. From \eqref{bounds},  
	\begin{equation} \label{SLLN_mart}
	\mathbb{E}\big[\left(Y_{n+1}^M-Y_{n}^M\right)^2\big]  \leq k^2 \frac{R^2}{c^2} \mathbb{E}\left[ \tau_n^2 \right]+k^2\mathbb{E}  \Big[\Big( \sum_{i=1}^{\tau_n} T_{\beta_n+i} \Big)^2\Big]+ 2kc_{\tilde{\delta}}\mathbb{E}  \Big[ \sum_{i=1}^{\tau_n} T_{\beta_n+i} \Big]+c_{\tilde{\delta}}^2.
	\end{equation}
	We aim to show the right-hand side of \eqref{SLLN_mart} is bounded by a constant independent of $n$. From Wald's equation \cite[Theorem 4.8.6]{durrett2019probability}, we have that $\mathbb{E}  \left[ \sum_{i=1}^{\tau_n} T_{\beta_n+i} \right]$ is bounded by a constant. For the second term, we use Wald's second equation \cite[Exercise 4.8.4]{durrett2019probability} for \textit{i.i.d.} random variables $\{\tilde{T}_{i}\}_{i=1}^n$ defined as $\tilde{T}_{i}:= T_{\beta_n+i}-\frac{1}{\lambda},$
	with $\mathbb{E} [ \tilde{T}_{i}  ]=0$ for all $i$. We take $\tilde{S}_n:=\sum_{i=1}^n \tilde{T}_{i}.$
	From Wald's second equation, for stopping time $\tau_n$ with finite expectation, $	\mathbb{E} \big[\tilde{S}_{\tau_n}^2\big]=\frac{1}{\lambda^2} \mathbb{E} \left[\tau_n\right]$. In addition, from the definition of $\tilde{S}_n$, we have $\mathbb{E} \big[\tilde{S}_{\tau_n}^2\big]=\mathbb{E} \big[  \big( \sum_{i=1}^{\tau_n} T_{\beta_n+i} - \frac{\tau_n}{\lambda}\big)^2 \big].$
Finally, we bound the second term on the right-hand side of \eqref{SLLN_mart}  with a constant as below
\begin{align}
    \mathbb{E} \Big[ \Big( \sum_{i=1}^{\tau_n} T_{\beta_n+i} \Big)^2\Big]&=\frac{1}{\lambda^2} \mathbb{E} [\tau_n]+\frac{2}{\lambda}\mathbb{E}  \Big[\tau_n\sum_{i=1}^{\tau_n} T_{\beta_n+i} \Big]-\frac{1}{\lambda^2} \mathbb{E} \Big[\tau_n^2\Big] \nonumber\\
    &\leq \frac{1}{\lambda^2} \mathbb{E}  [\tau_n]+\frac{1}{\lambda}\mathbb{E}  \Big[ \sum_{i=1}^{\tau_n} 2\tau_nT_{\beta_n+i} \Big] 
    \leq \frac{1}{\lambda^2} \mathbb{E}  [\tau_n]+\frac{1}{\lambda}\mathbb{E}  \Big[ \sum_{i=1}^{\tau_n} T_{\beta_n+i}^2 \Big]+\frac{1}{\lambda}\mathbb{E} \Big[\tau_n^3\Big].  \label{ET2}
\end{align}
	The last line  uses inequality $2xy \leq x^2+y^2$. We argued that  the moments of $\tau_n$ are bounded by the moments of the first hitting time to $0$ of a finite-state irreducible Markov chain  found by sampling system $Q^{(n)}$, or $\zeta_n$, and thus, are finite. Hence, the first and third terms  of \eqref{ET2} are bounded by a constant. 
 By Wald's equation,  the second term is also bounded by a constant. In conclusion,  \eqref{ET2} is bounded by a constant independent of $n$. Similarly, the first term on the right-hand side of \eqref{SLLN_mart}  is also bounded by a constant.
	Now, we verify the condition of \Cref{thm:SLLN} as follows  
	\begin{equation*}
	\sum_{n=1}^{\infty} \frac{\mathbb{E}\big[\left(Y_{n}^M-Y_{n-1}^M\right)^2\big] }{n^{2}} \leq c_5 \sum_{n=1}^{\infty}  \frac{1}{n^2}<\infty,
	\end{equation*}
	and the conditions of \Cref{thm:SLLN} are satisfied. Thus, by \Cref{thm:SLLN}, 
	$\lim_{n \to +\infty} \frac{Y_n^M}{n}=0$ 
    \emph{a.s.}
\Halmos \endproof

We now present the main result of this subsection in \Cref{thm:Convergence_mu>c/R_multi}, 
which proves the asymptotic optimality of policy $\pi_{\mathrm{Alg1}}$ 
for any $\mu>0$  for the multi-server  queueing system.  
The proof of this theorem is based on the submartingale (or supermartingale) property of the sequence $\{Y_n\}_{n=0}^{\infty}$. 

\begin{Theorem}
	\label{thm:Convergence_mu>c/R_multi}
		 Consider the multi-server Erlang-B queueing system with $k$ servers and service rate $\mu$. For any $\mu \in (0,+\infty)$,  policy $\pi_{\mathrm{Alg1}}$ converges to the best-in-class policy $\pi^*$. Specifically, for $\mu \in (c/R,+\infty)$, 
$Y_n$ converges to $+\infty$ \emph{a.s.}
	and the proposed policy  admits all arrivals after a random finite time subject to availability. Similarly, for $\mu \in (0,c/R)$,  
$Y_n$ converges to $-\infty$ \emph{a.s.},
		 and after a random finite time, an arrival is only accepted with a probability that converges to 0 as $n \rightarrow +\infty$.
\end{Theorem}
	
\proof{Proof of \Cref{thm:Convergence_mu>c/R_multi}.}
	For $\mu \in (c/R,+\infty)$, by Doob's decomposition for submartingale $\{Y_n\}_{n=0}^{\infty}$ and Lemmas \ref{lem:incr_seq} and \ref{lem:mart_seq}, $\lim_{n \to +\infty} Y_n=+\infty$ \emph{a.s.}
	In \Cref{alg1}, $X_{S(\cdot)}$ determines the acceptance rule, and between arrival $\beta_n$ and $\beta_{n+1}$,  $X_{S(\cdot)}$ is either equal to $X_{\beta_n}=Y_n$ or $X_{\beta_{n+1}}=Y_{n+1}$. Hence, the sign of $Y_{n}$ and $Y_{n+1}$ determines the acceptance rule between arrival $\beta_n$ and $\beta_{n+1}$. Thus, after a finite time, as long as there is an available server, the arrival is accepted, and $\pi_{\mathrm{Alg1}}$ converges to the best-in-class policy $\pi^*$. 
	The same arguments apply for the regime of 	$\mu \in (0,c/R)$. 
\Halmos \endproof

 
\subsubsection{Finite-time Performance Analysis} \label{subsec:regret_analysis_multi_server}

In this section, we characterize the regret in terms of the submartingale (or supermartingale) sequence $\{Y_n\}_{n=0}^{\infty}$ and processes $\{Y_n^A\}_{n=0}^{\infty}$ and $\{Y_n^M\}_{n=0}^{\infty}$ found from Doob's decomposition. 
As the sign of $\{Y_n\}_{n=0}^{\infty}$ determines the acceptance rule, we provide an upper bound for the probability of the event that $Y_n$ has an undesirable sign. Without loss of generality, in describing the methodology we assume that $\mu \in (c/R,+\infty)$ and from Doob's decomposition and \Cref{lem:incr_seq},  
	\begin{equation} \label{eq:doob_less_mart}
	\Pr\left(Y_{n} \leq 0\right)=\Pr\left(Y_{n}^A+Y_{n}^M \leq 0\right) \leq \Pr\big(Y_{n}^M \leq -\tilde{\delta}_1n\big) \text{ for some $\tilde{\delta}_1>0$}.
	\end{equation}
 Thus, it suffices to bound $\Pr\big(Y_{n}^M \leq -\tilde{\delta}_1n\big)$, as done in \Cref{lem:multi_server_prob}.  
 The proof of \Cref{lem:multi_server_prob} given in \Cref{sec:proof_multi_server_prob},
verifies a conditional sub-exponential property for the martingale difference sequence $\{Y_{n+1}^M-Y_n^M\}_{n=0}^{\infty}$, and utilizes a Bernstein-type bound for martingale difference sequences.  

\begin{Lemma} \label{lem:multi_server_prob}
Consider a multi-server Erlang-B queueing system with service rate $\mu$ following  policy $\pi_{\mathrm{Alg1}}$. For $\mu \in (c/R,+\infty)$,  there exists a problem-dependent constant $c_3$  
such that 
\begin{equation*}
    \Pr \big(  Y_n^M \leq -\tilde{\delta}_1n  \big) \leq \exp \left(-c_3n\right),
\end{equation*}
and for any $\mu \in (0,c/R)$, there exists a positive problem-dependent constant $c_4$ 
such that 
\begin{equation*}
    \Pr \big(  Y_n^M \geq -\tilde{\delta}_2 n \big) \leq \exp \left(-c_4n\right).
\end{equation*}
\end{Lemma}

We first give an upper bound for the expected regret when 
$\mu>c/R$. 
In this regime, when $Y_n$ is positive, $\pi_{\mathrm{Alg1}}$ 
follows the best-in-class policy $\pi^*$. 
However, for non-positive $Y_n$, the arrival is only admitted with a given probability. 
We quantify the impact of the arrivals 
for which $Y_n$ is non-positive using {the exponentially decaying probability established in} \Cref{lem:multi_server_prob}. Finally, in \Cref{thm:regret_mu>c/R_multi}, we argue
that for the Erlang-B queueing system with $\mu \in (c/R,+\infty)$ and 
function $f(n)$ such that $\log(f)=o(n)$, finite regret is achieved.

\begin{Theorem}
	\label{thm:regret_mu>c/R_multi}
	Consider the multi-server Erlang-B queueing system with $k$ servers and service rate $\mu$. For any $\mu \in (c/R,+\infty)$ and (valid) function $f$ such that $\log(f)=o(n)$, the expected regret $\E \left[\mathcal{R};  \pi_{\mathrm{Alg1}} \left(n\right) \right] $ under policy $\pi_{\mathrm{Alg1}}$ is upper bounded by a constant independent of $n$. 
\end{Theorem}

\proof{Proof of \Cref{thm:regret_mu>c/R_multi}.}
	 Let $K_n$ be the number of arrivals rejected after or at $\beta_n+\tau_n$ and before the first acceptance, $\beta_{n+1}$, i.e., $     K_n=\min \left\{i \geq 0 : A_{\beta_n+\tau_n+i}=1\right\}=\beta_{n+1}-\beta_n-\tau_n.$
	Note that if $Y_{n}>0$, the proposed policy will accept all arrivals from $\beta_{n-1}+\tau_{n-1}$ up to $\beta_{n}+\tau_n$ (subject to availability). 
 In this case, $\beta_{n-1}+\tau_{n-1}= \beta_n$.  But, if $Y_{n} \leq 0$, the arrivals are accepted with a certain probability and can contribute to the expected regret. Thus, we upper bound the regret as below
	\begin{align*} 
	\E \left[\mathcal{R} \left(n\right) ;  \pi_{\mathrm{Alg1}} \right]  &\leq \E \left[ \tau_0 \right]+
	\mathbb{E} \Big[\sum_{i=1}^{\infty} (\tau_i+K_{i-1}) \mathbbm{1} \{Y_{i} \leq 0\}\Big] \nonumber =\sum_{i=0}^{\infty}\mathbb{E} \left[ \tau_i \mathbbm{1} \left\{Y_{i} \leq 0\right\}\right] +\sum_{i=1}^{\infty}\mathbb{E} \left[K_{i-1} \mathbbm{1} \left\{Y_{i} \leq 0\right\}\right] \nonumber \\
	& \leq \sum_{i=0}^{\infty}\mathbb{E} \left[ \tau_i \Bc Y_{i} \leq 0\right] \Pr \left(Y_{i} \leq 0\right)+\sum_{i=1}^{\infty} f\left(i\right)\Pr\left(Y_{i} \leq 0\right)
	\\ &\leq \sum_{i=0}^{\infty}\mathbb{E} \left[ \tau_i \Bc Y_{i} \leq 0\right] \exp \left(-c_3i\right)+\sum_{i=1}^{\infty} f\left(i\right) \exp \left(-c_3i\right). 
	\end{align*}
 In the second line, we used the fact that given $Y_i \leq 0$, $K_{i}$ is geometric with $\E[K_i]\leq f(i)$.
 The last inequality follows from \eqref{eq:doob_less_mart} and \Cref{lem:multi_server_prob}.  In \Cref{lem:tau_finite_mean}, we argued that $\mathbb{E} [ \tau_i | Y_{i-1} \leq 0]$ is bounded by a constant.  Hence, for any function $f$ with $\log(f)=o(n)$, the expected regret is finite.
\Halmos \endproof

Next, we present the finite-time performance guarantee 
when $\mu <c/R$. In this regime, the expected regret consists of 
two terms. The first term arises from the arrivals for which 
$Y_n>0$, and we 
use the exponentially decaying probability of 
\Cref{lem:multi_server_prob} to bound this term.  The second term results from the arrivals accepted with a given probability when $Y_n\leq 0$. 
 We will use \Cref{lem:geom_log_up_bound} presented below to address this term;  proof is given in \Cref{sec:proof_geom_log_up}. In conclusion, \Cref{thm:regret_mu<c/R_multi} proves a polynomial in $\log(n)$ upper bound for the expected regret in the case of $\mu \in (0,c/R)$. 

\begin{Lemma} \label{lem:geom_log_up_bound}
     Let $ f(n)=\exp(n^{1-\epsilon})$ and $d= \ceil{ 3(\log^{\frac{1}{1-\epsilon}} (n+1) )}$ for a fixed $\epsilon \in (0,1)$. Then, for independent geometric random variables $\{y_i\}_{i=1}^n$  with corresponding success probabilities $\{f(i )^{-1}\}_{i=1}^n$, the sum $\sum_{i=d}^{n-1}i\Pr(y_1+\cdots+y_i <n , y_1+\cdots+y_{i+1} \geq n)$ is bounded by a constant determined by $\epsilon$. 
\end{Lemma}


\begin{Theorem} 
	\label{thm:regret_mu<c/R_multi}
	Consider the multi-server Erlang-B queueing system with $k$ servers and service rate  $\mu \in (0,c/R)$. For  $f(n)=\exp\left(n^{1-\epsilon}\right) $, the expected regret 
	under  policy $\pi_{\mathrm{Alg1}}$ is $\E \left[\mathcal{R} \left(n\right);  \pi_{\mathrm{Alg1}} \right] = O\big( \log^\frac{1}{1-\epsilon} (n)\big).$
\end{Theorem}

\proof{Proof of \Cref{thm:regret_mu<c/R_multi}.}
	In this case, the expected regret up to arrival $n$  equals the expected number of arrivals accepted from the first $n$ arrivals. Hence, we have 
\begin{align} 
	\E \Big[\mathcal{R} \left(n\right) ;  \pi_{\mathrm{Alg1}}\Big] &= \mathbb{E} \Big[\sum_{i=0}^{n-1} \mathbbm{1} \left\{A_i=1\right\}\Big] \nonumber \\
	& =\mathbb{E}\Big[ \sum_{i=0}^{n-1} \mathbbm{1} \left\{A_i=1,X_{S\left(i\right)}>0\right\}\Big]+\mathbb{E}\Big[ \sum_{i=0}^{n-1} \mathbbm{1} \left\{A_i=1,X_{S\left(i\right)}\leq 0\right\}\Big].\label{regret_part}
\end{align}
We first upper bound the first term using \eqref{eq:doob_less_mart} and \Cref{lem:multi_server_prob} as follows  
\begin{align} \label{eq:constant_part}
	\mathbb{E}\Big[ \sum_{i=0}^{n-1} \mathbbm{1} \left\{A_i=1,X_{S\left(i\right)}>0\right\}\Big] 
	 \leq \sum_{i=0}^{\infty} \mathbb{E} \Big[ \mathbbm{1} \left\{Y_i>0\right\}\tau_{i}  \Big]
	\leq \sum_{i=0}^{\infty}  \mathbb{E} \Big[ \tau_{i} \Bc Y_i>0 \Big]\exp(-c_{4} i).
\end{align}
	By \Cref{lem:tau_finite_mean}, the above summation is bounded by a constant $c_p$.  Next, we upper bound the second term of \eqref{regret_part}. { As defined before, $\tau_i$ is the first $j > \beta_i$ such that $N_{\beta_i+j}=0$ and  $K_i$ is equal to $\beta_{i+1}-\beta_i-\tau_i$, i.e., the number of rejected arrivals before arrival $\beta_{i+1}$ and after or at $\beta_{i}+\tau_i$. If $X_{\beta_{i}+\tau_i} \leq 0$, then  $K_i$ is  geometric with parameter $1/\alpha_{\beta_{i}+\tau_i}$. We define $G(i)$ as the index of the first accepted arrival after $i-1$ arrivals, or 
 $G(i) :=\min_m \big\{m\geq 0:\sum_{j=0}^{m}(\tau_j+K_j) \geq i\big\}$. We also take $F(i)$ to be the smallest $m$ such that the sum of the first $m+1$ geometric trials exceeds $i-1$, i.e., $F(i) := \min_m \big\{m\geq 0:\sum_{j\in B_m}(K_j +1 )\geq i\big\}$, where $B_m=\{j: 0\leq j \leq m, X_{\beta_{j}+\tau_j} \leq 0\}$. From these definitions, it follows that ${G(i) \leq F(i)}$.
 }
 The second term of \eqref{regret_part} is less than or equal to the expected number of times an arrival $i< n$ with  $X_{S(i)}\leq 0$ is accepted until arrival $\beta_{G(n)+1}$. Therefore, we have  
	\begin{align} 
	&\mathbb{E}\Big[ \sum_{i=0}^{n-1} \mathbbm{1} \left\{A_i=1,X_{S\left(i\right)}\leq 0\right\}\Big]  \leq \mathbb{E} \Big[ \sum_{i=0}^{G\left(n\right) }\tau_i \mathbbm{1} \left\{X_{\beta_i}\leq 0\right\}\Big] \leq \mathbb{E} \Big[ \sum_{i=0}^{F\left(n\right) }\tau_i \mathbbm{1} \left\{X_{\beta_i}\leq 0\right\}
 \Big] \nonumber \\
&	 \leq  \sum_{j=0}^{n-1} \mathbb{E} \Big[\sum_{i=0}^{F\left(n\right) } \tau_i \Bc F\left(n\right)=j\Big]\Pr\left(F\left(n\right)=j\right)\nonumber \\
& \leq c_{\tau}\sum_{j=0}^{d} (j+1)\Pr\left(F\left(n\right)=j\right)+c_{\tau}\sum_{j=d+1}^{n-1}(j+1)\Pr\Big(\sum_{i=1}^{j-1}y_i < n , \sum_{i=1}^{j}y_i \geq n\Big) \label{reg3_4}  \\
& \leq c_{\tau} \E\left[(F(n)+1) \mathbbm{1}\{F(n)\leq d \} \right]+c_{\tau}\sum_{j=d}^{n-2}(j+2)\Pr\Big(\sum_{i=1}^{j}y_i < n , \sum_{i=1}^{j+1}y_i \geq n\Big)
,\label{reg3} 
	\end{align}
	where $\{y_i\}_{i=1}^n$ are defined in \Cref{lem:geom_log_up_bound}, 
 $d= \ceil{ 3(\log^{\frac{1}{1-\epsilon}}(n+1) )}$,  $c_{\tau}$ is {found using \Cref{lem:tau_finite_mean} and is proportional to
	    $\sum_{j=0}^{k}\frac{\lambda^j}{\mu^j j!}$.
 Furthermore, \eqref{reg3_4} follows from the fact that the event $\{F(n)=j\}$ is equivalent to the event $\{\sum_{i=1}^{j-1}y_i < n , \sum_{i=1}^{j}y_i \geq n\}$.
 From \Cref{lem:geom_log_up_bound}, \eqref{reg3} is bounded by $c_{\tau}(d+3+c_{\epsilon})$, where $c_{\epsilon}$ is a constant determined by $\epsilon$.  }Finally, from \eqref{eq:constant_part} and \eqref{reg3}, \Cref{thm:regret_mu<c/R_multi} follows. 
\Halmos \endproof

\begin{Remark} \label{rem:tradeoff}
There is an exploration-exploitation trade-off in selecting
$f(n)$ on the two sides of $\mu=c/R$. When admitting is optimal, we want 
$f(n)$ to increase to  
infinity as slow as possible. 
Also, based on the proof of \Cref{thm:regret_mu>c/R_multi},  for our current bound, 
we cannot take $f(n)$ to grow exponentially fast since its exponent needs to depend on unknown 
$\mu$ to ensure constant regret. Conversely, when blocking all arrivals is optimal, we need $f(n)$ to converge to infinity as fast as possible.
As the learning algorithm needs to be agnostic about the parameter regime, $ f(n)=\exp\left(n^{1-\epsilon}\right)$ is a good choice: it ensures constant regret in one regime and polynomial regret in $\log(n)$ in the other. 
\end{Remark}

We next consider a decreasing sequence of $\epsilon$ values by choosing $\epsilon_n:=\frac{\varepsilon}{\sqrt{1+\log (n+1)}}$ for $n\geq 1$, where $\varepsilon \in (0,1)$.  The algorithm corresponding to the exploration function  $ f(n)=\exp\left(n^{1-\epsilon_n}\right)$ is asymptotically optimal from  \Cref{thm:Convergence_mu>c/R_multi}.  To determine the regret when $\mu>c/R$, we observe that $\log(f)=o(n)$ and the regret in this regime remains finite. 
For the case of $\mu<c/R$, we are able to reduce the order of regret further to $\log(n)$, as shown in \Cref{cor:regret_mu_less_c_multi} with proof in \Cref{sec:proof_cor_regret_mu_less_c/R}.

\begin{Corollary}
	\label{cor:regret_mu_less_c_multi}
	Consider the multi-server Erlang-B queueing system with $k$ servers and service rate  $\mu \in (0,c/R)$. For  $f(n)=\exp\left(n^{1-\epsilon_n}\right) $ where $\epsilon_n=\frac{\varepsilon}{\sqrt{1+\log (n+1)}}$ for all $n\geq 1$ and $\varepsilon \in (0,1)$, the expected regret  
	under  policy $\pi_{\mathrm{Alg1}}$ 
	is $\E \left[\mathcal{R} \left(n\right);\pi_{\mathrm{Alg1}} \right] = O\big( \log (n)\big)$.
\end{Corollary}

\begin{Remark}
    For some parameters, our problem setting overlaps with the setting of \cite{zhang2022learning}: when $\mu\leq c/R$ and $k=1$, our setting can be viewed as learning in an $M/M/1$ system with the optimal admission threshold of $0$, and when $c/R<\mu \leq h(\lambda, c/R)<+\infty$ (for a function $h$), our setting corresponds to an $M/M/1$ system with an optimal threshold of $1$.
    However, our work samples the system only at arrivals, in contrast to \cite{zhang2022learning} which samples the system at all times (so service times of departed jobs are known).
    Despite observing 
    less information, our proposed policy exhibits the same regret behavior as \cite{zhang2022learning} as shown in \Cref{cor:regret_mu_less_c_multi} and \Cref{thm:regret_mu>c/R_multi}.
    \end{Remark}


{\color{blue}

\section{Multi-server Queueing Model with a Finite Waiting Room} \label{sec:queue_sys_fin_buffer}

\subsection{Problem Formulation}\label{sec:problem_form_MB}

For $0 \leq j \leq k + N$, let $\eta(j)$ denote the stationary probability of having $j$ jobs in the $M/M/k/k+N$ queueing system, under a policy that admits all arrivals whenever capacity allows. The long-term average reward according to the reward function $K(a,s)=a(R-cs)$ is given as
\begin{align}
\begin{split}
	\limsup\limits_{n\rightarrow\infty}\frac{1}{n}\sum_{i=0}^{n-1}  \mathbb{E}[K(A_i,\sigma_i)]
		& = \sum_{j=0}^{k-1} \eta(j)\left(R-\frac{c}{\mu}\right) + \sum_{j=k}^{k+N-1} \eta(j)\left(R-\frac{c}{\mu} - \frac{c(j-k+1)}{k\mu}\right) \\
        & = \sum_{j=0}^{k-1} \eta(j)\left(R-\frac{c}{\mu}\right) + \sum_{j=k}^{k+N-1} \eta(j)\left(R-\frac{c(j+1)}{k\mu}\right),
\end{split}
        \label{eq:av_cost_multi} 
\end{align}	
where the terms in the second summation include the waiting time in the queue before service as well. Moreover, the stationary distribution $\eta(j)$ is given as follows:
\begin{equation*}
\eta(j)=\begin{dcases}
	\frac{(k \rho)^j}{j!} \pi_0,  \hspace{1.5cm} \text{for } j=0,\ldots,k-1; \\
	\frac{k^k \rho^j}{k!} \pi_0,  \hspace{1.5cm} \text{for } j=k,\ldots,k+N;
\end{dcases}
\end{equation*}
where $\rho=\frac{\lambda}{k\mu}$ and $\pi_0=\left(   \sum_{i=0}^{k} \frac{(k\rho)^i}{i!}+ \frac{(k\rho)^k}{k!}  \sum_{i=k+1}^{k+N} \rho^{i-k} \right)^{-1}$.
We need to compare \eqref{eq:av_cost_multi} with zero to find whether it is better to admit all arrivals (subject to room) or reject all. Thus, we need to compare the following term---RHS of \eqref{eq:av_cost_multi} divided by $\pi_0$---with zero:
\begin{align}
&\left(R-\frac{c}{\mu}\right)   \sum_{j=0}^{k-1}	\frac{(k \rho)^j}{j!} +\frac{k^{k}}{k!}\sum_{j=k}^{k+N-1}\left(R-\frac{c(j+1)}{k\mu}\right) \rho^j \nonumber \\
&=\left(R-\frac{k c  \rho}{\lambda}\right)   \sum_{j=0}^{k-1}	\frac{(k \rho)^j}{j!} +\frac{k^{k}}{k!}\sum_{j=k}^{k+N-1}\left(R-\frac{ c(j+1)\rho}{\lambda}\right) \rho^j\nonumber \\
&= \sum_{j=0}^{k-1}	\left(R -\frac{cj}{\lambda}\right)\frac{(k\rho)^j}{j!} - \frac{ck^k\rho^k}{\lambda (k-1)!}  +\frac{k^{k}}{k!}\sum_{j=k+1}^{k+N-1}\left(R-\frac{ cj}{\lambda}\right) \rho^j + \frac{k^k}{k!} \left(R \rho^k - \frac{c(N+k)}{\lambda}\rho^{k+N}\right)\nonumber \\
&=\sum_{j=0}^{k-1}	\left(R-\frac{cj }{\lambda}\right)	\frac{(k \rho)^j}{j!} +\frac{k^{k}}{k!}\sum_{j=k}^{k+N-1} \left(R-\frac{cj}{\lambda}\right) \rho^j -\frac{ck^k(N+k)}{\lambda k!} \rho^{k+N}, \label{eq:c/R_eq_multi}
\end{align}
where in the penultimate step we combine terms in both summations with a shift in the second term. By Descartes' rule of signs, \eqref{eq:c/R_eq_multi} has exactly one positive root, denoted as \(\rho^*\), which gives the threshold service rate \(\mu^* = \tfrac{\lambda}{k \rho^*}\). Note that, in the Erlang-B queueing system, we previously identified \(\mu^*\) as \(c/R\) in \Cref{sec:Proposedpolicy}.

For better readability, we start by reminding the reader about the notation: for $i\ge 1$, $N_i$ is the number of customers in the system seen by the $i^{\mathrm{th}}$ arrival, $A_i\in \{0,1\}$ is the not admit/admit decision made for the $i^{\mathrm{th}}$ arrival, $t_i$ is inter-arrival time between the $(i-1)^{\mathrm{th}}$ and $i^{\mathrm{th}}$ arrivals, $m_i$ is the number of departures in $t_i$, and $n_i$ is the number of customers in the system after $t_i$.

Next, we compute the transition probabilities $p \left(m_i,n_i,t_i;\mu\right)$ in order to construct the log-likelihood function. For $1 \leq n_{i-1}+a_{i-1} \leq k$, the transition probabilities, given $m_i+n_i=N_{i-1}+A_{i-1}$, are provided in \eqref{prob_base} (as they're equivalent to the $N=0$ case).
On the other hand, if $k<n_{i-1}+a_{i-1} \leq k+N$ and $n_i \geq k$, we have 
\begin{align}\label{eq:prob_term2}
	p \left(m_i,n_i,t_i;\mu\right)= \frac{(k \mu t_i)^{m_i}}{m_i!} \exp(-k\mu t_i),
\end{align} 
corresponding to the probability of having exactly $m_i$ points of a Poisson process of rate $k\mu$ in an interval of length $t_i$ (since all the $k$ servers are always busy). Finally, if $k<n_{i-1}+a_{i-1} \leq k+N$ and $n_i<k$, defining $s_i=n_{i-1}+a_{i-1}$, we have
\begin{align} 
\begin{split}
	& p \left(m_i,n_i,t_i;\mu\right)\\
    & =\int_{r=0}^{t_i} \frac{(k \mu)^{s_i-k}r^{s_i-k-1} \exp(-k\mu r)}{(s_i-k-1)!}  \binom{k}{n_i} \exp(-n_i \mu (t_i-r)) (1- \exp(-\mu (t_i-r)))^{k-n_i} dr, 
\end{split} \label{eq:prob_term3}
\end{align} 
where the inner integration represents the probability of observing $s_i-k$ departures by time $r$ leaving the system with exactly $k$ customers, which follows an Erlang distribution with rate $k\mu$ as in \eqref{eq:prob_term2} (since all $k$ servers are busy until then), together with the probability of exactly an additional $k-n_i$ departures from $k$ independent servers in the remaining $t_i-r$ time (as free servers need to idle after completing service). Then, the log-likelihood function is given as
\begin{align} 
\begin{split}
		l\left(\mathcal{H}_n;\mu\right) &=\sum_{i=1}^{n}\mathbb{I}\{1 \leq n_i+m_i \leq k \} \left( m_i\log(1-\exp(-\mu t_i))  -n_i\mu t_i \right) \\
  &+\sum_{i=1}^{n}\mathbb{I}\{k+1 \leq n_i+m_i ,k \leq n_i \} \left(  m_i \log(\mu)  -k\mu t_i \right) \\
	&+\sum_{i=1}^{n}\mathbb{I}\{k+1 \leq n_i+m_i , n_i\leq k-1 \}   \log(\tilde p \left(m_i,n_i,t_i;\mu\right)), 
\end{split}\label{eq:log_likelihood_multi}
\end{align}
where 
\begin{align*}
\tilde p \left(m_i,n_i,t_i;\mu\right)
=  \exp(-n_i \mu t_i)\int_{r=0}^{t_i}  \mu^{s_i-k}r^{s_i-k-1} (\exp(-\mu r)- \exp(-\mu t_i))^{k-n_i} dr.
\end{align*}
To determine the service rate that maximizes the log-likelihood function, we first need to compute its derivative with respect to $\mu$. First, note that the derivative of $\tilde{p} \left(m_i, n_i, t_i; \mu\right)$ is given by
\begin{align*}
	&\tilde p' \left(m_i,n_i,t_i;\mu\right)\\
	&=-n_it_i\tilde p \left(m_i,n_i,t_i;\mu\right)+\frac{s_i-k}{\mu}\tilde p \left(m_i,n_i,t_i;\mu\right)+ \\
	&(k-n_i)\exp(-n_i \mu t_i)\int_{r=0}^{t_i}  \mu^{s_i-k}r^{s_i-k-1}  (-r\exp(-\mu r)+t_i\exp(-\mu t_i))(\exp(-\mu r)- \exp(-\mu t_i))^{k-n_i-1} dr.
\end{align*}
Then, the derivative of $  \log(\tilde p \left(m_i,n_i,t_i;\mu\right))$ is 
\begin{align}
\begin{split}
	& \frac{\tilde p' \left(m_i,n_i,t_i;\mu\right)}{\tilde p\left(m_i,n_i,t_i;\mu\right)} \\ & =-n_it_i 
	+\frac{s_i-k}{\mu}+\frac{(k-n_i)\int_{r=0}^{t_i}  r^{s_i-k-1}  (-r\exp((t_i-r)\mu)+t_i)(\exp((t_i-r)\mu )- 1)^{k-n_i-1} dr}{\int_{r=0}^{t_i}  r^{s_i-k-1} (\exp((t_i-r)\mu)- 1)^{k-n_i} dr}. 
\end{split}\label{eq:tilde_p_term}
\end{align}
Setting $\psi_{1,i}=k-n_i$ and $\psi_{2,i}=s_i-k$, we simplify the last two terms in \eqref{eq:tilde_p_term} as follows
\begin{align}
\begin{split}
	& \frac{\psi_{2,i}}{\mu}
	+\frac{\psi_{1,i}\int_{r=0}^{t_i}  r^{\psi_{2,i}-1}  (-r\exp((t_i-r)\mu)+t_i)(\exp((t_i-r)\mu )- 1)^{\psi_{1,i}-1} dr}{\int_{r=0}^{t_i}  r^{\psi_{2,i}-1} (\exp((t_i-r)\mu)- 1)^{\psi_{1,i}} dr}= \\
	&\scalebox{0.80}{$\dfrac{\psi_{2,i} \int_{r=0}^{t_i}  r^{\psi_{2,i}-1} (\exp((t_i-r)\mu)- 1)^{\psi_{1,i}} dr+\mu\psi_{1,i}\int_{r=0}^{t_i}  r^{\psi_{2,i}-1}  (-r\exp((t_i-r)\mu)+t_i)(\exp((t_i-r)\mu )- 1)^{\psi_{1,i}-1} dr}{\mu\int_{r=0}^{t_i}  r^{\psi_{2,i}-1} (\exp((t_i-r)\mu)- 1)^{\psi_{1,i}} dr}.$ }
\end{split}\label{eq:simplify_multi}
\end{align}
Using integration by parts, given $\psi_{1,i}, \psi_{2,i} \geq 1$, we have
\begin{align*}
\psi_{2,i} \int_{r=0}^{t_i}  r^{\psi_{2,i}-1} (\exp((t_i-r)\mu)- 1)^{\psi_{1,i}} dr=\mu \psi_{1,i}\int_{r=0}^{t_i}  r^{\psi_{2,i}} \exp((t_i-r)\mu)(\exp((t_i-r)\mu)- 1)^{\psi_{1,i}-1} dr
\end{align*}
Plugging it back in \eqref{eq:simplify_multi}, we can simplify \eqref{eq:simplify_multi} further to get
\begin{align}\label{eq:simplify_der}
	\frac{\psi_{1,i} t_i\int_{r=0}^{t_i}  r^{\psi_{2,i}-1} (\exp((t_i-r)\mu )- 1)^{\psi_{1,i}-1} dr}{\int_{r=0}^{t_i}  r^{\psi_{2,i}-1} (\exp((t_i-r)\mu)- 1)^{\psi_{1,i}} dr},
\end{align}
which would further help us to simplify \eqref{eq:tilde_p_term} as below
\begin{align}\label{eq:ratio_tilde_p}
	\frac{\tilde p' \left(m_i,n_i,t_i;\mu\right)}{\tilde p\left(m_i,n_i,t_i;\mu\right)}=-n_it_i+\frac{\psi_{1,i} t_i\int_{r=0}^{t_i}  r^{\psi_{2,i}-1} (\exp((t_i-r)\mu )- 1)^{\psi_{1,i}-1} dr}{\int_{r=0}^{t_i}  r^{\psi_{2,i}-1} (\exp((t_i-r)\mu)- 1)^{\psi_{1,i}} dr}.
\end{align}
Thus, from \eqref{eq:log_likelihood_multi}, the derivative of the log-likelihood function with respect to $\mu$ is given as
\begin{align}
	&l'\left(\mathcal{H}_n;\mu\right) \nonumber\\
 &=\sum_{i=1}^{n}\mathbb{I}\{1 \leq n_i+m_i \leq k \} \left(  \frac{m_it_i\exp(-\mu t_i)}{1-\exp(-\mu t_i)}  -n_i t_i \right)+\sum_{i=1}^{n}\mathbb{I}\{ k+1 \leq n_i+m_i,  n_i \geq k  \} \left( \frac{ m_i}{\mu}  -k t_i \right)\nonumber \\
	&+\sum_{i=1}^{n}\mathbb{I}\{k+1 \leq n_i+m_i , n_i\leq k-1 \}  \left(-n_it_i+\frac{\psi_{1,i} t_i\int_{r=0}^{t_i}  r^{\psi_{2,i}-1} (\exp((t_i-r)\mu )- 1)^{\psi_{1,i}-1} dr}{\int_{r=0}^{t_i}  r^{\psi_{2,i}-1} (\exp((t_i-r)\mu)- 1)^{\psi_{1,i}} dr}\right) \label{eq:der_log_ll}.
\end{align}
To demonstrate that the log-likelihood function has a unique positive maximum, we first establish that the log-likelihood function is concave, or equivalently, that the transition probabilities $p\left(m_i, n_i, t_i; \mu\right)$ are log-concave with respect to $\mu$. It is easy to see that the probability terms \eqref{prob_base} and \eqref{eq:prob_term2} are log-concave. To argue that \eqref{eq:prob_term3} is log-concave, it suffices show that if a function $f(\mu,r):\mathbb{R}_+^2 \rightarrow \mathbb{R}_+$ is log-concave in $\mu$ for all $r$, then $\int_{0}^t f(\mu,r)dr$ is also log-concave for any positive $t$. For the result to hold it suffices to show that $\frac{ \int_{0}^t f'(\mu,r)dr}{\int_{0}^t f(\mu,r)dr}  $ is decreasing with respect to $\mu$. From log-concavity of $f$, we have that $\frac{f'(\mu,r)}{f(\mu,r)}$ is decreasing with respect to $\mu$. Thus, for $\mu_1 \leq \mu_2$ and fixed $r_1,r_2 \in (0,t)$, we have $\frac{f(\mu_1,r_1)}{f'(\mu_1,r_1)} >\frac{f(\mu_2,r_1)}{f'(\mu_2,r_1)}$ and $\frac{f(\mu_1,r_2)}{f'(\mu_1,r_2)} >\frac{f(\mu_2,r_2)}{f'(\mu_2,r_2)}$. As a result,
\begin{equation} \label{eq:dcr_f_g_int}
\frac{f(\mu_1,r_1)+f(\mu_1,r_2)}{f'(\mu_1,r_1)+f'(\mu_1,r_2)} >\frac{f(\mu_2,r_1)+f(\mu_2,r_2)}{f'(\mu_2,r_1)+f'(\mu_2,r_2)},
\end{equation} 
and by the definition of Riemann integral, the function $\frac{\int_{r=0}^{t} f dr}{\int_{r=0}^{t} f' dr}$ is also decreasing with respect to $\mu$. After showing the concavity of the log-likelihood function, it follows that its derivative is decreasing and has at most one non-negative zero. From \eqref{eq:der_log_ll},
\begin{align*}
&\lim_{\mu \to 0 } \; l'\left(\mathcal{H}_n;\mu\right)= +\infty, &&\lim_{\mu \to + \infty } l'\left(\mathcal{H}_n;\mu\right)= - \sum_{i=1}^n \min(n_i,k) t_i.
\end{align*}
As a result, the derivative of the log-likelihood function has exactly one positive zero, which maximizes the concave log-likelihood function---denote it as $\hat{\mu}_n$.
Let $\psi_1=k-n$ and $\psi_2=n+m-k$. We define functions $g$ and $h$ as
\begin{align}
	g\left(t,m,n,\mu\right)&:=\mathbb{I}\{1 \leq n+m \leq k \}  \frac{mt\exp(-\mu t)}{1-\exp(-\mu t)}+  \mathbb{I}\{ k+1 \leq n+m, n \geq k\} \frac{ m}{\mu} \nonumber 
	\\ &+\mathbb{I}\{k+1 \leq n+m , n\leq k-1 \}  \frac{\psi_1 t\int_{r=0}^{t}  r^{\psi_2-1} (\exp((t-r)\mu )- 1)^{\psi_1-1} dr}{\int_{r=0}^{t}  r^{\psi_2-1} (\exp((t-r)\mu)- 1)^{\psi_1} dr},  \label{eq:g_multi}\\
    \begin{split}
	h\left(t,m,n,\mu\right)&:=\left(  \mathbb{I}\{1 \leq n+m \leq k \}+\mathbb{I}\{k+1 \leq n+m , n\leq k-1 \}    \right)nt +\mathbb{I}\{k+1 \leq m+n, n \geq k\} kt\\
    & = \min(n,k)t.
    \end{split}\label{eq:h_multi}
\end{align} 
 We can represent function $l'\left(\mathcal{H}_n;\mu\right)$ at arrival $n$ as 
\begin{equation}
	l'\left(\mathcal{H}_n;\mu\right) =\sum_{i=1}^{n} 	\frac{ p' \left(m_i,n_i,t_i;\mu\right)}{p\left(m_i,n_i,t_i;\mu\right)}= \sum_{i=1}^{n} g\left(t_i,m_i,n_i,{\mu}\right)-\sum_{i=1}^{n} h\left(t_i,m_i,n_i,{\mu}\right), \label{eq:der_Log_LL}
\end{equation}
where the summation $\sum_{i=1}^{n}  h\left(t_i,m_i,n_i,\mu\right)$ is a non-negative constant independent of $\mu$ and $\sum_{i=1}^{n} g\left(t_i,m_i,n_i,{\mu}\right)$ is a non-negative and decreasing function of $\mu$. Moreover, the maximum likelihood estimate $\hat{\mu}_n$ is a solution to the following equation:
\begin{equation} \label{eq:MLE_new_multi} 
	\sum_{i=1}^{n} g\left(T_i,M_i,N_i,\hat{\mu}_n\right)=\sum_{i=1}^{n} h\left(T_i,M_i,N_i,\hat{\mu}_n\right).
\end{equation}
From the above discussion, at arrival $n$ we again have the following two cases:
\begin{enumerate}[wide, labelindent=0pt]
	\item  
	$\sum_{i=1}^{n} g\left(T_i,M_i,N_i,\mu^*\right)>\sum_{i=1}^{n} h\left(T_i,M_i,N_i,\mu^*\right)$ implies that $\hat{\mu}_n>\mu^*$.
	\item 
	$\sum_{i=1}^{n} g\left(T_i,M_i,N_i,\mu^*\right)\leq \sum_{i=1}^{n} h\left(T_i,M_i,N_i,\mu^*\right)$ implies that $\hat{\mu}_n\leq \mu^*$.
\end{enumerate}

\subsection{Analysis}\label{subsec:anal_ms_buffer}

\subsubsection{Asymptotic Optimality}\label{subsec:asymp_optimal_multi_server_MB}
In this subsection, we will use the same notation as in \Cref{sec:multi_server_queueing_model_sec}.  To prove asymptotic optimality of our proposed algorithm, we will repeat the arguments of \Cref{subsec:asymp_optimal_multi_server} and show that random variable $|X_n|$ ($X_n$ defined in \eqref{state_MC_original_klarger}) converges to $\infty$ with the (limiting) sign of $X_n$ determined by the sign of $\mu-\mu^*$. Similar to the queueing system discussed in \Cref{lem:tau_finite_mean}, we can argue that all moments of random variable $\tau_n$ are bounded by a constant independent of $n$ by coupling the queueing system that follows \Cref{alg1} with system $Q^{(n)}$ that accepts all arrivals (subject to availability) and noting that the latter leads to a finite state Markov chain, which is irreducible and geometrically ergodic.  We now need to show that the process $\{W_{n,m}\}_{m=0}^{\infty}$, defined as in \eqref{eq:def_W}, is a submartingale or supermartingale sequence (based on the sign of $\mu-\mu^*$).
\begin{Lemma} \label{lem:MB_mart}
	Fix $n \geq 0$. For $\mu\in (\mu^*,+\infty)$, the stochastic process $\{W_{n,m}\}_{m=0}^{\infty}$ forms a submartingale  sequence with respect to the filtration  $\{\mathcal{G}_{n,m}\}_{m=0}^{\infty}$ (defined in \Cref{lem:W_sub}).
	For $\mu\in (0,\mu^*)$, the process $\{W_{n,m}\}_{m=0}^{\infty}$ is a supermartingale with respect to filtration
	$\{\mathcal{G}_{n,m}\}_{m=0}^{\infty}$.
\end{Lemma}
\proof{Proof of \Cref{lem:MB_mart}.}
WLOG,  we assume $\mu \in (\mu^*,+\infty)$. We follow the proof of \Cref{lem:W_sub} and  first show 
$\mathbb{E} \left[\left| W_{n,m}\right|\right]<\infty$. It suffices to show that for every $i$, the expectation $\mathbb{E} \left[\left| D_i\right|\right]=\mathbb{E} \left[\left| g\left(T_i,M_i,N_i,\mu^*\right)-h\left(T_i,M_i,N_i,\mu^*\right)\right|\right]$ is finite; see proof of \Cref{lem:W_sub}. We first argue that for exponential interarrival time $T$ and bounded variables $m$ and $n$ (based on our system parameters), we have $\mathbb E[g\left(T,m,n,\mu^*\right)] <\infty$ and $\mathbb E[h\left(T,m,n,\mu^*\right)] <\infty$. From \eqref{eq:h_multi}, we can see that $\mathbb E[h\left(T,m,n,\mu^*\right)] <\infty$.   Furthermore, to check that $\mathbb E[g\left(T,m,n,\mu^*\right)] <\infty$, we first note that $\frac{mt\exp(-\mu^* t)}{1-\exp(-\mu^* t)}< \frac{m}{\mu^*}$. Thus, it remains to check that the last term in \eqref{eq:g_multi} has finite expectation with respect to $T$.
For $n \in \mathbb{Z}_+$ and $a>0$,
\begin{align*}
	\int_{r=0}^t r^n \exp(a(t-r)) dr=\frac{n!}{a^{n+1}}\left( \exp(at) -\sum_{i=0}^{n}   \frac{(at)^i}{i!}	\right)=\frac{n!}{a^{n+1}}\sum_{i=n+1}^{\infty}   \frac{(at)^i}{i!}.
\end{align*}
From this, we can rewrite the last term in \eqref{eq:g_multi}  as
\begin{align}
&\frac{\psi_1  (\psi_2-1)! t \sum_{j=1}^{\psi_1-1}(-1)^{\psi_1-1-j}  {\psi_1-1 \choose j} (\mu^* j)^{-\psi_2}  \left(\exp(j\mu^* t)- \sum_{i=0}^{\psi_2-1}   \frac{(j \mu^* t)^i}{i!}	\right) +(-1)^{\psi_1-1} \psi_1 \psi_2^{-1} t^{\psi_2+1} }{  (\psi_2-1)!\sum_{j=1}^{\psi_1}(-1)^{\psi_1-j}  {\psi_1 \choose j} (\mu^* j)^{-\psi_2}   \left(\exp(j\mu^* t)- \sum_{i=0}^{\psi_2-1}   \frac{(j \mu^* t)^i}{i!}	\right) +(-1)^{\psi_1} {\psi_2}^{-1} t^{\psi_2}} \label{eq:simp_first_term}\\
&=\frac{\psi_1  (\psi_2-1)! t \sum_{j=1}^{\psi_1-1}(-1)^{\psi_1-1-j}  {\psi_1-1 \choose j} (\mu^* j)^{-\psi_2}  \left( \sum_{i=\psi_2}^{\infty}   \frac{(j \mu^* t)^i}{i!}	\right) +(-1)^{\psi_1-1} \psi_1 \psi_2^{-1} t^{\psi_2+1} }{  (\psi_2-1)!\sum_{j=1}^{\psi_1}(-1)^{\psi_1-j}  {\psi_1 \choose j} (\mu^* j)^{-\psi_2}   \left( \sum_{i=\psi_2}^{\infty}   \frac{(j \mu^* t)^i}{i!}	\right) +(-1)^{\psi_1} {\psi_2}^{-1} t^{\psi_2}} \nonumber\\
&=\frac{t^{\psi_2+1}\psi_1  (\psi_2-1)!  \sum_{j=1}^{\psi_1-1}(-1)^{\psi_1-1-j}  {\psi_1-1 \choose j}  \left(\sum_{i=0}^{\infty}   \frac{(j \mu^* t)^i}{(i+\psi_2)!}	\right) +(-1)^{\psi_1-1} \psi_1 \psi_2^{-1} t^{\psi_2+1} }{ t^{\psi_2} (\psi_2-1)!\sum_{j=1}^{\psi_1}(-1)^{\psi_1-j}  {\psi_1 \choose j} \left( \sum_{i=0}^{\infty}   \frac{(j \mu^* t)^i}{(i+\psi_2)!}	\right)+(-1)^{\psi_1} {\psi_2}^{-1} t^{\psi_2}}\nonumber \\
&=\frac{\psi_1 t \left( (\psi_2-1)!  \sum_{j=1}^{\psi_1-1}(-1)^{\psi_1-1-j}  {\psi_1-1 \choose j}  \left(\sum_{i=0}^{\infty}   \frac{(j \mu^* t)^i}{(i+\psi_2)!}	\right) +(-1)^{\psi_1-1} {\psi_2}^{-1}  \right)}{  (\psi_2-1)!\sum_{j=1}^{\psi_1}(-1)^{\psi_1-j}  {\psi_1 \choose j} \left( \sum_{i=0}^{\infty}   \frac{(j \mu^* t)^i}{(i+\psi_2)!}	\right)+(-1)^{\psi_1} {\psi_2}^{-1}}\label{eq:simp_last_term}
\end{align}
In \eqref{eq:simp_last_term}, the denominator and its first $\psi_1 - 1$ derivatives vanish at $t = 0$---because they are proportional to the value and derivatives of $((t - 1) + 1)^{\psi_1}$ at $t = 0$, as seen via the binomial expansion. The $\psi_1^\text{th}$ derivative at $t = 0$ is equal to $\frac{(\psi_1!)^2 (\psi_2 - 1)! (\mu^*)^{\psi_1}}{(\psi_1 + \psi_2)!}$. Similarly, for the numerator, its value and its first $\psi_1 - 1$ derivatives at $t = 0$ are zero, while the $\psi_1$-th derivative at $t = 0$ is equal to $\frac{\psi_1!(\psi_1-1)!(\psi_2-1)!(\mu^*)^{\psi_1-1}}{(\psi_1+\psi_2-1)!}$. Consequently, \eqref{eq:simp_last_term} is finite at $t=0$. Furthermore, since the denominator is nonnegative (as evident from \eqref{eq:g_multi}) and converges to zero as $t \to \infty$, it remains bounded, and thus,  $\mathbb E[g\left(T,m,n,\mu^*\right)] <\infty$, which means the expectation $\mathbb{E} \left[\left| D_i\right|\right]$ is also finite.

We now assume that the true service rate $\mu$ is greater than $\mu^*$ (the solution to \eqref{eq:c/R_eq_multi}). To complete the proof, from  \eqref{eq:exp_W_submart} we need to argue that $\mathbb{E}[X_{i+1}-X_{i}\Bc X_{i},N_{i},\alpha_i ,A_i]=\mathbb{E} \left[ g\left(T_{i+1},M_{i+1},N_{i+1},\mu^*\right)-h\left(T_{i+1},M_{i+1},N_{i+1},\mu^*\right) \Bc X_{i},N_{i},\alpha_i ,A_i\right]$ is non-negative for all $i$. For that, from \eqref{eq:der_Log_LL} we need to show 
\begin{equation*}
\mathbb{E}\Big[  \frac{ p' \left(M_{i+1},N_{i+1},T_{i+1};\mu^*\right)}{p\left(M_{i+1},N_{i+1},T_{i+1};\mu^*\right)}  \Bc X_{i},N_{i},\alpha_i ,A_i  \Big]>0.
\end{equation*}
We have 
\begin{align*}
&\mathbb{E}\Big[  \frac{ p' \left(M_{i+1},N_{i+1},T_{i+1};\mu^*\right)}{p\left(M_{i+1},N_{i+1},T_{i+1};\mu^*\right)}  \Bc X_{i},N_{i},\alpha_i ,A_i,T_{i+1}  \Big]\\&=  \sum_{m=0}^{N_i+A_i}  \frac{ p' \left(m,N_i+A_i-m,T_{i+1};\mu^*\right)}{p\left(m,N_i+A_i-m,T_{i+1};\mu^*\right)}p\left(m,N_i+A_i-m,T_{i+1};\mu\right).
\end{align*}
We first note that if $\mu=\mu^*$, we have 
\begin{align} \label{eq:likelihood_zero}
\mathbb{E}\Big[  \frac{ p' \left(M_{i+1},N_{i+1},T_{i+1};\mu^*\right)}{p\left(M_{i+1},N_{i+1},T_{i+1};\mu^*\right)}  \Bc X_{i},N_{i},\alpha_i ,A_i,T_{i+1}  \Big]=  \sum_{m=0}^{N_i+A_i}  { p' \left(m,N_i+A_i-m,T_{i+1};\mu^*\right)}=0.
\end{align}
In the arguments following \eqref{eq:ratio_tilde_p}, we showed that for fixed $m,n,t$,  the ratio $\frac{ p' \left(m,n,t;\mu\right)}{p\left(m,n,t;\mu\right)}$ is non-increasing with respect to $\mu$. Thus, if $\mu>\mu^*$, we have that 
\begin{align*}
\frac{ p' \left(m,n,t;\mu\right)}{p\left(m,n,t;\mu\right)}\leq \frac{ p' \left(m,n,t;\mu^*\right)}{p\left(m,n,t;\mu^*\right)}.
\end{align*}
As a result, for $\mu>\mu^*$, from \eqref{eq:likelihood_zero}
\begin{align}\label{eq:p_frac_MB}
	\mathbb{E}\Big[  \frac{ p' \left(M_{i+1},N_{i+1},T_{i+1};\mu^*\right)}{p\left(M_{i+1},N_{i+1},T_{i+1};\mu^*\right)}  \Bc X_{i},N_{i},\alpha_i ,A_i  \Big]\geq  0,
\end{align}
and the martingale property of $\{W_{n,m}\}_{m=0}^{\infty}$ is established. 
\Halmos
\endproof
From the martingale property of  process $\{W_{n,m}\}_{m=0}^{\infty}$ and following the exact arguments of \Cref{thm:martingale}, we can show that process $\{Y_n\}_{n=0}^{\infty}$  is a submartingale or supermartingale sequence (depending on the sign of $\mu-\mu^*$) with respect to filtration $\{\mathcal{F}_n\}_{n=0}^{\infty}$ (defined in \Cref{thm:martingale}). Thus, from Doob's decomposition, we get $Y_n=Y_n^A+Y_n^M$,
where $Y_n^M$ is a martingale sequence, and $Y_n^A$ is a predictable and almost surely increasing (or decreasing) sequence with $Y_0^A=0$. Next, in parallel with  \Cref{lem:incr_seq} and \Cref{lem:mart_seq}, we present and prove the following two lemmas. 

\begin{Lemma} \label{lem:incr_seq_MB}
For $\mu \in (\mu^*,+\infty)$, there exists a positive problem-dependent constant $\tilde{\delta}_1$ such that the process $\{Y_n^A\}_{n=0}^{\infty}$ from Doob's decomposition of $\{Y_n\}_{n=0}^{\infty}$ satisfies
    $Y_n^A \geq \tilde{\delta}_1  n$ 
    \emph{a.s.}, 
 and for $\mu \in (0,\mu^*)$, there exists a negative constant $\tilde{\delta}_1$ such that the 
 process $\{Y_n^A\}_{n=0}^{\infty}$ satisfies
    $Y_n^A \leq \tilde{\delta}_1 n$ 
    \emph{a.s.}
\end{Lemma}
\proof{Proof of \Cref{lem:incr_seq_MB}}
WLOG,  we assume $\mu \in (\mu^*,+\infty)$. In \eqref{eq:der_log_ll} and the arguments following it, we showed that  for fixed $m,n,t$ such that $m+n \geq 1$, the ratio $\frac{ p' \left(m,n,t;\mu\right)}{p\left(m,n,t;\mu\right)}$ is strictly decreasing with respect to $\mu$. By repeating the same arguments as in \eqref{eq:likelihood_zero}-\eqref{eq:p_frac_MB}, we can show that when $N_i+A_i\geq 1$, the expectation $\mathbb{E}\Big[  \frac{ p' \left(M_{i+1},N_{i+1},T_{i+1};\mu^*\right)}{p\left(M_{i+1},N_{i+1},T_{i+1};\mu^*\right)}  \Bc X_{i},N_{i},\alpha_i ,A_i  \Big]$ is positive.  Define $\tilde{\delta}_1$ as
\begin{equation*}
	  \tilde{\delta}_1 := \min_{n,a \text{ s.t. } n+a>0 } \mathbb{E}\Big[  \frac{ p' \left(M_{i+1},N_{i+1},T_{i+1};\mu^*\right)}{p\left(M_{i+1},N_{i+1},T_{i+1};\mu^*\right)}  \Bc N_{i}=n,A_i=a  \Big]>0.
\end{equation*}
By repeating the same arguments as {in the proof of} \Cref{lem:incr_seq}, we have 
\begin{align} 
	Y_n^A&=\sum_{m=0}^{n-1} \left( \mathbb{E} \left[W_{m,\tau_m} \Bc \mathcal{F}_{m}\right] -Y_{m} \right)=\sum_{m=0}^{n-1}\mathbb{E} \left[W^A_{m,\tau_m} \Bc \mathcal{F}_{m}\right]\nonumber\\
 &= \sum_{m=0}^{n-1}\mathbb{E}\left[\sum_{i=0}^{\tau_m-1}\mathbb{E} \left[W_{m,i+1}-W_{m,i}\Bc  \mathcal{G}_{m,i}\right]\Bc \mathcal{F}_{m}\right]=  \sum_{m=0}^{n-1}\mathbb{E}\left[\sum_{i=0}^{\tau_m-1}\mathbb{E} \left[D_{\beta_m+i+1}\Bc  \mathcal{G}_{m,i}\right]\Bc \mathcal{F}_{m}\right], \label{eq:incr_seq_MB}
	\end{align} where the last line follows from \eqref{eq:def_W}. 
 In the proof of \Cref{lem:mart_seq_MB}, we showed that for $i\in \mathbb{N} $, 
\begin{equation*}
\mathbb{E}[D_{i+1}\Bc X_{i},N_{i},\alpha_i ,A_i]=	\mathbb{E}\Big[  \frac{ p' \left(M_{i+1},N_{i+1},T_{i+1};\mu^*\right)}{p\left(M_{i+1},N_{i+1},T_{i+1};\mu^*\right)}  \Bc X_{i},N_{i},\alpha_i ,A_i  \Big],
\end{equation*} which is greater than $ \tilde{\delta}_1 $ if $N_i+A_i \geq 1$. Since $A_{\beta_m}=1$, at least one term in the inner summation of  \eqref{eq:incr_seq_MB} exceeds  $ \tilde{\delta}_1$ and  we can conclude that $Y_n^A \geq  \tilde{\delta}_1n$.
\Halmos
\endproof


\begin{Lemma} \label{lem:mart_seq_MB}
The martingale process $\{Y_n^M\}_{n=0}^{\infty}$ found by Doob's decomposition of $\{Y_n\}_{n=0}^{\infty}$ satisfies
	\begin{align*}
	\lim_{n \to \infty} \frac{Y_n^M}{n}=0. \quad \emph{a.s.}
	\end{align*}
\end{Lemma}
\proof{Proof of \Cref{lem:mart_seq_MB}}
Similar to \Cref{lem:mart_seq}, We derive upper and lower bounds for the martingale difference sequence $Y_{n+1}^M-Y_n^M$. We have $Y_{n+1}^M-Y_{n}^M  = Y_{n+1}-Y_{n}-\left(Y_{n+1}^A-Y_{n}^A\right)$ and similar to \eqref{upper_bound}, we can derive an upper bound as $ Y_{n+1}^M-Y_{n}^M \leq \sum_{i=1}^{\tau_n} g\Big(T_{\beta_n+i},M_{\beta_n+i},N_{\beta_n+i},\mu^*\Big)  $. In the proof of \Cref{lem:MB_mart}, we argued that function $g(t,m,n,\mu^*)$ is bounded with respect to $t$. As variables $m$ and $n$ can take finitely many values, we can see that function $g(t,m,n,\mu^*)$ has a finite maximum $c_g$ with respect to variables $t$, $m$, and $n$.
For the lower bound, similar to \eqref{lower_bound}, we have 
\begin{align*}
    Y_{n+1}^M-Y_{n}^M &\geq -\sum_{i=1}^{\tau_n} h\Big(T_{\beta_n+i},M_{\beta_n+i},N_{\beta_n+i},\mu^*\Big)-\left(Y_{n+1}^A-Y_{n}^A\right) \\
    &\geq -k\sum_{i=1}^{\tau_n} T_{\beta_n+i} - \mathbb{E}\left[\sum_{i=0}^{\tau_n-1}\mathbb{E} \left[D_{\beta_n+i+1}\Bc  \mathcal{G}_{n,i}\right]\Bc \mathcal{F}_{n}\right] \tag*{(from \eqref{eq:incr_seq_MB})} \\
    &\geq -k\sum_{i=1}^{\tau_n} T_{\beta_n+i} - c_g \mathbb{E} [\tau_n \Bc \mathcal{F}_{n}].
\end{align*}
Finally, we have 
\begin{equation}\label{eq:upp_low_bnd}
    -k\sum_{i=1}^{\tau_n} T_{\beta_n+i} - c_{\tilde \delta}\leq Y_{n+1}^M-Y_{n}^M \leq c_g \tau_n.
\end{equation}
The rest of proof follows from the same arguments as \Cref{lem:mart_seq} and verifying the conditions of \Cref{thm:SLLN} (the strong law of large numbers for martingale sequences) for the martingale sequence $Y_n^M$ with $r=1$.
\Halmos
\endproof
From the above lemmas, asymptotic optimality of our proposed policy for the multi-server  queueing system with a finite buffer is proved as stated below. 
\begin{Theorem}
	Consider an $M/M/k/k+N$ queueing system with $k$ servers, buffer of size $N$, and service rate $\mu$. For any $\mu \in (0,+\infty)$,  policy $\pi_{\mathrm{Alg1}}$  admits all arrivals after a random finite time subject to availability. Similarly, for $\mu \in (0,\mu^*)$, after a random finite time, an arrival is only accepted with a probability that converges to 0 as $n \rightarrow +\infty$.
\end{Theorem}

\subsubsection{Finite-time Performance Analysis}\label{subsec:regret_analysis_multi_server_MB}
 In this subsection, we demonstrate that the regret bounds established in \Cref{thm:regret_mu>c/R_multi} and \Cref{cor:regret_mu_less_c_multi} can be extended to the multi-server queueing system with a finite buffer. To this end, we first prove an analog of \Cref{lem:multi_server_prob} for the buffered setting. Using the upper and lower bounds derived in \eqref{eq:upp_low_bnd}, we replicate the proof of \Cref{lem:multi_server_prob} (presented in \Cref{sec:proof_multi_server_prob}) to obtain the following lemma.

\begin{Lemma} 
	Consider a multi-server queueing system with finite buffer and service rate $\mu$ following  policy $\pi_{\mathrm{Alg1}}$. For $\mu \in (\mu^*,+\infty)$,  there exists a problem-dependent constant $c_5$  
	such that 
	\begin{equation*}
		\Pr \big(  Y_n^M \leq -\tilde{\delta}_1n  \big) \leq \exp \left(-c_5n\right),
	\end{equation*}
	and for any $\mu \in (0,\mu^*)$, there exists a positive problem-dependent constant $c_6$ 
	such that 
	\begin{equation*}
		\Pr \big(  Y_n^M \geq -\tilde{\delta}_2 n \big) \leq \exp \left(-c_6n\right).
	\end{equation*}
\end{Lemma}

From this lemma, the regret bounds of \Cref{thm:regret_mu<c/R_multi}, \Cref{cor:regret_mu_less_c_multi}, and \Cref{thm:regret_mu>c/R_multi}  follow using the same arguments. Specifically, when the service rate $\mu$ exceeds the boundary value $\mu^*$, \Cref{alg1} incurs a constant regret. Moreover, for the exploration function defined in \Cref{cor:regret_mu_less_c_multi} a $O(\log(n))$ regret arises  when $\mu <\mu^*$. 

\begin{Theorem}
    	Consider an $M/M/k/k+N$ queueing system with $k$ servers, buffer of size $N$, and service rate $\mu$.  For any $\mu \in (c/R,+\infty)$ and (valid) function $f$ such that $\log(f)=o(n)$, the expected regret $\E \left[\mathcal{R} \left(n\right) ;  \pi_{\mathrm{Alg1}}\right] $ under policy $\pi_{\mathrm{Alg1}}$ is upper bounded by a constant independent of $n$. 
\end{Theorem}
\begin{Theorem}
            	Consider an $M/M/k/k+N$ queueing system with $k$ servers, buffer of size $N$, and service rate $\mu$.  For  $f(n)=\exp\left(n^{1-\epsilon_n}\right) $ where $\epsilon_n=\frac{\varepsilon}{\sqrt{1+\log (n+1)}}$ for all $n\geq 1$ and $\varepsilon \in (0,1)$, the expected regret  
	under  policy $\pi_{\mathrm{Alg1}}$ 
	is $\E \left[\mathcal{R} \left(n\right);\pi_{\mathrm{Alg1}} \right] = O\big( \log (n)\big)$.
\end{Theorem}

}
\section{Simulation-based Numerical Results}
\label{sec:experimental results}

\begin{figure*}[t] 
	\centering
	\subfloat[$ \mu \in (c/R,+\infty)$  \label{fig:sim_diff_mu_a}]{\includegraphics[width=0.4\linewidth]{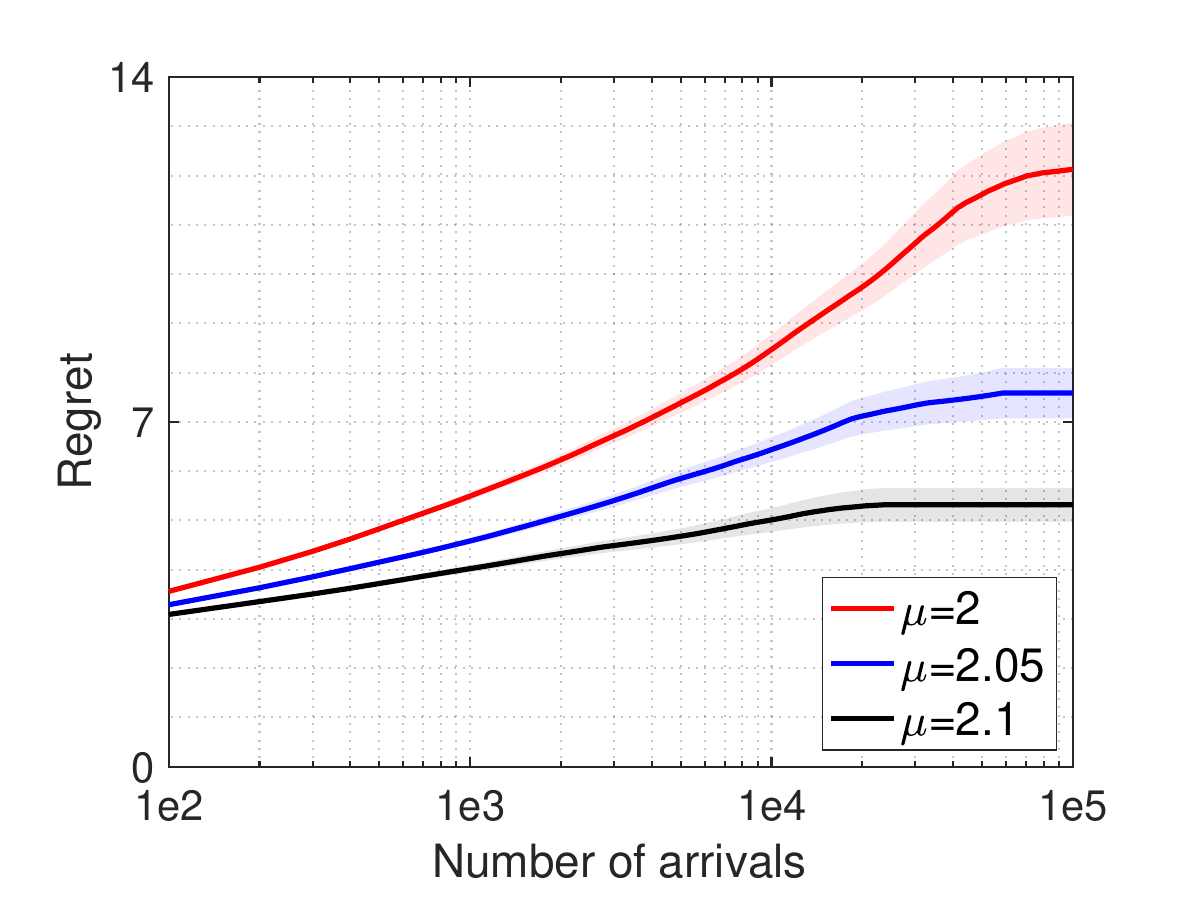}} \hspace{1cm}
	\subfloat[$ \mu \in (0,c/R)$  \label{fig:sim_diff_mu_b}]
 {\includegraphics[width=0.4\linewidth]{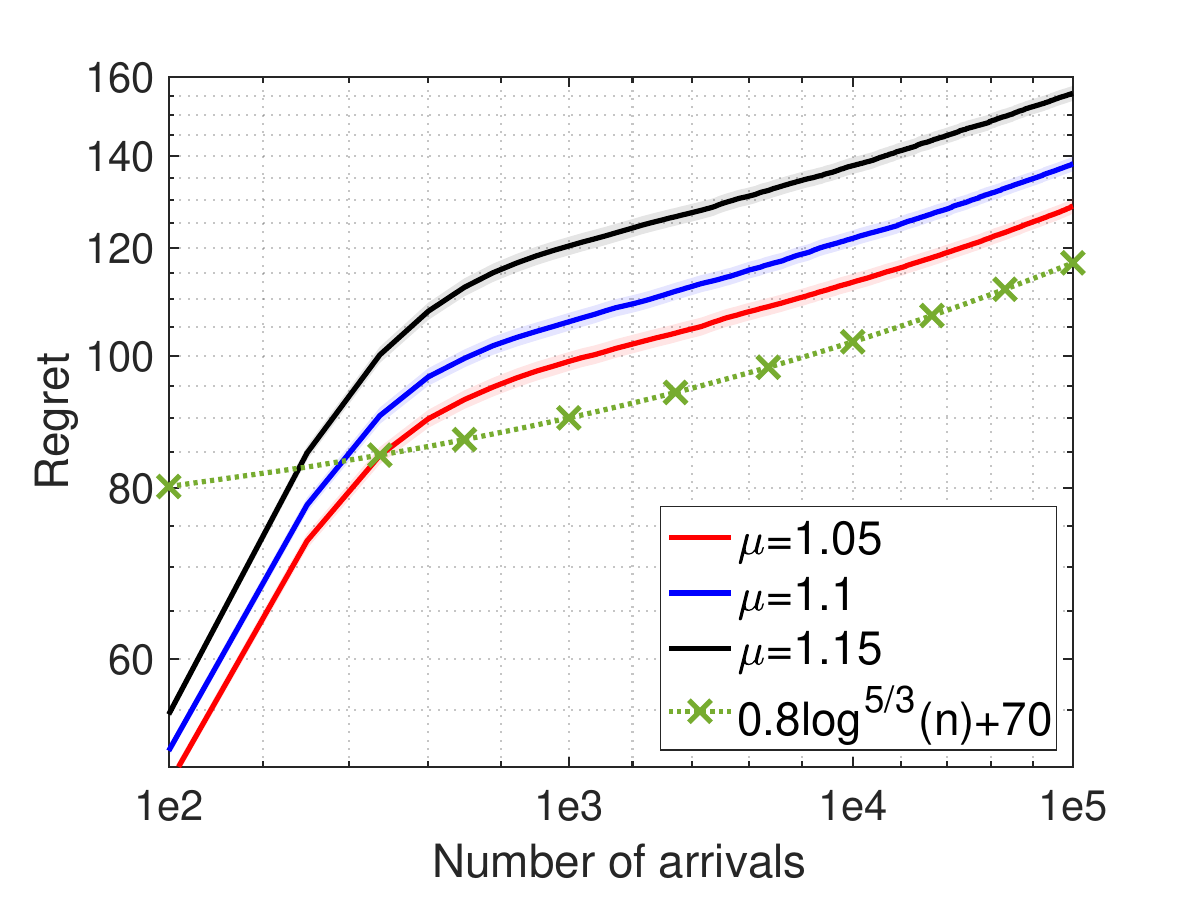}}
	\caption{Variations of regret for different service rates in a 5 server system with $\lambda=5$, $c/R=1.3$, $\epsilon=0.4$, $\frac{1}{1-\epsilon}=5/3$, and $f(n)=\exp\left(n^{1-\epsilon}\right)$ following \Cref{alg1}.} 
	\label{fig:sim_diff_mu}
	\end{figure*}

 In this section, we empirically evaluate the performance of  policy   $\pi_{\mathrm{Alg1}}$. We calculate the regret by finding the difference in the number of sub-optimal actions taken by $\pi_{\mathrm{Alg1}}$ compared to the optimal policy with the knowledge of the true service rate. The regret is averaged over 2500 simulation runs and plotted versus the number of incoming jobs. From our simulations, it can be observed that the proposed policy achieves finite regret for $\mu >c/R$, as predicted by our analysis. Further, the finite-time performance in the other regime corroborates our theoretical bound. We demonstrate the finite-time performance under various service rates and compare the performance of $\pi_{\mathrm{Alg1}}$ against the dispatching scheme that updates the acceptance rule at every arrival. Furthermore, we compare the performance of 
 \Cref{alg1} with two RL algorithms: R-learning and Thompson sampling. 
 In the plots of this section, we use a logarithmic scale for the x-axis when $\mu>c/R$ to display the variations clearly. Moreover, 
 when $\mu<c/R$, we plot  $\log\log(x)$ versus $\log(y)$ as the regret is bounded by a polynomial in $\log(n)$ 
 and this axes scaling provides a clearer depiction of the regret. Furthermore, the shaded regions in all plots indicate the  $\pm \sigma$ area of the mean regret.

\Cref{fig:sim_diff_mu} shows the regret performance for different service rates in a system with $5$~servers, $\lambda=5$, $c/R=1.3$, and $f(n)=\exp\left(n^{0.6}\right)$. 
We can see that the regret grows as the service rate approaches the boundary value  $c/R$ (from either direction). In addition, as the gap between the service rate and the boundary value narrows, the regret converges more slowly to its final value when $\mu>c/R$. The results of Figures \ref{fig:sim_diff_mu_a} and \ref{fig:sim_diff_mu_b} corroborate the theoretical bounds of Theorems \ref{thm:regret_mu>c/R_multi} and \ref{thm:regret_mu<c/R_multi}.

\begin{figure*}[t] 
	\centering
	\subfloat[$ \mu \in (c/R,+\infty)$  \label{fig:compare_acc_all_a}]{\includegraphics[width=0.4\linewidth]{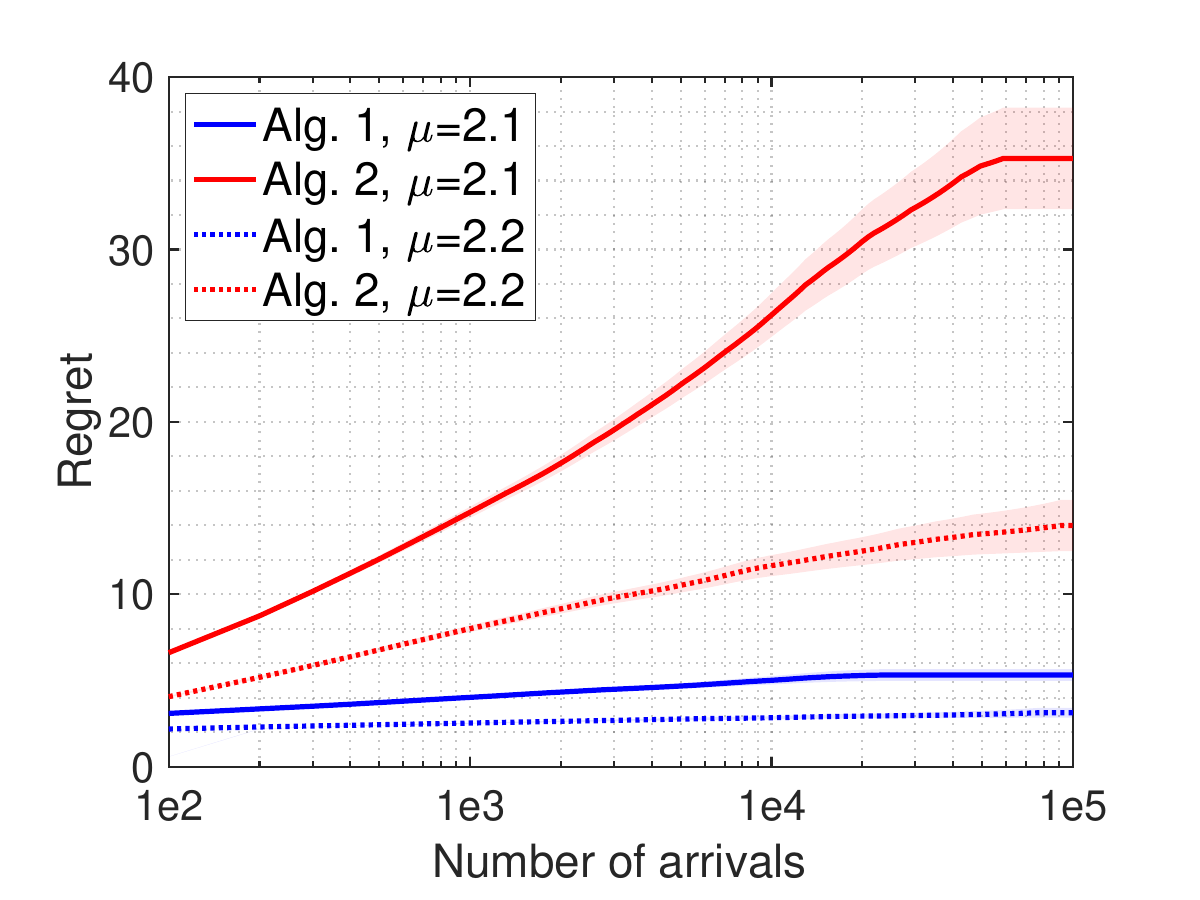}} \hspace{1cm}
	\subfloat[$ \mu \in (0,c/R)$  \label{fig:compare_acc_all_b}]{\includegraphics[width=0.4\linewidth]{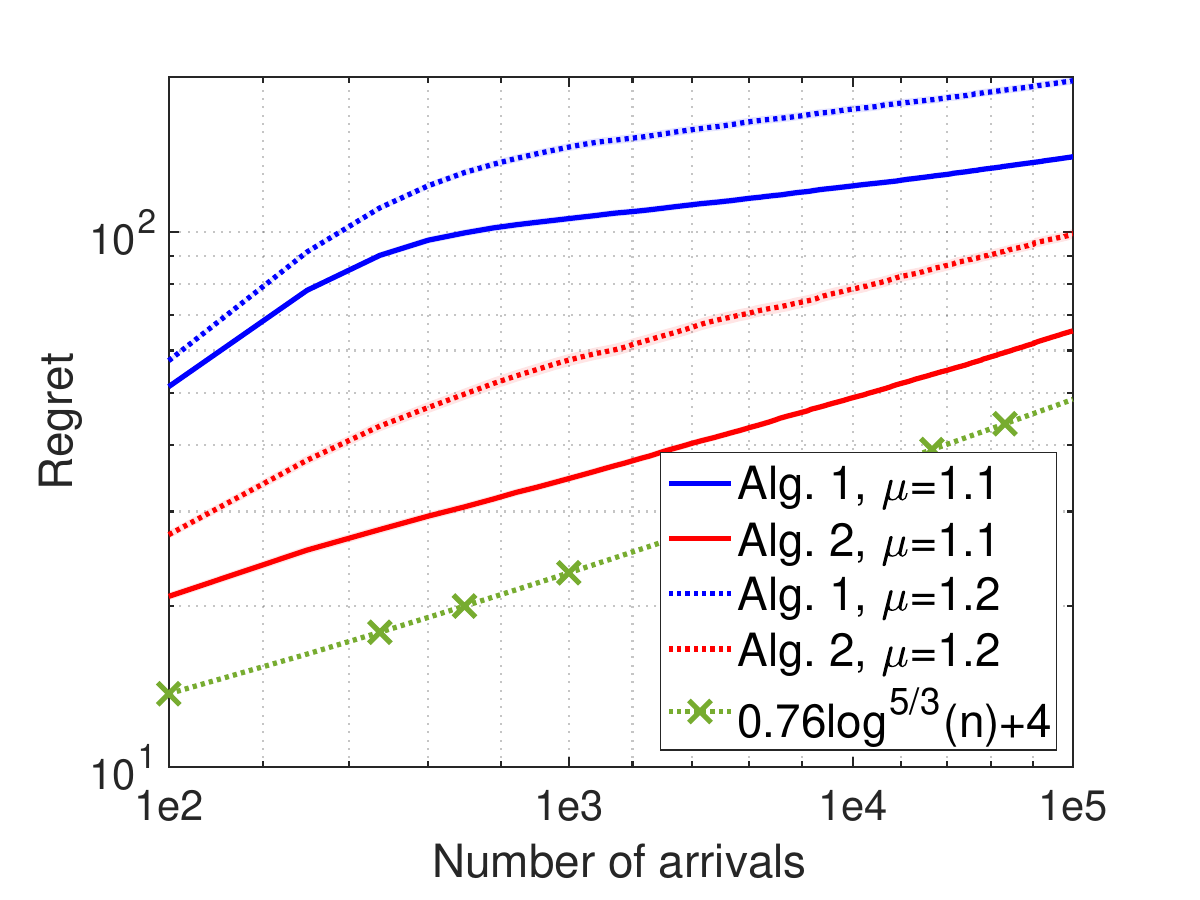}}
	\caption{Comparison of regret performance of \Cref{alg1} against Algorithm 2 
 in a 5 server system with $\lambda=5$, $c/R=1.3$, $\epsilon=0.4$, $\frac{1}{1-\epsilon}=5/3$, and $f(n)=\exp\left(n^{1-\epsilon}\right)$.} 
	\label{fig:compare_acc_all}
	\end{figure*}

In \Cref{fig:compare_acc_all}, we compare the performance of \Cref{alg1} with an algorithm that updates the policy parameters at every arrival, called Algorithm 2. The problem parameters $\lambda, k, c, R,\epsilon$ are the same as the setting of \Cref{fig:sim_diff_mu}. 
In  Algorithm 2, the admission probability decays faster than \Cref{alg1}, resulting in less exploration and better regret performance when $\mu<c/R$. From \Cref{fig:compare_acc_all_a}, \Cref{alg1} outperforms Algorithm 2 for $\mu >c/R$ due to its slower decaying admission probability and the greater number of arrivals accepted. Another intuitive justification is that \Cref{alg1} updates the policy parameters after observing a collection of arrivals, not prematurely after one sample, and the resulting averaging (and variance reduction) is useful in this regime. 

	\begin{figure*}[t] 
	\centering
	\subfloat[$\mu=2.5, \mu \in (c/R,+\infty)$ \label{fig:compare_RL_a}]{\includegraphics[width=0.4\linewidth]{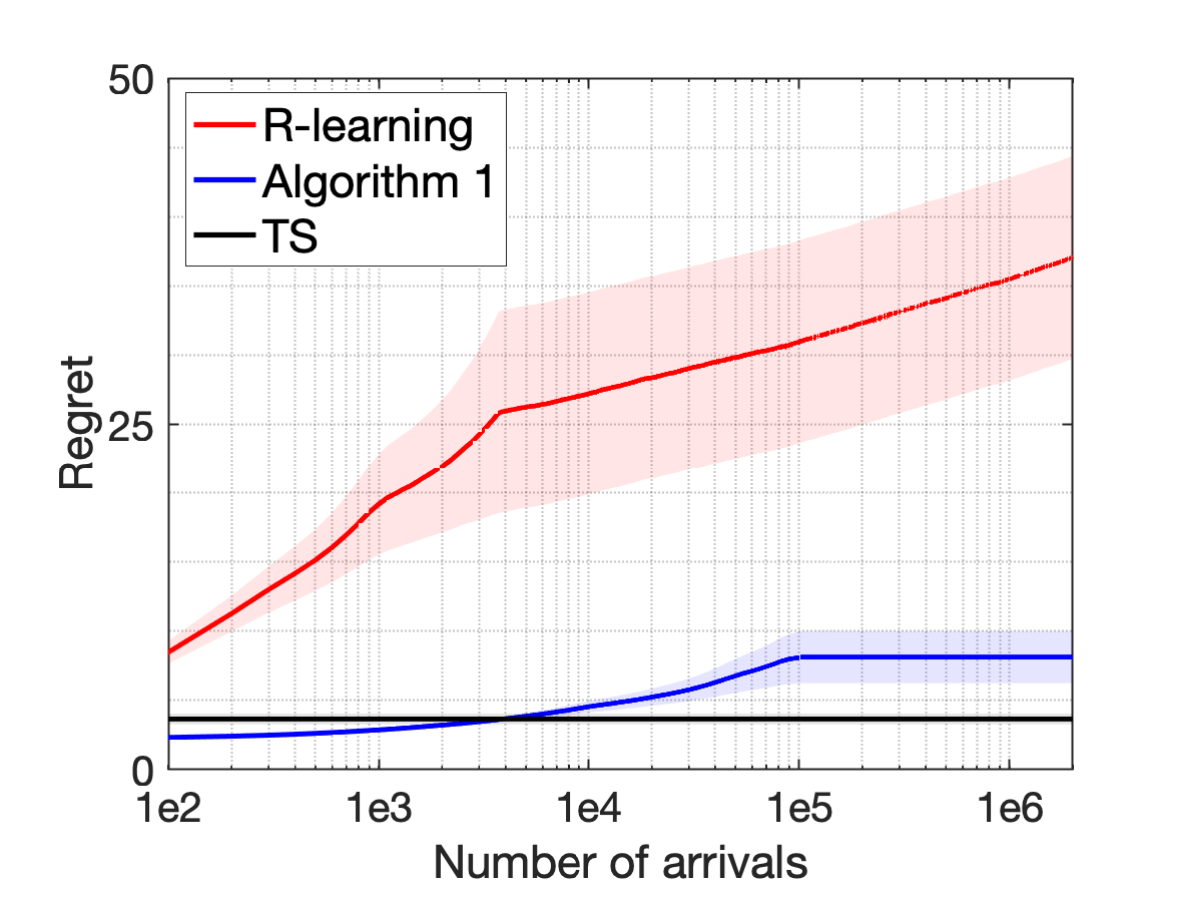}} \hspace{1cm}
	\subfloat[$\mu=1.05, \mu \in (0, c/R)$  \label{fig:compare_RL_b}]{\includegraphics[width=0.4\linewidth]{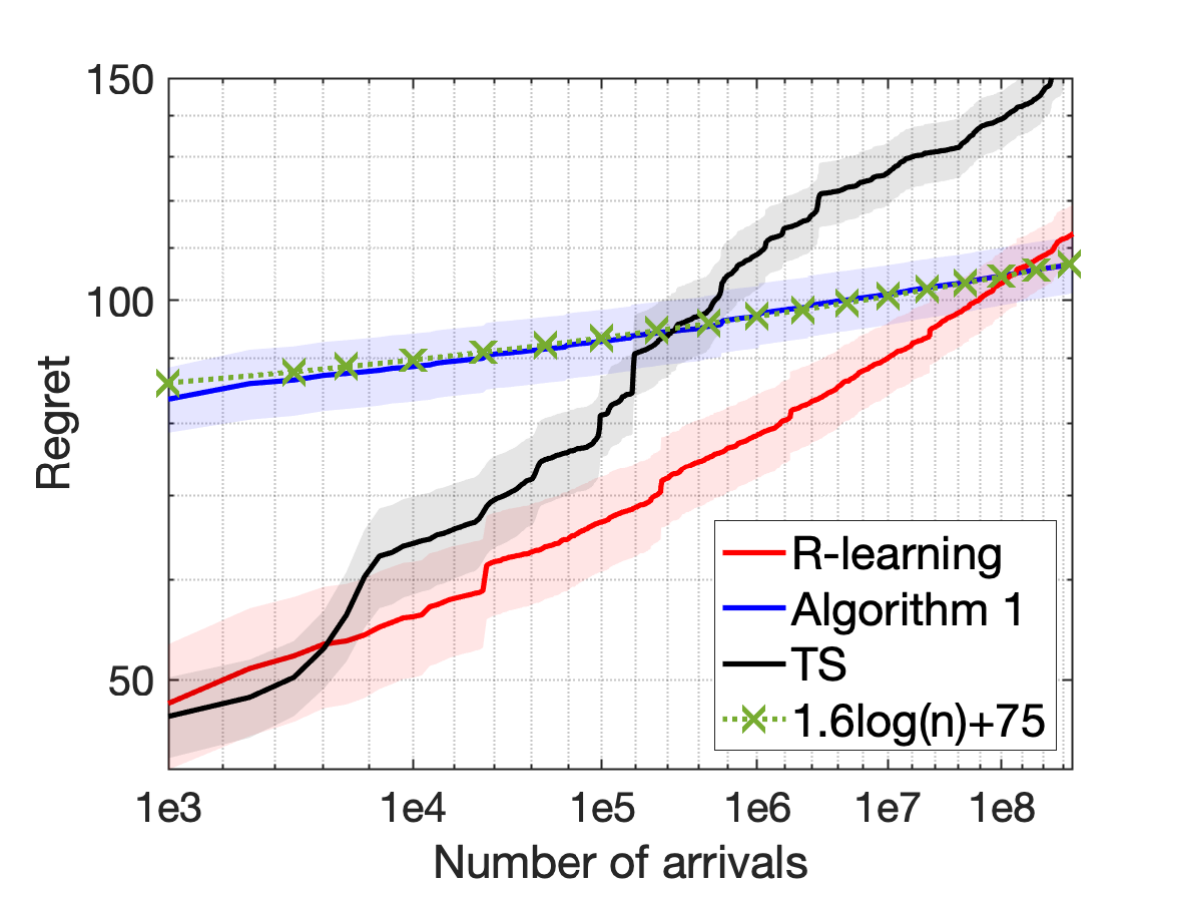}}
	\caption{Comparison of regret performance of \Cref{alg1} against RL algorithms in a 5 server system with $\lambda=5$, $c/R=1.3$, $\varepsilon=0.2$, and $f(n)=\exp\left(n^{1-\epsilon_n}\right)$.}  
	\label{fig:compare_RL}
	\end{figure*}

In \Cref{fig:compare_RL}, we compare the performance of \Cref{alg1} with two other algorithms: R-learning (\cite{sutton2018reinforcement}) and Thompson sampling~(\cite{gopalan2015thompson}). We consider a system with $k=5$, $\lambda=5$, and $c/R=1.3$. We also assume $f(n)=\exp\left(n^{1-\epsilon_n}\right)$ with $\epsilon_n=\frac{\varepsilon}{\sqrt{1+\log (n+1)}}$ and $\varepsilon=0.2$.  As noted in \Cref{sec:introduction}, the R-learning algorithm assumes that the service times are known ahead of the time when an arrival is accepted.   
 Despite not observing the service times, \Cref{fig:compare_RL} depicts that \Cref{alg1} outperforms R-learning in both regimes. Furthermore, empirically R-learning seems to have growing regret in both regimes. To implement the Thompson sampling algorithm, we use a uniform prior distribution defined on the two-point support $\{\mu_1,\mu_2\}$, where $\mu_1=\frac{c}{2R}<\frac{c}{R}$ and $\mu_2=\frac{3c}{2R}>\frac{c}{R}$, and update the posterior using $\eqref{eq:likelihood}$ upon every arrival.
As shown in \Cref{fig:compare_RL_a}, when $\mu>c/R$, the  Thompson sampling algorithm has a better final regret value compared to our algorithm, but both algorithms have constant regret. However, 
when $\mu < c/R$, \Cref{alg1} outperforms Thompson sampling; empirically, the asymptotic behavior of regret of both algorithms seem similar. We end by noting that theoretical analysis characterizing the regret performance for R-learning and Thompson sampling algorithms is not available in the literature.

\begin{figure*}[t] 
	\centering
	\subfloat[$\mu=2.2, \mu \in (c/R,+\infty)$ \label{fig:compare_en_a}]{\includegraphics[width=0.4\linewidth]{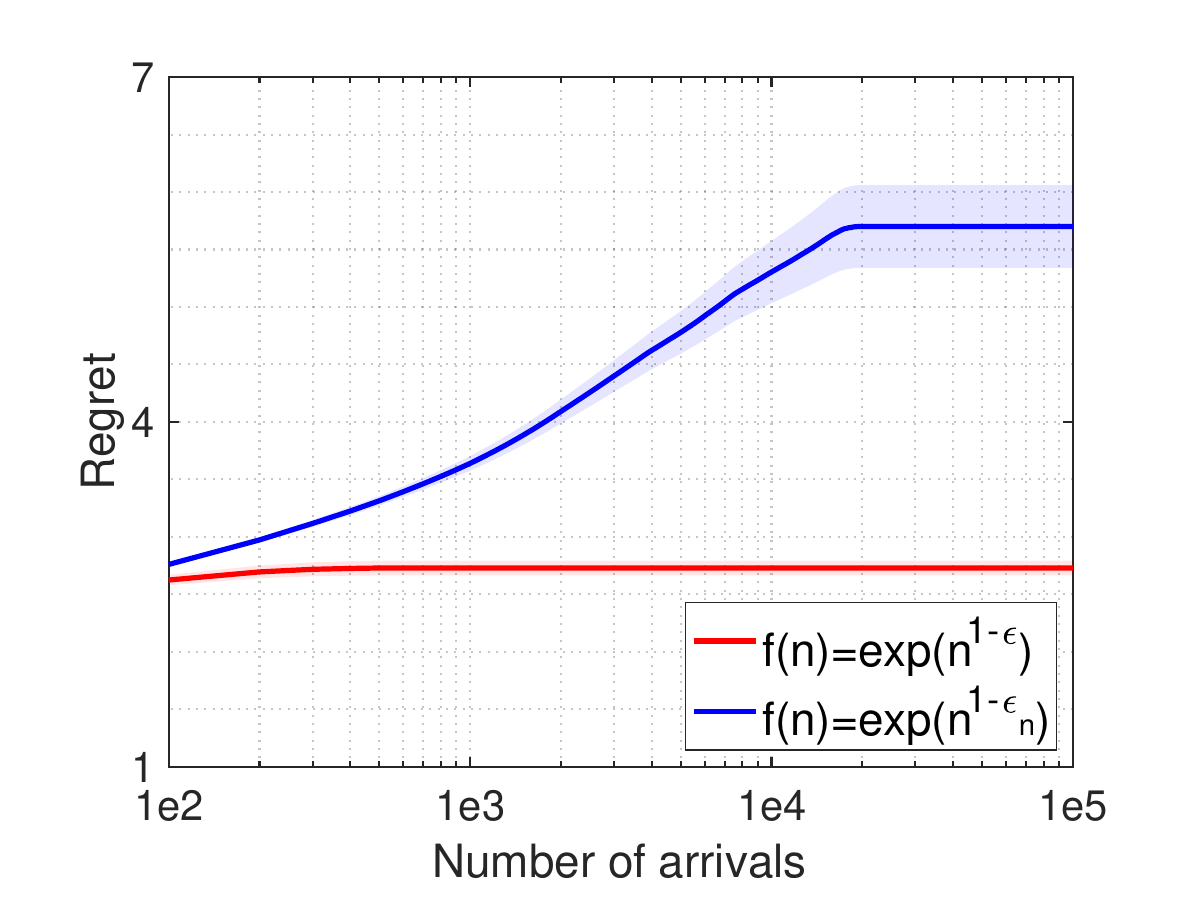}} \hspace{1cm}
	\subfloat[$\mu=1.1, \mu \in (0, c/R)$  \label{fig:compare_en_b}]{\includegraphics[width=0.4\linewidth]{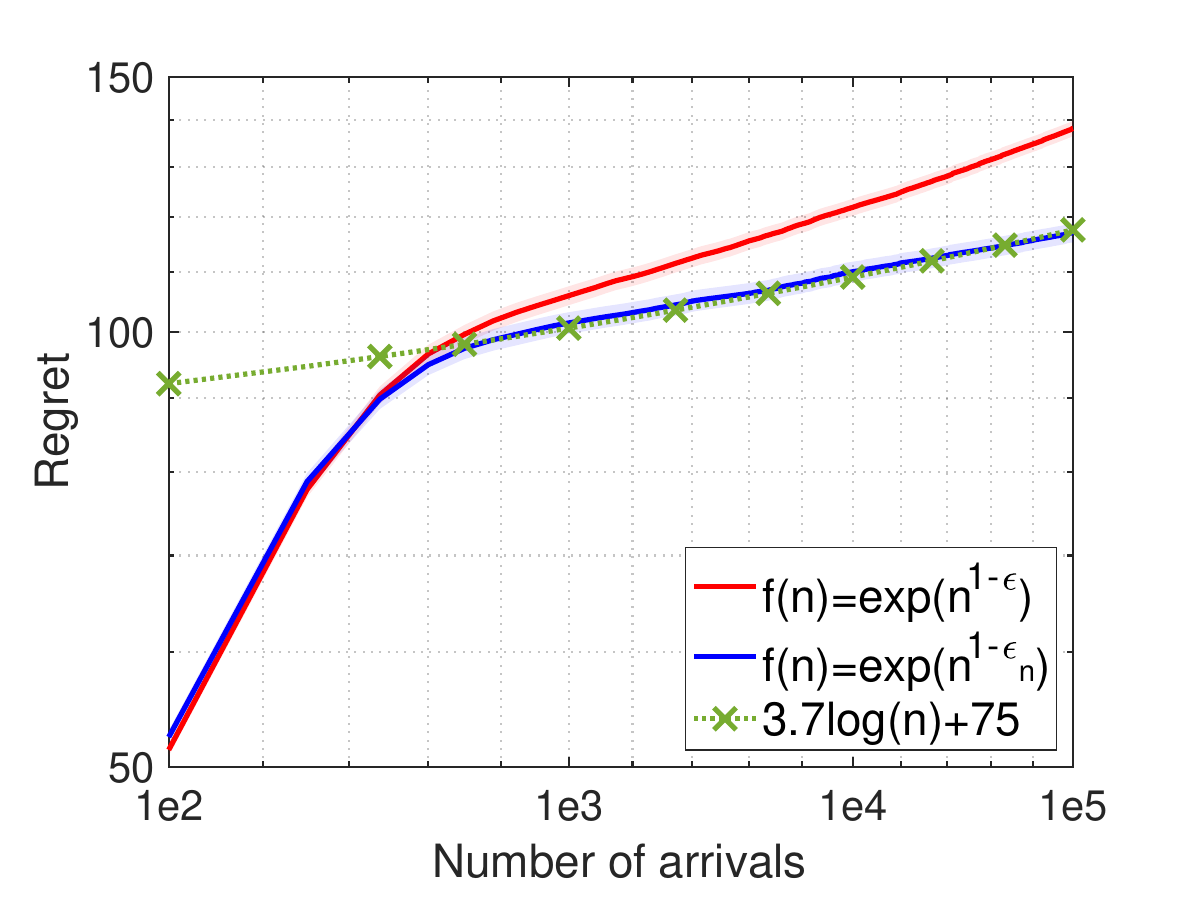}}
	\caption{Comparison of regret performance of \Cref{alg1} for different functions $f(n)$ in a 5 server system with $\lambda=5$, $c/R=1.3$, and $\epsilon=\varepsilon=0.55$.} 
	\label{fig:compare_en}
	\end{figure*}

In \Cref{fig:compare_en}, we compare the performance of \Cref{alg1} in a $5-$server system with $\lambda=5$ and $c/R=1.3$ for two different exploration functions $f(n)=\exp\left(n^{1-\epsilon}\right)$ and $f(n)=\exp\left(n^{1-\epsilon_n}\right)$, where $\epsilon_n=\frac{\varepsilon}{\sqrt{1+\log (n+1)}}$ and $\epsilon=\varepsilon=0.55$. 
In \Cref{cor:regret_mu_less_c_multi}, employing $f(n)=\exp\left(n^{1-\epsilon_n}\right)$ allows us to improve the order of the expected regret from $O(\log^\frac{1}{1-\epsilon} (n))$ to $O(\log (n))$.
This improvement is shown in the numerical results of \Cref{fig:compare_en_b}.
Since $\epsilon_n$ decreases with $n$, the arrival acceptance due to exploration decreases faster, leading to slightly inferior performance when $\mu>c/R$, as shown in \Cref{fig:compare_en_a}.

		\begin{figure*}[t]
	\centering
	\subfloat[$\mu=2.6, \mu \in (c/R,+\infty)$  \label{fig:sample_a}]{\includegraphics[width=0.4\linewidth]{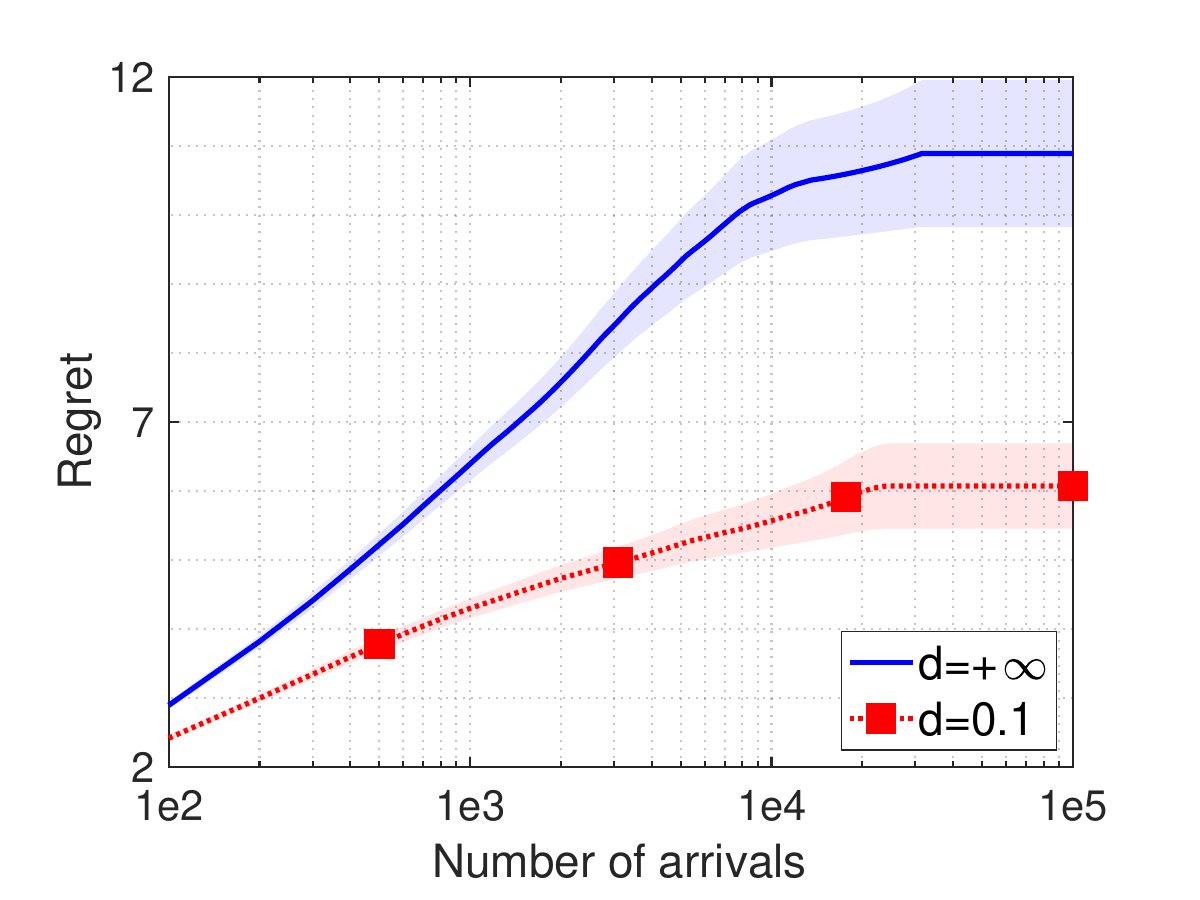}}\hspace{1cm}
	\subfloat[$\mu=0.6, \mu \in (0, c/R)$   \label{fig:sample_b}]{\includegraphics[width=0.4\linewidth]{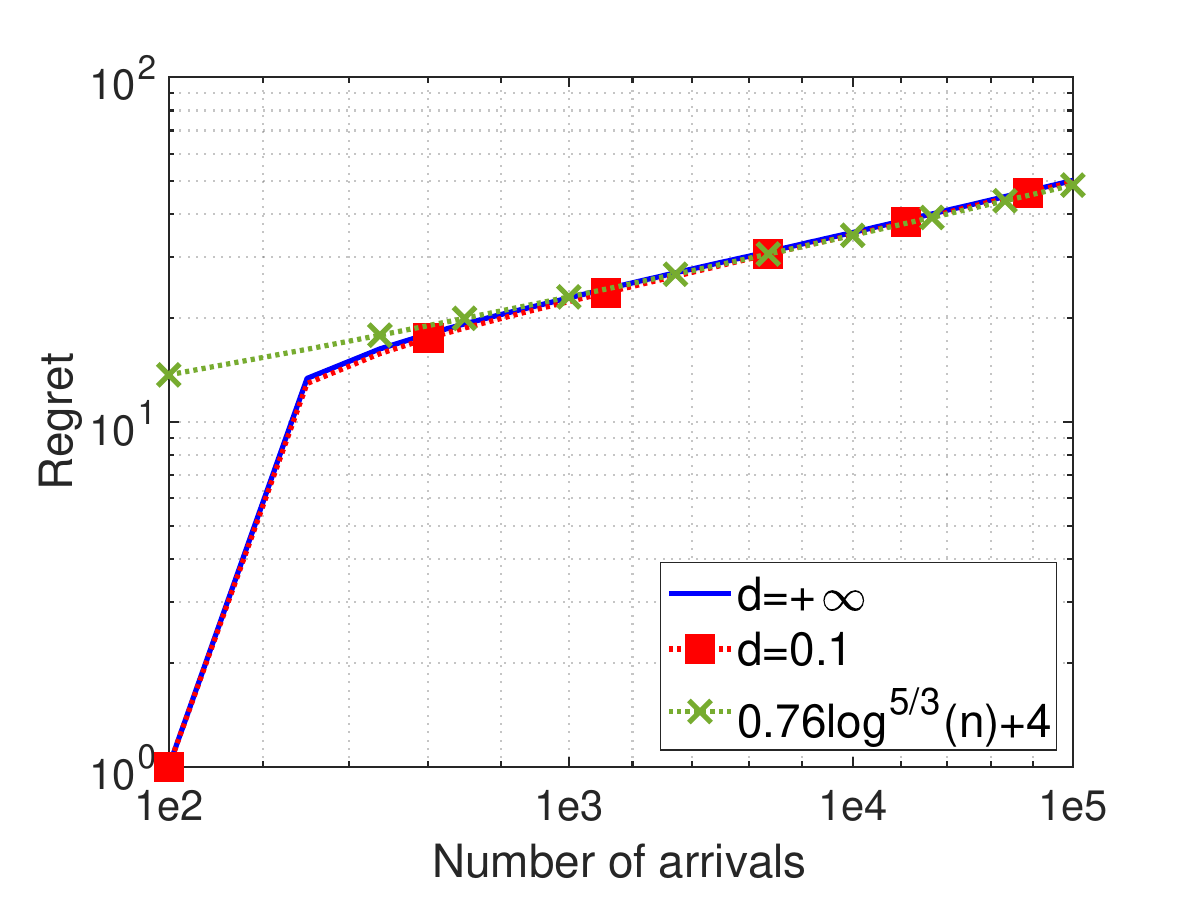}}
	\caption{Regret performance 
	for different sampling durations in a 2 server system with $\lambda=2$, $c/R=1.5$, $\epsilon=0.4$, $\tfrac{1}{1-\epsilon}=\tfrac{5}{3}$, and $f(n)=\exp\left(n^{1-\epsilon}\right)$.} 
	 \label{fig:sample}
	\end{figure*}

We next discuss a variant of our setting in which we can sample the system at other instances rather than only 
at the arrivals. One feasible approach is to  
modify the learning process 
as follows. Set a fixed sampling duration $d$. At each sampling time $t$, update functions $g$ and $h$  and the admittance probability accordingly.
From any sampling time $t$, if an arrival occurs before $d$ units of time, sample the system at the arrival and decide admission according to updated parameters. Otherwise, if $d$ units of time pass without an arrival, sample the system at $t+d$. After a new sampling is done, repeat the previous steps.
Note that (as a rule of thumb) for sampling to contribute to the learning, sampling duration $d$ should be less than  $1/\lambda$; setting $d=+\infty$ corresponds to policy $\pi_{\mathrm{Alg1}}$.   In \Cref{fig:sample}, in a $2-$server system with $\lambda=2$, $c/R=1.5$, $f(n)=\exp\left(n^{1-\epsilon}\right)$, and $\epsilon=0.4$, we depict the performance of the sampling scheme.
When $\mu>\lambda$, the performance of \Cref{alg1} can be improved by sampling; see \Cref{fig:sample_a}. However, as shown in \Cref{fig:sample_b}, 
when sampling according to the arrival rate is fast enough, performance does not improve with additional sampling. Moreover, \Cref{fig:sample} suggests that an adaptive sampling scheme might achieve the best trade-off.

\begin{figure*}[t]
    \centering
    	\subfloat[$k=N=5, \mu=2.2, \rho=0.455 \in (0,\rho^*)$  \label{fig:new_high_a}]{\includegraphics[width=0.4\linewidth]{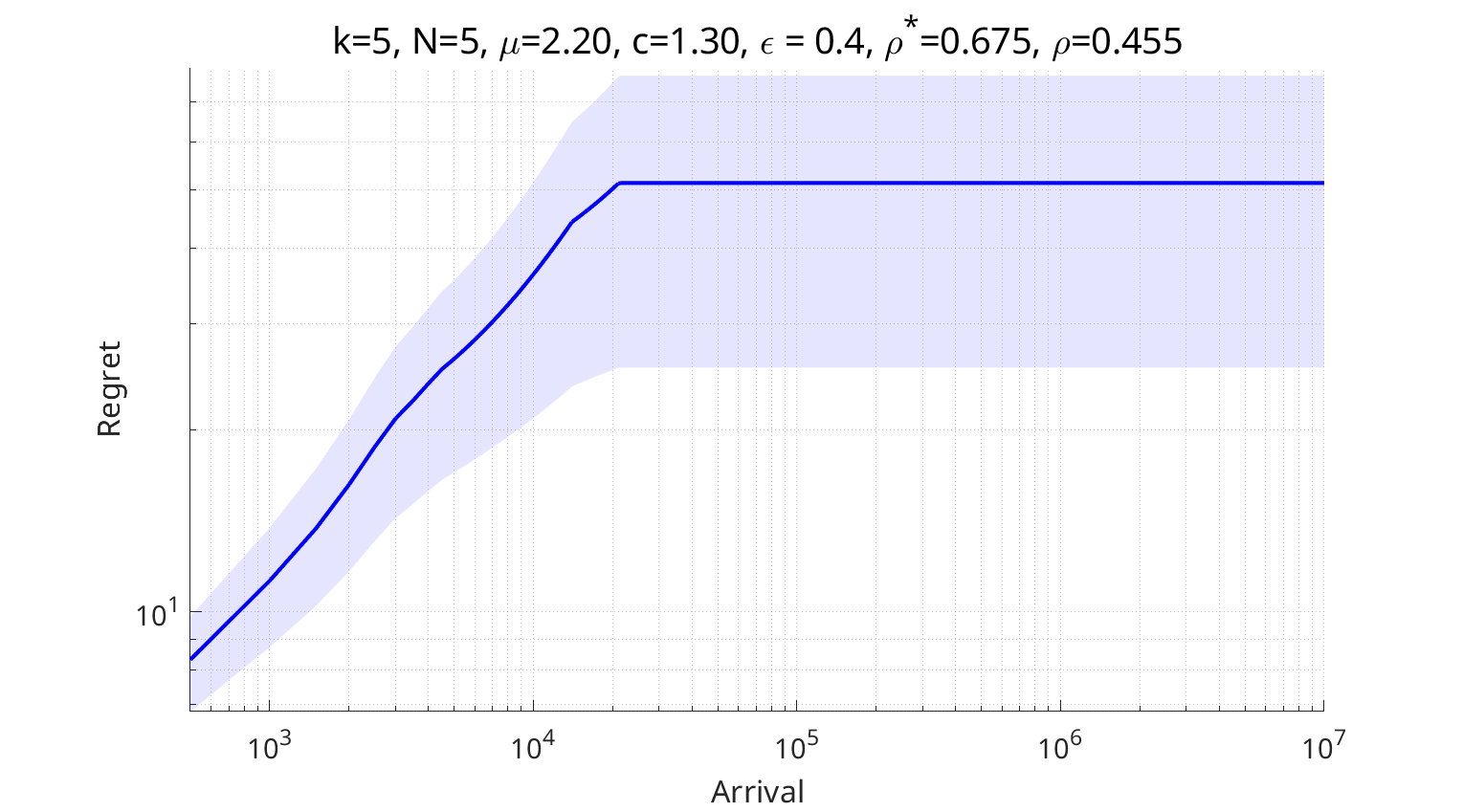}}\hspace{1cm}
	\subfloat[$k=N=5, \mu=2, \rho = 0.5 \in (0, \rho^*)$   \label{fig:new_high_b}]{\includegraphics[width=0.4\linewidth]{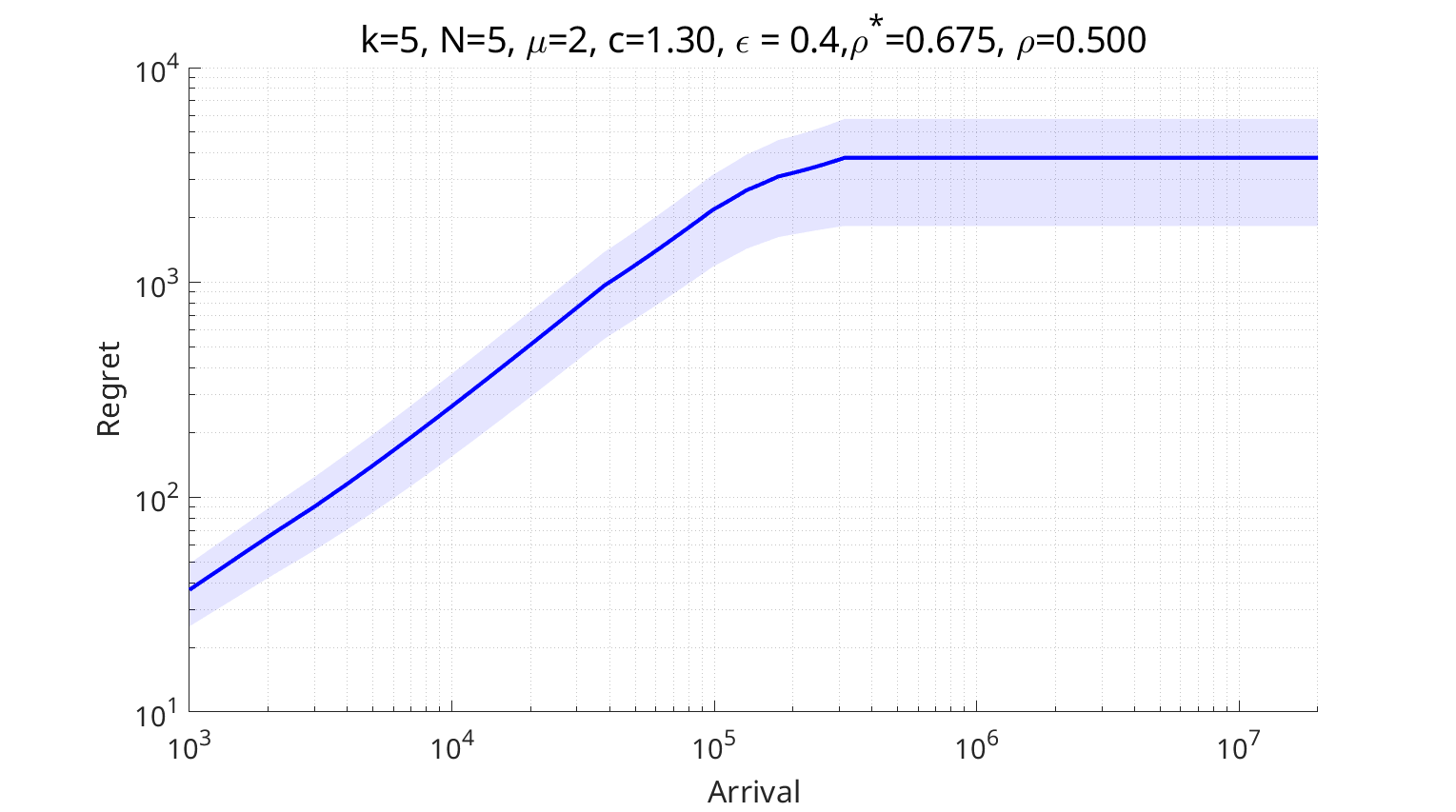}}
	\caption{{\color{blue}Regret performance versus the number of arrivals
	for two different service rates---both with $\rho$ not exceeding $\rho^*$---, and for a $k=5$ server system with $N=5$ spaces of waiting room.}} 
	 \label{fig:new_high}
\end{figure*}

\begin{figure*}[t]
    \centering
    	\subfloat[$k=N=5, \mu=1.25, \rho=0.8 \in (\rho^*,+\infty)$  \label{fig:new_low_a}]{\includegraphics[width=0.4\linewidth,height=1.5in]{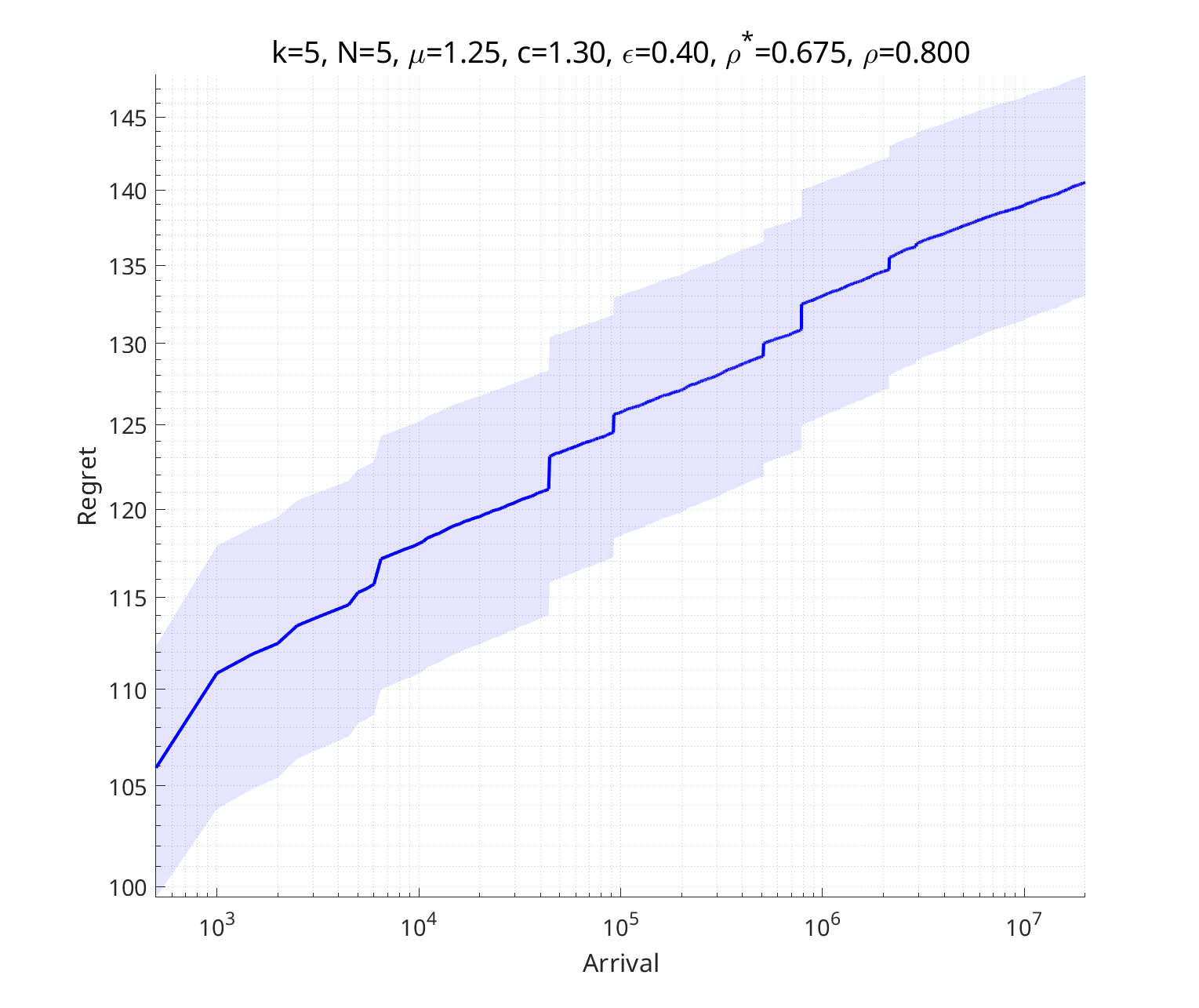}}\hspace{1cm}
	\subfloat[$k=N=5, \mu=1.05, \rho=0.953 \in (\rho^*, +\infty)$   \label{fig:new_low_b}]{\includegraphics[width=0.4\linewidth,height=1.5in]{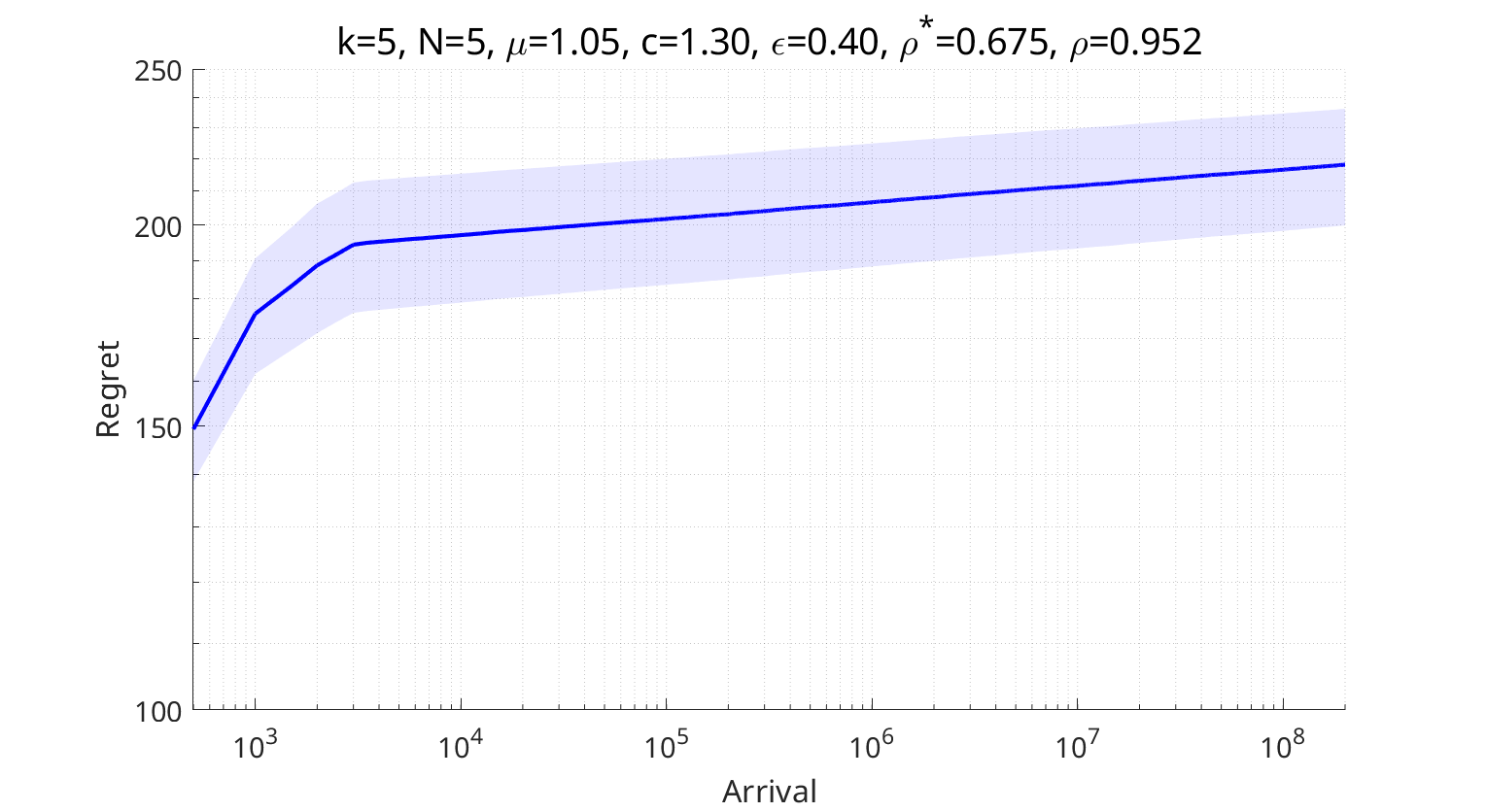}}
	\caption{{\color{blue} Regret performance versus the number of arrivals
	for two different service rates---both with $\rho$ not exceeding $\rho^*$---, and for a $k=5$ server system with $N=5$ spaces of waiting room.}}
	 \label{fig:new_low}
\end{figure*}

{\color{blue} Finally, in \Cref{fig:new_high} and \Cref{fig:new_low}, we provide simulation results obtained by running our algorithms for systems with a finite, non-zero waiting room---both figures are with $5$ servers and waiting room $N=5$ places. As with the no-waiting room case, for the Erlang-B systems, for high service rates, the regret is constant---see \Cref{fig:new_high} (\Cref{fig:new_high_a} and \Cref{fig:new_high_b})---, and for low service rates, the regret grows logarithmically in the number of arrivals---see \Cref{fig:new_low} (\Cref{fig:new_low_a} and \Cref{fig:new_low_b}).}

\section{Conclusions and Future Work}	\label{sec:discussion-and-conclusion}

In conclusion, we studied the problem of learning-based optimal admission control of an {\color{blue} $M/M/k/k+N$ service system  with unknown service rate where $N\geq 0$ and finite (the Erlang-B blocking system is when $N=0$).} 
We showed that the extreme contrast in the optimal control schemes in different parameter regimes---quickly converging to always admitting arrivals if room versus quickly rejecting all arrivals---makes learning challenging. With the system being sampled only at arrivals, we designed a dispatching policy based on 
ML estimation of the unknown service rate
followed by using the certainty equivalent law with forced exploration. We proved asymptotic optimality of our policy, 
and established finite-time guarantees for {\color{blue} specific parameter settings: constant regret when $\mu > c/R$ for $N=0$ or $\rho^* > \rho$ for $N>0$, and logarithmic regret when $\mu < c/R$ for $N=0$ or $\rho^* < \rho$ for $N>0$}. Through simulations, we also showed that our policy achieves a good trade-off of the regret over all parameter regimes.

We plan to study the following in future work. First, 
we proved a $\log(n)$ upper bound for the regret when $\mu <c/R$. 
One direction is to explore lower bounds in this regime; 
we conjecture that {\color{blue}the tight lower bound} is $\Omega(\log(n))$. {\color{blue} We expect this to be true based on our experimental results and the fact that it is consistent with the lower bound on the asymptotic growth of the regret found in the literature \cite{agrawal1989certainty,agrawal1989asymptotically,borkar1979adaptive,gopalan2015thompson,graves1997asymptotically,kumar1982optimal,kumar2015stochastic,lai1995machine,mandl1974estimation} on learning in unknown stochastic dynamic systems under the assumption that the transition kernels of the underlying controlled Markov chains are strictly bounded away from $0$.}

Another future research direction is to allow for different sampling and update 
schemes (including by an independent Poisson process) and theoretically analyze the regret. Yet another direction is to extend our results to other service-time distributions, as the optimal admission control policy is unchanged due to the insensitivity~(\cite{kelly2011reversibility,srikant2013communication}) of the Erlang-B system. {\color{blue} However, generalizing to the $M/G/k/k$ system will need new ideas as the current sampling at arrivals paradigm results in a partially observed/hidden Markov process---the hidden state is the remaining service time of each customer in service---for non-exponential service times instead of a fully observed Markov process as in the case of exponential service times. Owing to this (optimal) admission control questions are known to be hard with the sampling procedure/information structure that we have considered---see \cite{oz2022optimal} for a recent discussion of this.} 

\begin{APPENDICES} 

\section{Analysis of the Single-server Erlang-B Queueing System}
\label{app:Appendix_single_server}

\subsection{\texorpdfstring{\Cref{lem:infinite_acceptance}}{Lemma 1}}
\label{sec:proof_lem_inf_acc}

\begin{Lemma} \label{lem:infinite_acceptance}
In a single-server Erlang-B queueing system, the number of accepted arrivals following policy $\pi_{\mathrm{Alg1}}$  is almost surely infinite.
\end{Lemma}
\proof{Proof of \Cref{lem:infinite_acceptance}.}
Let $A$ be the event that the system stops accepting new arrivals after some finite arrival, $A_1$ the event that the server is always busy after some finite arrival, $A_2$ the event that the server is available after some finite arrival but rejects all subsequent arrivals according to Line \ref{prob_line} of \Cref{alg1}, and $A_{2,m}$ as the event that for the first time at arrival $m$, the server is available but rejects all arrivals. We have 
\begin{equation} \label{MC_welldefined}
\Pr \left( A \right)=\Pr \left( A_1 \right)+\Pr \left( A_2 \right)=\Pr \left( A_2 \right)
= \sum_{m=0}^\infty \Pr \left(  A_{2,m} \right)
\leq  \sum_{m=0}^\infty \lim_{n \to +\infty}  \Big( 1- \frac{1}{f(m)} \Big)^n =0,   
\end{equation}
where the inequality follows from the fact that for $n \geq m$, we have $\alpha_n=\alpha_{m} \leq m$,
which means the acceptance probability is fixed after arrival $m$, as no other arrivals are accepted. From \eqref{MC_welldefined}, we conclude that almost surely an infinite number of arrivals are accepted following \Cref{alg1}.  \Halmos
\endproof

\subsection{Proof of \texorpdfstring{\Cref{lem:geom_log_up_bound}}{Lemma 6}}
\label{sec:proof_geom_log_up}

\proof{Proof of \Cref{lem:geom_log_up_bound}.}
We first bound the probability term $\Pr\big(\sum_{j=1}^i y_j < n , \sum_{j=1}^{i+1} y_j  \geq n\big)$ using the probability of the first event. We take $p_i=1-q_i=\exp\left(-i^{1-\epsilon}\right)$ and then use the Chernoff bound to get
	\begin{align} \label{chernoff}
	\Pr\left(y_1+\cdots+y_i < n , y_1+\cdots+y_{i+1} \geq n\right)\leq \Pr\left(y_1+\cdots+y_i\leq n  \right) \leq   \min_{t \geq0} e^{tn} \prod_{j=1}^{i} \frac{p_j}{e^t-\left(1-p_j\right)}.
	\end{align}
	Take $b= \ceil{ \left(\log \left(n+1\right)\right)^{\frac{1}{1-\epsilon}}}$ and $t \geq 0$ such that $e^t= \frac{n+1}{n}q_i.$
	From \eqref{chernoff}, for $i \geq d \geq b$ we have
	\begin{align} 
	\Pr\left(y_1+\cdots+y_i\leq n  \right)  &\leq   \left(\frac{n+1}{n}\right)^n q_i^n \prod_{j=1}^{i} \frac{p_j}{\frac{1}{n}\left(1-p_i\right)+\left(p_j-p_i\right)} \nonumber \leq \left(\frac{n+1}{n}\right)^n q_i^n \prod_{j=1}^{i} p_j \prod_{j=1}^{b} \frac{1}{p_j-p_i} \prod_{j=b+1}^{i} \frac{n}{1-p_i} \nonumber \\
	& \leq \left(\frac{n+1}{n}\right)^n q_i^{n-\left(i-b\right)} n^{i-b} \prod_{j=b+1}^{i} p_j \prod_{j=1}^{b} \frac{1}{1-\exp\left(-\left(i^{1-\epsilon}-j^{1-\epsilon}\right)\right)}.  \label{chern_upp}
	\end{align}
	Since $q_i \leq 1$ and $n\geq i-b$, we have $\left(\frac{n+1}{n}\right)^n q_i^{n-\left(i-b\right)} \leq e.$ 
	By concavity and gradient inequality, for $1 \leq j \leq i$,  we have $	i^{1-\epsilon}-j^{1-\epsilon} \geq \frac{1-\epsilon}{i^\epsilon}  \left(i-j\right)$. Using this inequality and setting  $\kappa:=\ceil {{i^\epsilon}/{(1-\epsilon)}}$, we have
	\begin{align} 
	\prod_{j=1}^{b} \frac{1}{1-\exp\left(-\left(i^{1-\epsilon}-j^{1-\epsilon}\right)\right)} 
 &\leq \prod_{j=1}^{b} \frac{1}{1-\exp\left(-\frac{1-\epsilon}{i^\epsilon}  \left(i-j\right) \right)} 
 \leq \prod_{t=1}^{\infty} \frac{1}{1-\exp\left(-\left(\frac{1-\epsilon}{i^\epsilon} \right) t \right)} \nonumber  \\
 &\leq  \prod_{t=1}^{\kappa-1} \frac{1}{1-\exp\left(-\left(\frac{1-\epsilon}{i^\epsilon} \right) t \right)} \prod_{t=\kappa}^{\infty} \frac{1}{1-\exp\left(-\frac{1}{\kappa} t  \right)} \nonumber \\ 
	&\leq  \prod_{t=1}^{\kappa-1} \frac{1}{1-\exp\left(-\left(\frac{1-\epsilon}{i^\epsilon} \right) t \right)} \prod_{j=1}^{\infty} \prod_{t=j\kappa}^{\left(j+1\right)\kappa-1} \frac{1}{1-\exp\left(-\frac{1}{\kappa} t  \right)} \nonumber  \\
	&\leq  \prod_{t=1}^{\kappa-1} \frac{1}{1-\exp\left(-\left(\frac{1-\epsilon}{i^\epsilon} \right) t \right)} \prod_{j=1}^{\infty}  \left(\frac{1}{1-\exp\left(-j \right)}\right)^\kappa \nonumber
 \leq (c_u)^\kappa   \prod_{t=1}^{\kappa-1} \frac{1}{1-\exp\left(-\left(\frac{1-\epsilon}{i^\epsilon} \right) t \right)}.  
	\end{align}
	The last inequality is true as follows. For $a_j=\left(\exp\left(j\right)-1\right)^{-1}$, using the fact that $1+x\leq \exp(x)$, we have 
	\begin{align*}
	    \prod_{j=1}^{\infty} \frac{1}{1-\exp\left(-j \right)}=\prod_{j=1}^{\infty} \left( 1+a_j \right) \leq \exp \Big(  \sum_{j=1}^\infty a_j  \Big)=c_u,
	\end{align*}
	For $1 \leq t \leq \kappa-1$, we have $\frac{1-\epsilon}{i^\epsilon} t \leq \frac{1-\epsilon}{i^\epsilon} \left(\kappa -1\right) <1 ,$
	and $1-\exp\left(-x\right) \geq x/2$ for $x \leq 1$. Therefore, we can write $1-\exp\big(-\big(\frac{1-\epsilon}{i^\epsilon} \big) t \big) \geq \frac{1}{2} \frac{1-\epsilon}{i^\epsilon}t.$
 As a result, we can further simplify the second product term in \eqref{chern_upp} as follows,
	\begin{equation} \label{final_upp}
	\prod_{j=1}^{b} \frac{1}{1-\exp\left(-\frac{1-\epsilon}{i^\epsilon}  \left(i-j\right) \right)} 
	\leq (c_u)^\kappa  \prod_{t=1}^{\kappa-1} 2\frac{i^\epsilon}{\left(1-\epsilon\right)t} 
	\leq (c_u)^\kappa  2^{\kappa-1} \frac{1}{\left(\kappa-1\right)!} \left(\frac{i^\epsilon}{1-\epsilon} \right)^{\kappa-1} . 
	\end{equation}  
 For $x>0$ and $k \in \mathbb{N}$, $x^k/k! \leq \exp(x)$. Thus, $\frac{ ec_u^\kappa  2^{\kappa-1}}{\left(\kappa-1\right)!\left(1-\epsilon\right)^{\kappa-1}} \leq e c_u \exp(\frac{2c_u}{1-\epsilon})=:c_e$, which is an $\epsilon-$dependent constant. Next we upper bound  the term $ \prod_{j=b+1}^{i} p_j $ using integral lower bound as below:
	\begin{equation} \label{sum}
	\left(b+1\right)^{1-\epsilon}+\ldots+i^{1-\epsilon} \geq \frac{1}{2-\epsilon}\left(i^{2-\epsilon}-b^{2-\epsilon}\right).
	\end{equation}
    Thus, using the above discussion, we simplify \eqref{chern_upp} to get
	\begin{equation} \label{final_upper_prob}
	\Pr\left(y_1+\cdots+y_i\leq n  \right)\leq c_e \exp\Big(-\frac{1}{2-\epsilon}\left(i^{2-\epsilon}-b^{2-\epsilon}\right)\Big)
	n^{i-b} i^{\epsilon\left(\kappa-1\right)}.
	\end{equation}
	We upper bound the summation given in the statement of \Cref{lem:geom_log_up_bound}. From \eqref{final_upper_prob} and using the fact that $d\geq b$, 
	\begin{align*} 
	\sum_{i=d}^{n}i\Pr\left(y_1+...+y_i\leq n\right)  &\leq c_e\sum_{i=d}^{n}i \exp\left(-\frac{1}{2-\epsilon}\left(i^{2-\epsilon}-b^{2-\epsilon}\right)\right)
	(n+1)^{i-b} i^{\epsilon\left(\kappa-1\right)}  \nonumber  \\
	& \leq c_e (n+1)^{-b}\exp\left(\frac{b^{2-\epsilon}}{2-\epsilon}\right) \sum_{i=d}^{\infty}i \exp \left(-\frac{i^{2-\epsilon}}{2-\epsilon}+i\log(n+1)+\frac{\epsilon}{1-\epsilon}\log\left(i\right)i^\epsilon\right) \nonumber\\
	&\leq \tilde{c}_e \exp \left( -b \log(n+1)+\frac{b^{2-\epsilon}}{2-\epsilon} \right) 
\nonumber\\ 
&\leq \tilde{c}_e \exp \Big( -b (b-1)^{1-\epsilon}+\frac{b^{2-\epsilon}}{2-\epsilon} \Big)=\tilde{c}_e \exp \bigg( -b^{2-\epsilon} \bigg( \bigg(1-\frac{1}{b}\bigg)^{1-\epsilon} - \frac{1}{2-\epsilon} \bigg)\bigg),  
	\end{align*}
	where we have used $b= \ceil{ (\log^{\frac{1}{1-\epsilon}} \left(n+1\right))}$ in the last line. The third inequality holds as for $i \geq d $, the negative term inside the second exponential function is dominating. Further, as $n$ grows, $b$ converges to infinity; hence, in the final term, the exponential term converges to zero. Thus, we can bound the sum with a constant. 
\Halmos
\endproof

\subsection{Proof of \texorpdfstring{\Cref{cor:regret_mu_less_c_multi}}{Corollary 1}}
\label{sec:proof_cor_regret_mu_less_c/R}

\proof{Proof.}
 We follow the same arguments as in \Cref{thm:regret_mu<c/R_multi} to show a $O(\log(n))$ regret. As a parallel to \Cref{lem:geom_log_up_bound},  we  bound 
 $\sum_{i=\Tilde d}^{n-1}i\Pr\big(\sum_{j=1}^i y_j < n , \sum_{j=1}^{i+1} y_j  \geq n\big)$ for independent geometric random variables $\{y_i\}_{i=1}^n$  with success probability $\{f(i )^{-1}\}_{i=1}^n$ following similar arguments to  \Cref{lem:geom_log_up_bound}.  Denote the smallest $i $ that satisfies $i^{1-{\epsilon_i}} \geq \log(n+1)$ as $b$ and let $\Tilde d$ be the smallest integer $i$ such that $\log (n+1) \leq \frac{1}{3} i^{1-\epsilon_{b+1}}$.
We note that $i^{1-{\epsilon_i}}$ is increasing for  $i\geq 1$ as $\epsilon_i$ is a decreasing sequence.
Take $p_i=\exp\left(-i^{1-\epsilon_i}\right)$ and $t \geq 0$ such that $e^t= \frac{n+1}{n}(1-p_i),$ which exists for $i>b$.
	From \eqref{chern_upp}, for $i > b$,
	\begin{equation} 
		\Pr\left(y_1+\cdots+y_i\leq n  \right)  
		 \leq e n^{i-b} \prod_{j=b+1}^{i} p_j \prod_{j=1}^{b} \frac{1}{1-\exp\left(-\left(i^{1-\epsilon_i}-j^{1-\epsilon_j}\right)\right)}.  \label{chern_upp_cor}
	\end{equation}
	Moreover, for $1 \leq j \leq i$, by concavity and gradient inequality,  we have $\epsilon_j \geq \epsilon_i$ and 
	\begin{equation} \label{eq:bound_fcn_cor}
		i^{1-\epsilon_i}-j^{1-\epsilon_j} \geq 	i^{1-\epsilon_i}-j^{1-\epsilon_i} \geq \frac{1-\epsilon_i}{i^{\epsilon_i}}  \left(i-j\right).
	\end{equation} 
	We define $\kappa=\ceil {{i^{\epsilon_i}}/{(1-\epsilon_i)}}$ and using  \eqref{final_upp}, simplify the second product term in the RHS  of \eqref{chern_upp_cor}  to get
	\begin{equation} \label{final_upp_cor}
		 \prod_{j=1}^{b} \frac{1}{1-\exp\left(-\left(i^{1-\epsilon_i}-j^{1-\epsilon_j}\right)\right)} \leq \prod_{j=1}^{b} \frac{1}{1-\exp\left(-\frac{1-\epsilon_i}{i^{\epsilon_i}}  \left(i-j\right) \right)}
		\leq c_u^\kappa  2^{\kappa-1} \frac{1}{\left(\kappa-1\right)!} \left(\frac{i^{\epsilon_i}}{1-\epsilon_i} \right)^{\kappa-1} . 
	\end{equation}  
	Furthermore, using an integral lower bound, we find an upper bound for the term $ \prod_{j=b+1}^{i} p_j $:
	\begin{equation} \label{sum_cor}
		\left(b+1\right)^{1-\epsilon_{b+1}}+\ldots+i^{1-\epsilon_i} \geq	\left(b+1\right)^{1-\epsilon_{b+1}}+\ldots+i^{1-\epsilon_{b+1}} \geq  \frac{1}{2-\epsilon_{b+1}}\left(i^{2-\epsilon_{b+1}}-b^{2-\epsilon_{b+1}}\right).
	\end{equation}
	Using \eqref{final_upp_cor}, \eqref{sum_cor}, and the fact that $\frac{ ec_u^\kappa  2^{\kappa-1}}{\left(\kappa-1\right)!\left(1-\epsilon_i\right)^{\kappa-1}} \leq e c_u \exp(\frac{2c_u}{1-\varepsilon})=:c_e$, we simplify \eqref{chern_upp_cor} to get
	\begin{equation} \label{final_upper_prob_cor}
		\Pr\left(y_1+\cdots+y_i\leq n  \right)\leq c_e \exp\Big(- \frac{1}{2-\epsilon_{b+1}}\left(i^{2-\epsilon_{b+1}}-b^{2-\epsilon_{b+1}}\right)\Big)
		n^{i-b} i^{\epsilon_i\left(\kappa-1\right)}.
	\end{equation}
	Finally, we can bound $\sum_{i=d}^{n-1}i\Pr\left(y_1+\cdots+y_i <n , y_1+\cdots+y_{i+1} \geq n\right)$ using \eqref{final_upper_prob_cor} as follows
	\begin{align} 
		\sum_{i=\Tilde d}^{n}i\Pr(y_1+...+y_i\leq n) 
		&\leq c_e 	(n+1)^{-b}\exp\Big(\frac{b^{2-\epsilon_{b+1}}}{2-\epsilon_{b+1}}\Big) \sum_{i=\Tilde d}^{\infty}i \exp \Big(\frac{-i^{2-\epsilon_{b+1}}}{2-\epsilon_{b+1}}+i\log(n+1)+\frac{\epsilon_i}{1-\epsilon_i}\log(i)i^{\epsilon_i}\Big) \nonumber\\
   &\leq c_e (n+1)^{-b}\exp\Big(\frac{b^{2-\epsilon_{b+1}}}{2-\epsilon_{b+1}}\Big) \sum_{i=\Tilde d}^{\infty}i \exp \Big(\frac{-i^{2-\epsilon_{b+1}}}{2-\epsilon_{b+1}}+\frac{i^{2-\epsilon_{b+1}}}{3}+\frac{\epsilon_i}{1-\epsilon_i}\log(i)i^{\epsilon_i}\Big) \nonumber \\
		& \leq c_e (n+1)^{-b}\exp\Big(\frac{b^{2-\epsilon_{b+1}}}{2-\epsilon_{b+1}}\Big), \label{final_term_cor}
	\end{align}
where the second line follows from  $\log 	(n+1) \leq \frac{1}{3}(\Tilde  d)^{1-\epsilon_{b+1}}\leq \frac{1}{3} i^{1-\epsilon_{b+1}}$ for $i \geq \Tilde d$.
As the negative term inside the second exponential function is the dominating term, we can bound the summation with a constant independent of $n$.  From the definition of $b$, we have $(b-1)^{1-\epsilon_{b-1}} < \log(n+1) \leq b^{1-\epsilon_b}$. Thus
\begin{align}
(n+1)^{-b}\exp\Big(\frac{b^{2-\epsilon_{b+1}}}{2-\epsilon_{b+1}}\Big) &=\exp\Big(b\Big(\frac{b^{1-\epsilon_{b+1}}}{2-\epsilon_{b+1}}-\log (n+1)\Big)\Big) \leq \exp\Big(b\Big(\frac{b^{1-\epsilon_{b+1}}}{2-\epsilon_{b+1}}-(b-1)^{1-\epsilon_{b-1}} \Big)\Big)\nonumber\\
& = \exp\Big(- b^{2-\epsilon_{b+1}}\Big({b^{\epsilon_{b+1}-\epsilon_{b-1}}}\Big(1-\frac{1}{b}\Big)^{1-\epsilon_{b-1}}-\frac{1}{2-\epsilon_{b+1}} \Big)\Big). \label{eq:rhs_bnd}
\end{align}
We note that as $b$ grows to infinity, the term $\big(1-\frac{1}{b}\big)^{1-\epsilon_{b-1}}$ converges to $1$, and the term $b^{2-\epsilon_{b+1}}$ converges to $\infty$. Since $\epsilon_{b+1}<\epsilon_{b-1}$, the term $b^{\epsilon_{b+1}-\epsilon_{b-1}}$ is less than $1$. However, we also note that for large enough $b$,
\begin{align*}
  1> b^{\epsilon_{b+1}-\epsilon_{b-1}} &= b^{\frac{\varepsilon}{\sqrt{1+\log(b+2)}}-\frac{\varepsilon}{\sqrt{1+\log (b)}}} 
     = \exp\Bigg(  \frac{\varepsilon \log(b)}{\sqrt{1+\log(b+2)}}-\frac{\varepsilon \log(b)}{\sqrt{1+\log(b) }}\Bigg) \\
    & >\exp(\sqrt{\log(b+2)-1 }-\sqrt{\log(b)+1}\big),
\end{align*} which follows from $\varepsilon<1$ and $\left(\log\left(b\right)\right)^2>\left(\log\left(b+2\right)\right)^2-1$ for sufficiently large $b$ (since $\left(\log\left(b+2\right)\right)^2-\left(\log\left(b \right)\right)^2$ converges to 0 as $b$ grows). 
Thus, $b^{\epsilon_{b+1}-\epsilon_{b-1}}$ converges to $1$ as $b$ increases without bound.
Using all of these, we can assert that the RHS of \eqref{eq:rhs_bnd} goes to $0$ as $b$ increases to infinity, and so we can bound it by a constant independent of $n$. 
 Finally, by repeating the arguments of \Cref{thm:regret_mu<c/R_multi}, the expected regret is upper bounded by a linear function of $\Tilde d$ and we conclude that the expected regret is of the order $O(\log (n))$. 
 \Halmos \endproof
\section{Analysis of the Multi-server Erlang-B Queueing System} 
\label{Appendix_sec_5}

\subsection{\texorpdfstring{\Cref{lem:Sampled_MC_wd_m}}{Lemma 7}}
\label{proof_MC_multi}

\begin{Lemma} \label{lem:Sampled_MC_wd_m}
In a multi-server Erlang-B queueing system following policy $\pi_{\mathrm{Alg1}}$, the number of accepted arrivals that find the system empty  is almost surely infinite.
\end{Lemma}

\proof{Proof.} 
 By observing Markov process $\{\tilde{X}_n\}_{n=0}^\infty$, we first argue that  the system becomes empty infinitely often following our proposed policy. By coupling the two systems, we get
 \begin{align*}
     &\Pr \left(\text{returns to state 0 at a finite time} \Bc N_{n}=0,X_{n}=x,\alpha_{n}=\alpha \right)\\ 
     & \qquad \geq 
     \Pr \left(\text{returns to state 0 at a finite time in a system that accepts all arrivals} \Bc N_{n}=0 \right)=1.
 \end{align*}
Thus,  state 0 is visited infinitely often. Let $A$ be the event that the system admits a finite number of arrivals at instances when the server is empty, $A_1$ be the event that the system admits a finite number of arrivals, 
and $A_2$ be the event that the system gets empty a finite number of times. We have $\Pr \left( A \right)\leq \Pr \left( A_1 \right)+\Pr \left( A_2 \right)=0, $
wherein $\Pr \left( A_1 \right)=0$ follows from  the same arguments as \Cref{lem:infinite_acceptance}. 
\Halmos
\endproof

	\subsection{\texorpdfstring{\Cref{lem:multi_server_prob}}{Lemma 13}}
\label{sec:proof_multi_server_prob}
 We first present the following lemma, which is used in the proof of \Cref{lem:multi_server_prob}.

\begin{Lemma}\cite[Theorem 2.19 ]{wainwright2019high}   \label{lemma:MDS}
	let $\{\left(D_i,\mathcal{F}_i\right)\}_{i=1}^\infty$ be a martingale difference sequence such that for $\nu_i,\alpha_i>0$, we have  $\mathbb{E}\big[\exp (\tilde{\lambda} D_i)\Bc \mathcal{F}_{i-1}\big]\leq \exp \big(\frac{\tilde{\lambda}^2\nu_i^2}{2}\big) $
	\emph{a.s.} for any $|\tilde{\lambda}|<1/{\alpha_i}$. Then the sum $\sum_{i=1}^{n}D_i$ satisfies the concentration inequality
\begin{equation*}
    \Pr \Big(\Big|\sum_{i=1}^{n}D_i\Big| \geq t\Big)\leq 2\exp \Big(-\min \Big( \frac{t^2}{2\sum_{i=1}^{n}\nu_i^2},\frac{t}{2\max\limits_{i=1,\ldots,n} \alpha_{i}}\Big)\Big).
\end{equation*}
\end{Lemma}

\proof{Proof of \Cref{lem:multi_server_prob}.}
Without loss of generality, we assume $\mu>c/R$. 
 	Note that $\tilde{\delta}_1$ and $\tilde{\delta}_2$ are as defined in \Cref{lem:incr_seq}.
	We define the martingale difference sequence $\{Y_n^D\}_{n=0}^\infty$ as $Y_n^D=Y_{n+1}^M-Y_{n}^M.$
	To verify the conditions of \Cref{lemma:MDS}, we argue that $\mathbb{E}\big[\exp (\tilde{\lambda} \left|Y_i^D\right|)\Bc \mathcal{F}_{i-1}\big]$ is bounded for some positive $\tilde{\lambda}$. We show this by proving $\mathbb{E}\big[\exp (\tilde{\lambda} Y_i^D)\Bc \mathcal{F}_{i-1}\big]$ and $\mathbb{E}\big[\exp (-\tilde{\lambda} Y_i^D)\Bc \mathcal{F}_{i-1}\big]$ are bounded for some positive $\tilde{\lambda}$. From 
 \eqref{bounds}, 
\begin{align}
\mathbb{E}\big[\exp (\tilde{\lambda} Y_i^D)\Bc \mathcal{F}_{i-1}\big]
\leq \mathbb{E}\big[\exp \big(\tilde{\lambda}k\frac{R}{c} \tau_i \big)\Bc \mathcal{F}_{i-1}\big] \label{sub-exp_pos_lambda_1}
\leq \mathbb{E}\big[\exp \big(\tilde{\lambda}k\frac{R}{c} \zeta_i \big)\big],
\end{align}
where $\zeta_i$ is the first passage time of state zero starting from zero in a finite-state irreducible Markov chain, and thus, sub-exponential. From \cite[Theorem 2.8.2]{vershynin2018high}, the moment generating function of $\zeta_i$ is bounded at some $\tilde{\lambda}_1$ independent of $i$, which leads to a finite bound. 
For $\mathbb{E}\big[\exp (-\tilde{\lambda} Y_i^D)\Bc \mathcal{F}_{i-1}\big]$, using \eqref{bounds}, 
	\begin{equation*} \label{sub-exp}
	\mathbb{E}\big[\exp (-\tilde{\lambda} Y_i^D)\Bc \mathcal{F}_{i-1}\big] \leq \mathbb{E}\Big[\exp \Big(\tilde{\lambda}  \Big(k\sum_{j=1}^{\tau_i} T_{\beta_i+j} +c_{\tilde{\delta}} \Big)\Big)\Bc \mathcal{F}_{i-1}\Big]  \leq 
	\mathbb{E}\Big[\exp \Big(\tilde{\lambda}  \Big(k\sum_{j=1}^{\zeta_i} T_{\beta_i+j} +c_{\tilde{\delta}} \Big)\Big)\Big]. 
	\end{equation*}  
	From the above inequality, it suffices to show $\sum_{j=1}^{\zeta_i} T_{\beta_i+j} $ is sub-exponential. From \cite[Theorem 2.8.2]{vershynin2018high}, we need to argue that for some positive $\tilde{\lambda}$, 
$	\mathbb{E}\Big[\exp \Big(\tilde{\lambda} \sum_{j=1}^{\zeta_i} T_{\beta_i+j} \Big)\Big] \leq 2.$	
	For $\tilde{\lambda}<\lambda$, we define the martingale sequence $\{M_{i,m}\}_{m=0}^\infty$ with respect to filtration $\{\mathcal{G}_{i,m}\}_{m=0}^{\infty}$ as
	\begin{align*}
	M_{i,m}=\frac{\exp \left(\tilde{\lambda} \sum_{j=1}^{m}T_{\beta_i+j}\right)}{\mathbb{E}\left[\exp \left(\tilde{\lambda} \sum_{j=1}^{m}T_{\beta_i+j}\right)\right]} =\frac{\exp \left(\tilde{\lambda} \sum_{j=1}^{m}T_{\beta_i+j}\right)}{\left(\frac{\lambda}{\lambda-\tilde{\lambda}}\right)^m}.
	\end{align*}
 The passage time $\zeta_i$ is a finite-mean stopping time for the martingale sequence $\{M_{i,m}\}_{m=0}^\infty$. Therefore, using the optional stopping theorem for non-negative supermartingale sequences,  we have  $\E\left[M_{i,\zeta_i}\right]\leq\E\left[M_{i,0}\right],$
	or $	\mathbb{E} \big[{\exp \big(\tilde{\lambda} \sum_{j=1}^{\zeta_i}T_{\beta_i+j}\big)\left(\frac{\lambda}{\lambda-\tilde{\lambda}}\right)^{-\zeta_i}}\big]\leq1.$
	Using the Cauchy-Bunyakovsky-Schwarz inequality, we have 
\begin{equation} \label{subexp_tau}
    \mathbb{E} \big[\exp\big(\frac{\tilde{\lambda}}{2}\sum_{j=1}^{\zeta_i}T_{\beta_i+j}\big)\big] \leq \sqrt{\mathbb{E}\big[ \big(\frac{\lambda}{\lambda-\tilde{\lambda}}\big)^{\zeta_i}\big]}
    = \sqrt{\mathbb{E} \big[\exp\big(\log \big(\frac{\lambda}{\lambda-\tilde{\lambda}}\big)\zeta_i\big)\big] } .
\end{equation}
	As $\zeta_i$ is a sub-exponential random variable, we can choose $ \tilde{\lambda}$ such that the RHS of \eqref{subexp_tau} is less than or equal to $2$ 
 and the  conditions of \Cref{lemma:MDS} are verified. Consequently, we apply \Cref{lemma:MDS} to conclude that 
 \begin{align*}
	\Pr\left(Y_n^M \leq -\tilde{\delta}_1n\right)=\Pr\Big(\sum_{i=0}^{n-1} \left(Y_{i+1}^M-Y_i^M \right) \leq -\tilde{\delta}_1n\Big)  \leq \exp \Big(-\min\Big(\frac{\tilde{\delta}_1^2 n^2}{2nv^2},\frac{\tilde{\delta}_1n}{2\alpha}\Big)\Big)=\exp\left(-c_3 n\right),
	\end{align*}  where  $\nu$ and $\alpha$ are positive constants independent of $n$.
\Halmos \endproof

\end{APPENDICES}

\bibliographystyle{informs2014} 
\bibliography{Bibliography} 

\end{document}